\documentclass[prd,nofootinbib]{revtex4}
\usepackage{graphicx}
\usepackage{amsmath,amssymb}
\usepackage{hyperref}
\usepackage[usenames]{color}
\usepackage{url}
\usepackage{epstopdf}
\usepackage{xcolor}

\makeatletter
\def\mathcolor#1#{\@mathcolor{#1}}
\def\@mathcolor#1#2#3{%
  \protect\leavevmode
  \begingroup
    \color#1{#2}#3%
  \endgroup
}
\makeatother

\hypersetup{
    colorlinks=true,
    linkcolor=black,
    citecolor=blue,
    urlcolor=black
}

\def\be{\begin{equation}}
\def\ee{\end{equation}}
\def\ba{\begin{eqnarray}}
\def\ea{\end{eqnarray}}
\def\nn{\nonumber}
\def\f{\frac}
\def\l{\left}
\def\r{\right}
\def\hub{{\cal H}}

\allowdisplaybreaks[3]

\begin{document}
\title{An Extended action for the effective field theory of dark energy: a stability analysis and a complete guide to the mapping at the basis of EFTCAMB}

\author{Noemi Frusciante$^{1}$, Georgios Papadomanolakis$^{2}$ and  Alessandra Silvestri$^{2}$}

\smallskip
\affiliation{$^{1}$ Sorbonne Universit$\acute{\text{e}}$s, UPMC Univ Paris 6 et CNRS, UMR 7095, Institut d'Astrophysique de Paris, GReCO, 98 bis bd Arago, 75014 Paris, France \\
\smallskip 
$^{2}$ Institute Lorentz, Leiden University, PO Box 9506, Leiden 2300 RA, The Netherlands}

\begin{abstract}
\vspace{0.3cm}
We present a generalization of the effective field theory (EFT) formalism for dark energy and modified gravity models to  include operators with higher order spatial derivatives. This allows the extension of the EFT framework  to a wider class of gravity theories such as Ho\v rava gravity. We present the corresponding extended action, both in the EFT and the Arnowitt-Deser-Misner (ADM) formalism, and proceed to work out a convenient mapping between the two, providing a self contained and general procedure to translate  a given model of gravity into the EFT language at the basis of the Einstein-Boltzmann solver EFTCAMB. Putting this mapping at work, we illustrate, for several interesting models of dark energy and modified gravity, how to express them in the ADM notation and then map them into the EFT formalism.  We also provide for the first time, the full mapping of GLPV models into the EFT framework. 
We next perform a thorough analysis of the physical stability of the generalized EFT action, in absence of matter components. We work out  viability conditions that correspond to the absence of ghosts and modes that propagate with a negative speed of sound  in the scalar and tensor sector, as well as the absence of tachyonic modes in the scalar sector.
Finally, we extend and generalize the phenomenological basis in terms of $\alpha$-functions introduced to parametrize Horndeski models, to cover all theories with higher order spatial derivatives included in our extended action. We elaborate on the impact of the additional functions on physical quantities, such as the kinetic term and the speeds of propagation for scalar and tensor modes.  
\end{abstract}

\maketitle

\tableofcontents
\section{Introduction}

The long standing problem of   cosmic acceleration, the spread of new theories of gravity and the unprecedented possibility to test them against cosmological data, in the past years have led to the search for a unifying framework  to describe  deviations from General Relativity (GR)~\cite{Sotiriou:2008rp,Silvestri:2009hh,DeFelice:2010aj,Clifton:2011jh,Tsujikawa:2013fta,Deffayet:2013lga,Joyce:2014kja, Koyama:2015vza,Bull:2015stt} on cosmological scales. An interesting proposal,  the effective field theory (EFT) of dark energy and modified gravity (DE/MG)~\cite{Gubitosi:2012hu,Bloomfield:2012ff,Gleyzes:2013ooa,Bloomfield:2013efa,Piazza:2013coa,Frusciante:2013zop,Gleyzes:2014rba,Perenon:2015sla}, was formulated recently, inspired by the EFT  of inflation, quintessence~\cite{Creminelli:2006xe,Cheung:2007st,Weinberg:2008hq,Creminelli:2008wc}  and large scale structure~\cite{Park:2010cw,Jimenez:2011nn,Carrasco:2012cv,Hertzberg:2012qn,Carrasco:2013mua,Porto:2013qua,Senatore:2014vja}. It represents a model independent  framework to describe the evolution of linear cosmological perturbations in all theories of gravity which introduce  an extra scalar degree of freedom (DoF) and have a well defined Jordan frame. Such framework is formulated at the level of the action, which is built in unitary gauge out of all operators that are invariant under the reduced symmetries of the system, i.e. time-dependent spatial diffeomorphisms, and are at most quadratic in perturbations around a Friedmann-Lemaitre-Robertson-Walker (FLRW) Universe. The outcome not only offers a model independent setup, but also a powerful unifying language, since most of the candidate models of DE/MG can be exactly mapped into the EFT language. The latter include  quintessence~\cite{Tsujikawa:2013fta}, f(R) gravity~\cite{DeFelice:2010aj}, Horndeski/Generalized Galileon (hereafter GG)~\cite{Horndeski:1974wa,Deffayet:2009mn}, Gleyzes-Langlois-Piazza-Vernizzi theories (GLPV)~\cite{Gleyzes:2014dya}, low-energy Ho\v rava gravity~\cite{Horava:2008ih,Horava:2009uw}. 

A powerful bridge between theory and the observational side has further been offered by the implementation of the EFT of DE/MG into the Einstein-Boltzmann solver CAMB/CosmoMC~\cite{CAMB,Lewis:1999bs,Lewis:2002ah}, which resulted in the publicly available patches EFTCAMB/EFTCosmoMC~\cite{Hu:2013twa,Raveri:2014cka,Hu:2014sea,Hu:2014oga,Frusciante:2015maa}  (\url{http://wwwhome.lorentz.leidenuniv.nl/~hu/codes/}).  The resulting solver,  evolves the full dynamics of linear scalar and tensor perturbations  without resorting to any approximation, such as the common quasi-static one. The equations are implemented in the EFT language, offering a powerful unifying setup. As a result, with the same code and hence same accuracy, the user can investigate both model independent departures from GR, as well as explore the dynamics in specific models, after they are mapped in the EFT language.  Many models of gravity are built-in in the most recent version of EFTCAMB, which, interestingly, allows also the use of parametrization alternatives to the EFT one, such as the parametrization in terms of $\alpha$-functions proposed in Ref.~\cite{Bellini:2014fua} to describe the Horndeski/GG models, which hereafter we will refer to as ReParametrized Horndeski (RPH). Let us notice that the latter has also been implemented in CLASS~\cite{Lesgourgues:2011re}, resulting in HiCLASS~\cite{Bellini:2015xja}. As discussed below, part of this paper is devoted to the extension of this basis. Let us conclude this brief overview of EFTCAMB, by noticing that an important feature is  the built-in  set of stability conditions that guarantee that the underlying theory of gravity explored at any time is viable. Since EFT of DE/MG is formulated at the level of the action, it is indeed possible  to identify powerful yet general conditions of theoretical viability; the latter are consequently enforced as theoretical \textit{priors} when using EFTCosmoMC, optimizing the exploration of the parameter space. 
Part of this paper is devoted, as we will describe, to the extension and generalization of such conditions. 

In the present work we propose an extension of the original EFT action for DE/MG~\cite{Gubitosi:2012hu,Bloomfield:2012ff}  by including extra operators with up to sixth order spatial derivatives acting on perturbations. This will allow us to cover a wider range of theories, e.g. Ho\v rava gravity~\cite{Horava:2008ih,Horava:2009uw}, as shown in Refs.~\cite{Kase:2014cwa,Gao:2014soa,Frusciante:2015maa}. The latter model has recently gained attention in the cosmological context~\cite{Calcagni:2009ar,Kiritsis:2009sh,Brandenberger:2009yt,Mukohyama:2009gg,Cai:2009dx,Chen:2009jr,Cai:2010hi,Carroll:2004ai,Zuntz:2008zz,Gao:2009ht,Wang:2009yz,Kobayashi:2009hh,Dutta:2009jn,Kobayashi:2010eh,Dutta:2010jh,Mukohyama:2010xz,Blas:2012vn,Audren:2013dwa,Audren:2014hza,Frusciante:2015maa}, as well as in the quantum gravity sector~\cite{Horava:2008ih,Horava:2009uw,Sotiriou:2010wn,Visser:2011mf,Barvinsky:2015kil}, since higher spatial derivatives have been shown to be relevant in building gravity models  exhibiting powercounting and renormalizable behaviour  in the ultra-violet regime~(UV)~\cite{Visser:2009fg,Visser:2009ys,Blas:2009qj}. 

We will  work out a very general recipe that can be directly applied to any gravity theory with one extra scalar DoF in order to efficiently map it into the EFT language,  once the corresponding Lagrangian  is written in the Arnowitt-Deser-Misner (ADM) formalism. We will pay particular attention to the different conventions by adapting all the calculations to the specific convention used in EFTCAMB, in order to provide a ready-to-use guide on the full mapping of models into this code. Such method has been already used in Refs.~\cite{Gleyzes:2013ooa,Kase:2014cwa} and here we will further extend it by including  the extra operators in our extended action. Additionally we will revisit some of the already known mappings in order to accommodate the EFTCAMB conventions.  Moreover, we will present for the first time the complete mapping of the covariant formulation of the GLPV theories~\cite{Gleyzes:2014dya,Gleyzes:2014qga} into the EFT formalism. 
Interestingly, we will  perform a detailed study of the stability conditions for the gravity sector of our extended EFT action. Stability analysis  for a restricted subset of EFT models can already be found in the literature~\cite{Gubitosi:2012hu,Bloomfield:2012ff,Gleyzes:2013ooa,Piazza:2013pua,Gleyzes:2014qga}.  This analysis will allow us to have a first glimpse at the viable parameter space of theories covered by the extended EFT framework and to obtain very general conditions to be implement in EFTCAMB.  In particular, we will compute the conditions necessary to avoid ghost instabilities and to guarantee a positive (squared) speed of propagation for scalar and tensor modes. We will also present the condition to avoid tachyonic instabilities in the scalar sector.   
Finally, we will  proceed to extend the RPH basis of Ref.~\cite{Bellini:2014fua} in order to include all the models of our generalized EFT action, which results in the definition of new functions. Finally, we will comment on the impact of these functions on the kinetic term and speeds of propagation of both scalar and tensor modes.  

In details, the paper is organized as follows. In Section~\ref{extendedEFT}, we propose a generalization of the EFT action for DE/MG that  includes all  operators with up to six-th order spatial derivatives. In Section~\ref{Sec:fromgeneraltoEFT}, we outline a general procedure to map any theory of gravity with one extra scalar DoF, and a well defined Jordan frame, into the EFT formalism. We achieve this through an interesting, intermediate step which consists of deriving  an equivalent action in the ADM formalism, in Section~\ref{EFTinADM},  and work out the mapping between the EFT and ADM formalism, in Section~\ref{Sec:mapping}. In order to illustrate the power of such method, in Section~\ref{Sec:examples} we provide some mapping examples: minimally coupled quintessence, f(R)-theory, Horndeski/GG, GLPV and Ho\v rava gravity.  In Section~\ref{Sec:stability}, we work out the physical stability conditions for the extended EFT action, guaranteeing the avoidance of ghost and tachyonic instabilities and positive speeds of propagation for tensor and scalar modes.
In Section~\ref{Sec:basis}, we extend the RPH basis to include the class of theories described by the generalized EFT action and we elaborate on the phenomenology associated to it.   The last two sections  are more or less independent, so the reader interested only in one of these can skip the other parts. Finally, in Section~\ref{Sec:conclusion}, we summarize and comment on our results.

\section{An extended EFT action}\label{extendedEFT}

The EFT framework for DE/MG models, introduced in Refs.~\cite{Gubitosi:2012hu,Bloomfield:2012ff}, provides a systematic and unified way to study the dynamics of linear perturbations in a wide range of DE/MG models characterized by an additional scalar DoF and for which there exists a well defined Jordan frame~\cite{Sotiriou:2008rp,DeFelice:2010aj,Clifton:2011jh,Tsujikawa:2013fta,Deffayet:2013lga,Koyama:2015vza}.  The  action is constructed in the unitary gauge as an expansion up to second order in perturbations around the FLRW background of  all operators that are invariant under time-dependent spatial-diffeomorphisms. Each of the latter appear in the action accompanied by a  time dependent coefficient. The choice of the unitary gauge implies that the scalar DoF is "eaten" by the metric, thus it does not appear explicitly in the action. It can be made explicit by  the  St$\ddot{\text{u}}$kelberg technique which, by means of an infinitesimal time-coordinate transformation,  allows one to restore the broken symmetry by introducing a new field describing the dynamic and evolution of the extra DoF. For a detailed description of this formalism we refer the readers to Refs.~\cite{Gubitosi:2012hu,Bloomfield:2012ff,Bloomfield:2013efa,Gleyzes:2013ooa,Gleyzes:2014rba}.   In this paper we will always work in the unitary gauge. 

The original EFT action introduced in Refs.~\cite{Gubitosi:2012hu,Bloomfield:2012ff}, and its follow ups in Refs.~\cite{Piazza:2013coa,Gleyzes:2013ooa,Gleyzes:2014rba,Perenon:2015sla}, cover most of the theories of cosmological interest, such as  Horndeski/GG~\cite{Horndeski:1974wa,Deffayet:2009mn}, GLPV~\cite{Gleyzes:2014dya} and low-energy Ho\v rava~\cite{Horava:2008ih,Horava:2009uw}. However, operators with higher order spatial derivatives are  not included. On the other hand, theories which exhibit higher than second order spatial derivatives in the field equations have been gaining attention in the cosmological context~\cite{Blas:2009qj,Kobayashi:2010eh,Piazza:2013coa,Kase:2014cwa,Gao:2014soa}, moreover,  they appear to be interesting models for quantum gravity as well~\cite{Horava:2008ih,Horava:2009uw,Visser:2009fg,Sotiriou:2010wn,Visser:2011mf,Barvinsky:2015kil}. As long as one deals with scales that are sufficiently larger than the non-linear cutoff, the  EFT formalism can be safely used to study these theories. In the following, we propose an extended EFT action that includes operators up to sixth order in spatial derivatives:
\begin{align}\label{EFTgeneral}
\mathcal{S}_{EFT}&=\int d^4x\sqrt{-g}\l[\frac{m_0^2}{2}(1+\Omega(t))R+\Lambda(t)-c(t)\delta g^{00}+\frac{M^4_2(t)}{2}(\delta g^{00})^2-\frac{\bar{M}^3_1(t)}{2}\delta g^{00}\delta K-\frac{\bar{M}^2_2(t)}{2}(\delta K)^2\nonumber\r.\\
&\l.-\frac{\bar{M}_3^2(t)}{2}\delta K^{\mu}_{\nu}\delta K^{\nu}_{\mu}+\f{\hat{M}^2(t)}{2}\delta g^{00}\delta\mathcal{R}+m^2_2(t)h^{\mu\nu}\partial_{\mu}g^{00}\partial_{\nu}g^{00}+\frac{\bar{m}_5(t)}{2}\delta\mathcal{R}\delta K+\lambda_1(t)(\delta\mathcal{R})^2\nonumber \r. \\
&\l.+\lambda_2(t)\delta\mathcal{R}^{\mu}_{\nu}\delta\mathcal{R}^{\nu}_{\mu} +\lambda_3(t)\delta\mathcal{R}h^{\mu\nu}\nabla_{\mu}\partial_{\nu}g^{00}+\lambda_4(t) h^{\mu\nu}\partial_{\mu}g^{00} \nabla^2 \partial_{\nu}g^{00}+\lambda_5(t)h^{\mu\nu}\nabla_{\mu}\mathcal{R}\nabla_{\nu}\mathcal{R}\r.\nonumber\\&
\l.+\lambda_6(t)h^{\mu\nu}\nabla_{\mu}\mathcal{R}_{ij}\nabla_{\nu}\mathcal{R}^{ij}+\lambda_7(t) h^{\mu\nu}\partial_{\mu} g^{00}\nabla^4 \partial_{\nu}g^{00}+\lambda_8(t)h^{\mu\nu}\nabla^2\mathcal{R}\nabla_{\mu}\partial_{\nu}g^{00}\r],
\end{align}
where $m_0^2$ is the Planck mass, $g$  is the determinant of the four dimensional metric $g_{\mu\nu}$, $h^{\mu\nu}=\l(g^{\mu\nu}+n^\mu n^\nu\r)$ is the spatial metric on constant-time hypersurfaces, $n_{\mu}$ is the normal vector to the constant-time hypersurfaces, $\delta g^{00}$ is the perturbation of the upper time-time component of the metric, $R$ is the trace of the four dimensional Ricci scalar,  $\mathcal{R}_{\mu\nu}$  is the three dimensional Ricci tensor and $\mathcal{R}$ is its trace, $K_{\mu\nu}$ is the extrinsic curvature and $K$ is its trace and $\nabla^2=\nabla_\mu\nabla^\mu$ with $\nabla_{\mu}$ being the covariant derivative constructed with $g_{\mu\nu}$. The coefficients $\{\Omega, \Lambda, c, M^4_2, \bar{M}^3_1,\bar{M}^2_2, \bar{M}_3^2, \hat{M}^2, m_2^2, \bar{m}_5, \lambda_i\}$ (with $i=1$ to $8$) are free functions of time and hereafter we will refer to them as \emph{EFT functions}.  $\{\Omega, \Lambda, c\}$ are usually called background EFT functions as they are the only ones contributing to both the background and linear perturbation equations, while the others enter only at the level of perturbations. Let us notice that the operators corresponding to $ \bar{m}_5,\lambda_{1,2}$  have already been considered in Ref.~\cite{Gleyzes:2013ooa}, while the remaining operators have been introduced by some of the authors of this paper in Ref.~\cite{Frusciante:2015maa}, where it is shown that they are necessary to map the high-energy Ho\v rava gravity action~\cite{Blas:2009qj} in the EFT formalism.

The EFT formalism offers a unifying approach to study large scale structure (LSS) in DE/MG models. Once  implemented into an Einstein-Boltzmann solver like CAMB~\cite{Lewis:1999bs}, it clearly provides a very powerful software with which to test gravity on cosmological scales. This has been achieved with the patches EFTCAMB/EFTCosmoMC, introduced in Refs.~\cite{Hu:2013twa,Raveri:2014cka} and publicly available at~\url{http://wwwhome.lorentz.leidenuniv.nl/~hu/codes/}.  This software can be used in two main realizations: the \emph{pure EFT} and the \emph{mapping EFT}. The former corresponds to an agnostic exploration of dark energy, where the user can turn on and off different EFT functions and explore their effects on the LSS. In the latter case instead, one specializes to a model (or a class of models, e.g. $f(R)$ gravity), maps it into the EFT functions and proceed to study the corresponding dynamics of perturbations. We refer the reader to Ref.~\cite{Hu:2014oga} for technical details of the code. 

There are some key virtues of EFTCAMB which make it a very interesting tool to constrain gravity on cosmological scales. One is the possibility of imposing powerful yet general conditions of stability at the level of the EFT action, which makes the exploration of the parameter space very efficient~\cite{Raveri:2014cka}. We will elaborate on this in Section~\ref{Sec:stability}. Another, is the fact that a vast range of specific models of DE/MG can be implemented \emph{exactly} and the corresponding dynamics of perturbations be evolved, in the same code, guaranteeing unprecedented accuracy and consistency. 

In order to use EFTCAMB in the mapping mode it is necessary to determine the expressions of the EFT functions corresponding to the given model. Several models are already built-in in the currently public version of EFTCAMB.  This paper offers a complete guide on how to map specific models and classes of models of DE/MG all the way into the EFT language at the basis of EFTCAMB, whether they are initially formulated in the ADM or covariant formalism; all this,  without the need of going through the cumbersome expansion of the models to quadratic order in perturbations around the FLRW background. 

\section{From a General Lagrangian in ADM formalism to the EFT framework}\label{Sec:fromgeneraltoEFT}
In this Section we use a general Lagrangian in the ADM formalism which covers the same class of theories described by the EFT action~(\ref{EFTgeneral}). This will allow us to make a parallel between the ADM and EFT formalisms, and to use the former as a convenient platform for a general mapping description of DE/MG theories into the EFT language.  In particular, in Section~\ref{generallagrangian}  we will expand a general ADM action up to second order in perturbations, in Section~\ref{EFTinADM} we will write the EFT action in ADM form and, finally, in Section~\ref{Sec:mapping} we will provide the mapping between the two.
  
\subsection{A General Lagrangian in ADM formalism}\label{generallagrangian}

Let us introduce the  3+1 decomposition of spacetime typical of the ADM formalism, for which the line element reads: 
\begin{equation}
ds^2=-N^2 dt^2+h_{ij}(dx^i+N^i dt)(dx^j+N^j dt)\,,
\end{equation}
where $N(t,x^i)$ is the lapse function, $N^i(t,x^i)$ the shift and $h_{ij}(t,x^i)$ is the three dimensional spatial metric. We also adopt the following definition of the normal vector to the hypersurfaces of constant time and the corresponding extrinsic curvature:
\begin{equation}\label{convention}
n_{\mu}=N\delta_{\mu 0}, \qquad K_{\mu\nu}=h^\lambda_\mu\nabla_{\lambda}n_{\nu}.
\end{equation}
The general Lagrangian  we use in this Section  has been proposed in Ref.~\cite{Kase:2014cwa} and can be written as follows:
\be\label{ADM_lagrangian}
L=L(N,{\mathcal R}, \mathcal{S}, K,  {\mathcal Z}, {\mathcal U},  {\mathcal Z}_1,  {\mathcal Z}_2, \alpha_1, \alpha_2, \alpha_3, \alpha_4, \alpha_5; t)\,,
\ee
where the above geometrical quantities are defined as follows:
\ba\label{ADM_quantities}
&&\mathcal{S}=K_{\mu\nu}K^{\mu\nu}\,,\,\,{\mathcal Z}={\mathcal R}_{\mu\nu}{\mathcal R}^{\mu\nu}\,,\,\,{\mathcal U}={\mathcal R}_{\mu\nu}K^{\mu\nu}\,,\,\,{\mathcal Z}_1=\nabla_i{\mathcal R}\nabla^i{\mathcal R}\,,\,\,{\mathcal Z}_2=\nabla_i{\mathcal R}_{jk}\nabla^i{\mathcal R}^{jk}\,,\nonumber\\
&&\,\,\alpha_1=a^ia_i\,,\,\,\alpha_2=a^i\Delta a_i\,,\,\,\alpha_3={\mathcal R}\nabla_ia^i\,,\,\,\alpha_4=a_i\Delta^2a^i\,,\,\,\alpha_5=\Delta {\mathcal R}\nabla_ia^i,
\ea
with $\Delta=\nabla_k\nabla^k$ and $a^i$  is the acceleration of the normal vector, $n^{\mu}\nabla_{\mu}n_{\nu}$. $\nabla_\mu$ and $\nabla_k$ are  the covariant derivatives constructed respectively with the four dimensional metric, $g_{\mu\nu}$ and the three metric, $h_{ij}$. 
 
The operators considered in the Lagrangian~(\ref{ADM_lagrangian}) allow to describe gravity theories with up to sixth order spatial derivatives, therefore the range of theories covered by such a Lagrangian is the same as the EFT action proposed in Section~\ref{extendedEFT}.   The resulting general action, constructed with purely geometrical quantities, is sufficient to cover most of the candidate models of modified gravity~\cite{Sotiriou:2008rp,DeFelice:2010aj,Clifton:2011jh,Tsujikawa:2013fta,Deffayet:2013lga,Koyama:2015vza}. 

We shall now proceed to work out the mapping of Lagrangian~(\ref{ADM_lagrangian}) into the EFT formalism.  The procedure that we will implement in the following retraces that of Refs.~\cite{Gleyzes:2013ooa,Kase:2014cwa}. However,  there are some tricky differences between the EFT language of Ref.~\cite{Gleyzes:2013ooa} and the one at the basis of EFTCAMB~\cite{Hu:2013twa,Raveri:2014cka}. Most notably the different sign convention for the normal vector, $n_{\mu}$, and the extrinsic curvature, $K_{\mu\nu}$ (see Eq. (\ref{convention})),  a different notation for the conformal coupling and the use of $\delta g^{00}$ in the action instead of $g^{00}$, which changes the definition of some EFT functions. It is therefore important that we present all details of the calculation as well as derive a final result which is compatible with EFTCAMB. In particular, the results of this Section account for the different convention for the normal vector.

We shall now expand the quantities in the Lagrangian~(\ref{ADM_lagrangian}) in terms of perturbations by considering for the background a  flat FLRW metric  of the form:
\be
ds^2=-dt^2+a(t)^2\delta_{ij}dx^idx^j,
\ee
where $a(t)$ is the scale factor. Therefore, we can  define: 
\ba \label{perturbations}
&&\delta K=3H+K\,,\qquad \delta K_{\mu\nu}=Hh_{\mu\nu}+K_{\mu\nu}\,, \qquad \delta \mathcal{S}=\mathcal{S}-3H^2=-2H\delta K+\delta K^{\mu}_{\nu}\delta K_{\mu}^{\nu}\,,\nn \\
&&\delta \mathcal{U}=-H\delta\mathcal{R}+\delta K^\mu_\nu \delta K^\nu_\mu\,,  \qquad \delta \alpha_1=\partial_i\delta N\partial^i \delta N\,, \quad \delta \alpha_2=\partial_i \delta N \nabla_k\nabla^k\partial^i\delta N\,, \qquad \delta \alpha_3=\mathcal{R}\nabla_i\partial^i\delta N\,,\nn\\
&&  \delta \alpha_4=\partial_i\delta N \Delta^2 \partial^i \delta N\,, \qquad  \delta \alpha_5=\Delta^2\mathcal{R}\nabla_i\partial^i\delta N\,, \quad
\delta \mathcal{Z}_1=\nabla_i\delta\mathcal{R}\nabla^i\delta\mathcal{R}\,,\qquad \delta \mathcal{Z}_2=\nabla_i\delta\mathcal{R}_{jk}\nabla^i\delta\mathcal{R}^{jk},
\ea  
where $H\equiv \dot{a}/a$ is the Hubble parameter and $\partial_\mu$ is the partial derivative w.r.t. the coordinate $x^{\mu}$. The operators $\mathcal{R}, \mathcal{Z}$ and  $\mathcal{U}$ vanish on a flat FLRW background, thus they contribute only to perturbations, and for convenience we can write $\mathcal{R}=\delta \mathcal{R}=\delta_1\mathcal{R}+\delta_2\mathcal{R}  ,\,\mathcal{Z}=\delta \mathcal{Z},\, \mathcal{U}=\delta \mathcal{U}$, where $\delta_1\mathcal{R}$ and $\delta_2 \mathcal{R}$ are the perturbations of the Ricci scalar respectively at first and second order.
We now proceed with a simple expansion of the Lagrangian~(\ref{ADM_lagrangian}) up to second order:
\ba\label{firstADMlag}
\delta L &=&\bar{L}+L_N \delta N+L_K \delta K+L_{\mathcal{S}} \delta \mathcal{S}+L_{\mathcal{R}}\delta \mathcal{R}+L_{\mathcal{U}}\delta \mathcal{U}+L_{\mathcal{Z}}\delta \mathcal{Z}+
\sum\limits_{i=1}^5 L_{\alpha_i} \delta \alpha_i + \sum\limits_{i=1}^2 L_{\mathcal{Z}_i} \delta \mathcal{Z}_i\nn\\
&&+\f{1}{2}\l(\delta N \f{\partial}{\partial N}+\delta K \f{\partial}{\partial K}+\delta \mathcal{S} \f{\partial}{\partial \mathcal{S}}+\delta\mathcal{R}\f{\partial}{\partial \mathcal{R}}+\delta \mathcal{U}\f{\partial}{\partial \mathcal{U}}\r)^2L +\mathcal{O}(3),
\ea 
where $\bar{L}$ is the Lagrangian evaluated on the background and $L_X=\partial L/ \partial X$ is the derivative of the Lagrangian w.r.t the quantity $X$. It can be shown that by considering the perturbed quantities in~(\ref{perturbations}) and, after some manipulations, it is possible to obtain  the following expression for the action up to second order in perturbations:
\ba \label{actionexpanded}
\mathcal{S}_{ADM}&=&\int{}d^4x\sqrt{-g}\l[\bar{L}+\dot{\mathcal{F}}+3H\mathcal{F}+(L_N-\dot{\mathcal{F}})  \delta N+\l(\dot{\mathcal{F}}+\f{1}{2}L_{NN}\r)(\delta N)^2+L_{\mathcal{S}} \delta K_\mu^\nu\delta K_\nu^\mu+\f{1}{2}\mathcal{A}(\delta K)^2+\mathcal{B}\delta N \delta K \r.\nn\\
&+&\l.\mathcal{C}\delta K \delta\mathcal{R}+\mathcal{D}\delta N\delta \mathcal{R}+\mathcal{E}\delta \mathcal{R}+\f{1}{2}\mathcal{G}(\delta \mathcal{R})^2 +L_\mathcal{Z}\delta \mathcal{R}^\mu_\nu\mathcal{R}_\mu^\nu+L_{\alpha_1}\partial_i\delta N\partial^i \delta N+ L_{\alpha_2}\partial_i \delta N \nabla_k\nabla^k\partial^i\delta N  \r.\nn \\
&+&\l. L_{\alpha_3}\mathcal{R}\nabla_i\partial^i\delta N  + L_{\alpha_4}\partial_i\delta N \Delta^2 \partial^i \delta N+ L_{\alpha_5}\Delta\mathcal{R}\nabla_i\partial^i\delta N+L_{\mathcal{Z}_1}\nabla_i\delta\mathcal{R}\nabla^i\delta\mathcal{R}+L_{\mathcal{Z}_2}\nabla_i\delta\mathcal{R}_{jk}\nabla^i\delta\mathcal{R}^{jk} \r]\,,
\ea
where: 
\begin{align}\label{Coefficientdefinitions}
\mathcal{A}&=L_{KK}+4H^2L_{\mathcal{S}\mathcal{S}}-4HL_{SK},\nn\\
\mathcal{B}&=L_{KN}-2HL_{\mathcal{S}N},\nn \\
\mathcal{C}&=L_{KR}-2HL_{\mathcal{S}R}+\f{1}{2}L_{\mathcal{U}}-HL_{K\mathcal{U}}+2H^2L_{\mathcal{S}\mathcal{U}}, \nn\\
\mathcal{D}&= L_{N\mathcal{R}}+\f{1}{2}\dot{L}_{\mathcal{U}}-HL_{N\mathcal{U}}, \nn\\
\mathcal{E}&=L_{\mathcal{R}}-\f{3}{2}HL_{\mathcal{U}}-\f{1}{2}\dot{L}_{\mathcal{U}}, \nn\\
\mathcal{F}&=L_K-2HL_{\mathcal{S}}, \nn\\
\mathcal{G}&=L_{\mathcal{R}\mathcal{R}}+H^2L_{\mathcal{U}\mathcal{U}}-2HL_{\mathcal{R}\mathcal{U}}.
\end{align}
Here and throughout the paper, unless stated otherwise, dots indicate derivatives w.r.t. cosmic time, $t$. The above quantities are general functions of time evaluated on the background. In order to obtain  action~(\ref{actionexpanded}), we have followed the same steps as in Refs.~\cite{Gleyzes:2013ooa,Kase:2014cwa}, however, there are some differences in the results due to the different convention that we use for the normal vector (Eq.~(\ref{convention})). As a result the differences stem from the terms which contain $K$ and $K_{\mu\nu}$. More details are in Appendix~\ref{KSperturbations}, where we derive the contribution of $\delta K$ and $\delta \mathcal{S}$, and in Appendix~\ref{Uperturbations}, where we explicitly comment and derive the perturbations generated by $\mathcal{U}$. 

Finally, we derive the modified Friedmann equations considering the  first order action, which  can be written as follows:
\be
\mathcal{S}_{ADM}^{(1)}=\int{}d^4x\l[\delta\sqrt{h}(\bar{L}+3H\mathcal{F}+\mathcal{\dot{F}})+a^3(L_N+3H\mathcal{F}+\bar{L})\delta N+a^3 \mathcal{E}\delta_1 \mathcal{R}\r],
\ee
where $\delta_1 \mathcal{R}$ is the contribution of the Ricci scalar at first order. Notice that we used  $\sqrt{-g}=N\sqrt{h}$, where $h$ is the determinant of the three dimensional metric. It is straightforward to show that by varying the above action w.r.t. $\delta N$ and $\delta \sqrt{h}$, one finds the  Friedmann equations: 
\begin{align}\label{Friedmann}
&L_N+3H\mathcal{F}+\bar{L}=0\,,\nn\\
&\bar{L}+3H\mathcal{F}+\mathcal{\dot{F}}=0.
\end{align}
Hence, the homogeneous part  of action~(\ref{actionexpanded}) vanishes after applying the Friedmann equations.

\subsection{The EFT action in ADM notation}\label{EFTinADM}
We shall now go back to the EFT action~(\ref{EFTgeneral}) and rewrite it in the ADM notation. This will allow us to easily compare it with action~(\ref{actionexpanded}) and obtain a general recipe to map an ADM action into the EFT language. To this purpose, an important step is to connect the  $\delta g^{00}$ used in this formalism with $\delta N$ used in the ADM formalism:
\be\label{linkg00N}
g^{00}=-\frac{1}{N^2}=-1+2\delta N-3(\delta N)^2+ ...\equiv-1+\delta g^{00}\,,
\ee
from which follows that $(\delta g^{00})^2=4(\delta N)^2$ at second order. Considering the Eqs.~(\ref{perturbations}) and~(\ref{linkg00N}), it is very easy to write the EFT action in terms of ADM quantities, the only term which requires a bit  of manipulation is $(1+\Omega(t))R$, which we will show in the following.  First, let us  use  the Gauss-Codazzi relation~\cite{Gourgoulhon:2007ue} which allows one to express the four dimensional Ricci scalar in terms of three dimensional quantities typical of ADM formalism:
\begin{equation}\label{GaussCodazzi}
R=\mathcal{R}+K_{\mu\nu}K^{\mu\nu}-K^2+2\nabla_{\nu}(n^{\nu}\nabla_{\mu}n^{\mu}-n^{\mu}\nabla_{\mu}n^{\nu})\,.
\end{equation}
Then, we can write: 
\begin{align}
\int d^4x\sqrt{-g}\f{m_0^2}{2}(1+\Omega)R&=\int d^4x\sqrt{-g}\f{m_0^2}{2}(1+\Omega)\l[\mathcal{R}+K_{\mu\nu}K^{\mu\nu}-K^2+2\nabla_{\nu}\l(n^{\nu}\nabla_{\mu}n^{\mu}-n^{\mu}\nabla_{\mu}n^{\nu}\r)\r] \,,\nn\\
&=\int d^4x\sqrt{-g}\frac{m_0^2}{2}(1+\Omega)\l[\mathcal{R}+\mathcal{S}-K^2+2\nabla_{\nu}\l(n^{\nu}K-a^{\nu}\r)\r]\,,\nn\\
&=\int d^4x\sqrt{-g}\l[\frac{m_0^2}{2}(1+\Omega)\l(\mathcal{R}+\mathcal{S}-K^2\r)+m_0^2\dot{\Omega}\frac{K}{N}\r]\,,
\end{align}
where in the last line we have used that $\nabla^{\nu}a_{\nu}=0$. Proceeding as usual and employing the relation~(\ref{DerTrick}), we obtain:
\begin{align}
\int d^4x\sqrt{-g}\f{m_0^2}{2}(1+\Omega)R&=\int d^4x\sqrt{-g}m_0^2\l\{\frac{1}{2}(1+\Omega)\mathcal{R}+3H^2(1+\Omega)+2\dot{H}(1+\Omega)+2H\dot{\Omega}+\ddot{\Omega}\r.\nonumber\\
&\l.+\l[H\dot{\Omega}-2\dot{H}(1+\Omega)-\ddot{\Omega}\r]\delta N-\dot{\Omega}\delta K\delta N+\frac{(1+\Omega)}{2}\delta K^{\mu}_{\nu}\delta K^{\nu}_{\mu}-\frac{(1+\Omega)}{2}(\delta K)^2\r.\nonumber\\
&\l.+\l[2\dot{H}(1+\Omega)+2H\dot{\Omega}+\ddot{\Omega}-3H\dot{\Omega}\r](\delta N)^2\r\}.
\end{align}

Finally, after combining terms correctly, we obtain the final form of the EFT action in the ADM notation, up to second order in perturbations:
\begin{align}\label{EFTADM}
\mathcal{S}_{EFT}&=\int d^4x\sqrt{-g}\l\{\frac{m_0^2}{2}(1+\Omega)\mathcal{R}+3H^2m_0^2(1+\Omega)+2\dot{H}m_0^2(1+\Omega)+2m_0^2H\dot{\Omega}+m_0^2\ddot{\Omega}+\Lambda \r.\nonumber\\
&\l.+\l[H\dot{\Omega}m_0^2-2\dot{H}m_0^2(1+\Omega)-\ddot{\Omega}m_0^2-2c\r]\delta N -(m_0^2\dot{\Omega}+\bar{M}^3_1)\delta K\delta N  
+\f{1}{2}\l[m_0^2(1+\Omega)-\bar{M}_3^2\r]\delta K^{\mu}_{\nu}\delta K^{\nu}_{\mu}\r.   \nonumber\\
&\l.-\f{1}{2}\l[m_0^2(1+\Omega)+\bar{M}^2_2\r](\delta K)^2 +\hat{M}^2\delta N\delta\mathcal{R} +\l[2\dot{H}m_0^2(1+\Omega)+\ddot{\Omega}m_0^2-Hm_0^2\dot{\Omega}+3c+2M^4_2\r](\delta N)^2 \r.\nonumber\\
&\l.+4m^2_2h^{\mu\nu}\partial_{\mu}\delta N\partial_{\nu}\delta N+\frac{\bar{m}_5}{2}\delta\mathcal{R}\delta K+\lambda_1(\delta\mathcal{R})^2+\lambda_2\delta\mathcal{R}^{\mu}_{\nu}\delta\mathcal{R}^{\nu}_{\mu}+2\lambda_3\delta\mathcal{R}h^{\mu\nu}\nabla_{\mu}\partial_{\nu}\delta N +4\lambda_4 h^{\mu\nu}\partial_{\mu}\delta N \nabla^2 \partial_{\nu}\delta N\r.\nonumber \\
&\l.+\lambda_5h^{\mu\nu}\nabla_{\mu}\mathcal{R}\nabla_{\nu}\mathcal{R}+\lambda_6h^{\mu\nu}\nabla_{\mu}\mathcal{R}_{ij}\nabla_{\nu}\mathcal{R}^{ij}+4\lambda_7 h^{\mu\nu}\partial_{\mu} \delta N\nabla^4 \partial_{\nu}\delta N+2\lambda_8h^{\mu\nu}\nabla^2\mathcal{R}\nabla_{\mu}\partial_{\nu}\delta N\r\}\,.
\end{align}
This final form of the action will be the starting point from which we will construct a general mapping  between the EFT and ADM  formalisms.

\subsection{The Mapping}\label{Sec:mapping}

We  now proceed to explicitly work out the mapping between the  EFT action~(\ref{EFTADM}) and the ADM one~(\ref{actionexpanded}).  The result will be a very convenient recipe in order to quickly map any model written in the ADM notation into the EFT formalism. In the next Section we will apply it to most of the interesting candidate models of DE/MG, providing a complete guide on how to go from covariant formulations all the way to the EFT formalism at the basis of the Einstein-Boltzmann solver EFTCAMB~\cite{Hu:2013twa,Raveri:2014cka}.

A direct comparison between actions~(\ref{actionexpanded}) and~(\ref{EFTADM}) allows us to straightforwardly identify the following:
\begin{align}
&\f{m_0^2}{2}(1+\Omega)=\mathcal{E} \,, \qquad -2c+m_0^2\l[-2\dot{H}(1+\Omega)-\ddot{\Omega}+H\dot{\Omega}\r]=L_N-\dot{\mathcal{F}}\,,\nn\\
&\Lambda+m_0^2\l[3H^2(1+\Omega)+2\dot{H}(1+\Omega)+2H\dot{\Omega}+\ddot{\Omega}\r]=\bar{L}+3H\mathcal{F}+\dot{\mathcal{F}}\,, \nn\\
&m_0^2\l[2\dot{H}(1+\Omega)-H\dot{\Omega}+\ddot{\Omega}\r]+2M^4_2+3c=\dot{\mathcal{F}}+\frac{L_{NN}}{2},\nonumber\\
&-m_0^2(1+\Omega)-\bar{M^2_2}= \mathcal{A},\quad \lambda_1=\f{\mathcal{G}}{2},\quad
-m_0^2\dot{\Omega}-\bar{M}_1^3=\mathcal{B},\nonumber\\
&\frac{\bar{m}_5}{2}=\mathcal{C},\quad \hat{M}^2=\mathcal{D},\qquad \frac{m_0^2}{2}(1+\Omega)-\frac{\bar{M}^2_3}{2}=L_{\mathcal{S}},\qquad 4m^2_2=L_{\alpha_1},\quad  \lambda_5=L_{\mathcal{Z}_1},\nonumber\\
&4\lambda_4=L_{\alpha_2},\qquad 2\lambda_3=L_{\alpha_3},\qquad 4\lambda_7=L_{\alpha_4}, \qquad 2\lambda_8=L_{\alpha_5},\qquad \lambda_2=L_{\mathcal{Z}},\quad\lambda_6= L_{\mathcal{Z}_2}.
\end{align}
It is now simply a matter of inverting these relations in order to obtain the desired general mapping results:
\begin{align}\label{Map}
&\Omega(t)=\frac{2}{m_0^2}\mathcal{E}-1,\qquad c(t)=\frac{1}{2}(\dot{\mathcal{F}}-L_N)+(H\dot{\mathcal{E}}-\ddot{\mathcal{E}}-2\mathcal{E}\dot{H}),\nonumber\\
&\Lambda(t)=\bar{L}+\dot{\mathcal{F}}+3H\mathcal{F}-(6H^2\mathcal{E}+2\ddot{\mathcal{E}}+4H\dot{\mathcal{E}}+4\dot{H}\mathcal{E})\,,\qquad \bar{M}^2_2(t)=-\mathcal{A}-2\mathcal{E},\nonumber\\
&M_2^4(t)=\frac{1}{2}\l(L_N+\frac{L_{NN}}{2}\r)-\frac{c}{2}, \qquad \bar{M}_1^3(t)=-\mathcal{B}-2\dot{\mathcal{E}},\qquad \bar{M}^2_3(t)=-2L_{\mathcal{S}}+2\mathcal{E},\nonumber\\
&m_2^2(t)=\f{L_{\alpha_1}}{4},\qquad \bar{m}_5(t)=2\mathcal{C},\qquad \hat{M}^2(t)=\mathcal{D},\qquad \lambda_1(t)=\frac{\mathcal{G}}{2},\nonumber\\
&\lambda_2(t)=L_{\mathcal{Z}},\qquad \lambda_3(t)=\frac{L_{\alpha_3}}{2},\qquad \lambda_4(t)=\frac{L_{\alpha_2}}{4},\qquad \lambda_5(t)=L_{\mathcal{Z}_1},\nonumber\\
&\lambda_6(t)=L_{\mathcal{Z}_2},\qquad \lambda_7(t)=\frac{L_{\alpha_4}}{4},\qquad \lambda_8(t)=\frac{L_{\alpha_5}}{2}.
\end{align}

Let us stress that the above definitions of the EFT functions are very useful  if one is interested in writing a specific action in EFT language. Indeed the only step required before applying~(\ref{Map}), is to write the action which specifies the chosen theory in ADM form, without the need of perturbing the theory and its action up to quadratic order.

The expressions of the EFT functions corresponding to a given model, and their time-dependence, are all that is needed in order to implement a specific model of DE/MG in EFTCAMB and have it solve for the dynamics of perturbations, outputting observable quantities of interest.  Since EFTCAMB uses the scale factor as the time variable and the Hubble parameter expressed w.r.t conformal time, one needs to convert the cosmic time $t$ in the argument of the functions in Eq.~(\ref{Map}) into the scale factor, $a$, their time derivatives into derivatives w.r.t. the scale factor and transform the Hubble parameter into the one in conformal time $\tau$, while considering it a function of $a$, see Ref.~\cite{Hu:2014oga}. This is a straightforward step and we will give some examples in Appendix~\ref{Conformal}. 

Let us conclude this Section looking at the equations for the background. Working with the EFT action, and expanding it to first order while using the ADM notation, one obtains:
\ba
\mathcal{S}_{EFT}^{(1)}&=&\int{}d^4x\l\{a^3\f{m_0^2}{2}\l(1+\Omega\r)\delta_1\mathcal{R}+\l[3H^2m_0^2(1+\Omega)+2\dot{H}m_0^2(1+\Omega)+2m_0^2H\dot{\Omega}+m_0^2\ddot{\Omega}+\Lambda\r]\delta\sqrt{h} \r.\nonumber\\
&+&\l.a^3\l[3H\dot{\Omega}m_0^2-2c+3H^2m_0^2(1+\Omega)+\Lambda\r]\delta N\r\}\,,
\ea
therefore the variation w.r.t. $\delta N$ and $\delta \sqrt{h}$ yields:
\ba\label{FriedmannEFT}
&&3H\dot{\Omega}m_0^2-2c+3H^2m_0^2(1+\Omega)+\Lambda=0 \,,\nn\\
&&3H^2m_0^2(1+\Omega)+2\dot{H}m_0^2(1+\Omega)+2m_0^2H\dot{\Omega}+m_0^2\ddot{\Omega}+\Lambda=0  \,.
\ea
Using the mapping~(\ref{Map}), it is easy to verify that these equations correspond to those in the ADM formalism~(\ref{Friedmann}). Once the mapping~(\ref{Map}) has been worked out, it is straightforward to obtain the Friedmann equations without having to vary the action for each specific model. 

\section{Model mapping examples}\label{Sec:examples}

Having derived the precise mapping between the  ADM formalism and the EFT approach in Section~\ref{Sec:mapping}, we proceed to apply it to some specific cases which are of cosmological interest, i.e. minimally coupled quintessence~\cite{Tsujikawa:2013fta}, $f(R)$ theory~\cite{DeFelice:2010aj}, Horndeski/GG~\cite{Horndeski:1974wa,Deffayet:2009mn}, GLPV~\cite{Gleyzes:2014dya} and  Ho\v rava gravity~\cite{Blas:2009qj}. The mapping of some of these theories is already present in the literature (see Refs.~\cite{Gubitosi:2012hu,Bloomfield:2012ff,Bloomfield:2013efa,Gleyzes:2013ooa,Gleyzes:2014rba,Frusciante:2015maa} for more details). However, since one of the main purposes of this work is to provide a self-contained and general recipe that can be used to easily implement a specific theory in EFTCAMB, we will present all the mapping of  interest, including those that are already in the literature due to the aforementioned differences in the definition of the normal vector and some of the EFT functions. Let us notice that the mapping of the GLPV Lagrangians in particular, is one of the new results obtained in this work.

\subsection{Minimally coupled quintessence}\label{5e}

As illustrated in Refs.~\cite{Gubitosi:2012hu,Bloomfield:2012ff,Gleyzes:2014rba}, the mapping of minimally coupled quintessence~\cite{Tsujikawa:2013fta} into EFT functions is very straightforward. The typical action for such a model is of the following form:
\be
\mathcal{S}_{\phi}=\int{}d^4x\sqrt{-g}\l[\f{m_0^2}{2}R-\frac{1}{2}\partial^{\nu}\phi\partial_{\nu}\phi-V(\phi)\r]\,,
\ee
where $\phi(t,x^i)$ is a scalar field and $V(\phi)$ is its potential. Let us proceed by rewriting the  second term in unitary gauge and in  ADM quantities: 
\be
-\frac{1}{2}g^{\mu\nu}\partial_{\mu}\phi\partial_{\nu}\phi \,\, \rightarrow \,\, -\f{\dot{\phi}^2_0(t)}{2} g^{00}\equiv \f{\dot{\phi}^2_0(t)}{2N^2}\,,
\ee
where  $\phi_0(t)$ is the field background value. Substituting back into the action we get, in the ADM formalism, the following action:
\ba
\mathcal{S}_{\phi}=\int{}d^4x\sqrt{-g}\l\{\f{m_0^2}{2}\l[\mathcal{R} +\mathcal{S}-K^2\r]+\f{1}{N^2}\f{\dot{\phi}^2_0(t)}{2}-V(\phi_0)\r\} \,,
\ea
where we have used the Gauss-Codazzi relation~(\ref{GaussCodazzi}) to express the four dimensional  Ricci scalar in terms of three dimensional quantities. Now, since the initial covariant action has been written in terms of ADM quantities, we can finally apply the results in Eqs.~(\ref{Map}) to get the EFT functions: 
\be\label{quintessencemapping}
\Omega(t)=0, \qquad c(t)=\f{\dot{\phi}^2_0}{2}, \qquad \Lambda(t)=\f{\dot{\phi}^2_0}{2}-V(\phi_0).
\ee
Notice that the other EFT functions are zero. In Refs.~\cite{Gubitosi:2012hu,Bloomfield:2012ff} the above mapping has been obtained directly from the covariant action while our approach follows more strictly the one adopted in Ref.~\cite{Gleyzes:2014rba}. However, let us notice that w.r.t. it, our results differ due to a different definition of the background EFT functions.\footnote{The background EFT functions adopted here are related to the ones in Ref.~\cite{Gleyzes:2014rba}, by  the following relations: 
\be\label{backgrounddifference}
1+\Omega(t)=f(t)\,, \qquad \Lambda(t)=-\tilde{\Lambda}(t)+c(t) \,, \qquad c(t)=\tilde{c}(t) \,.
\ee
where $f$ and tildes quantities correspond to the EFT functions in Ref.~\cite{Gleyzes:2014rba}. These differences are due to the fact that in our formalism we have in the EFT action the term $-c\delta g^{00}$ while in the other formalism the authors use  $-\tilde{c} g^{00}$, therefore an extra contribution to $\tilde{\Lambda}$ from this operator comes when using $g^{00}=-1+\delta g^{00}$. Instead the different definition of the conformal coupling function, $\Omega$, is due to numerical reasons related to the implementation of the EFT approach in CAMB. }

Moreover, in order to use them in EFTCAMB one need to convert them in conformal time $\tau$, therefore one has:
\be
c(\tau)=\hub^2\f{\phi^{\prime\,2}_0}{2}\,, \qquad \Lambda(\tau)=\hub^2\f{\phi^{\prime\,2}_0}{2} -V(\phi_0)\,,
\ee
where the prime indicates the derivative w.r.t. the scale factor, $a(\tau)$, and $\hub\equiv \f{1}{a}\f{da}{d\tau}$ is the Hubble parameter in conformal time. Minimally coupled quintessence models are already implemented in the public versions of EFTCAMB~\cite{Hu:2014oga}.

\subsection{$f(R)$ gravity}\label{fR}

The second example we shall illustrate is that of $f(R)$ gravity~\cite{Sotiriou:2008rp,DeFelice:2010aj}. The mapping of the latter into the EFT language was derived in Refs.~\cite{Gubitosi:2012hu,Gleyzes:2014rba}. Here, we present an analogous approach which uses the ADM formalism. Let us start with the action :
\be
\mathcal{S}_f=\int{}d^4x\sqrt{-g}\f{m_0^2}{2}\l[R+f(R)\r],
\ee
where $f(R)$ is a general function of the four dimensional Ricci scalar.

In order to map it into our EFT approach, we will proceed to expand this action around the  background value of the Ricci scalar, $R^{(0)}$. Therefore, we choose a specific time slicing where the constant time hypersurfaces coincide with uniform $R$ hypersurfaces. This allows us to truncate the expansion at the linear order because higher orders will always contribute one power or more of $\delta R$ to the equations of motion, which vanishes.  For a more complete analysis we refer the reader to Ref.~\cite{Gubitosi:2012hu} . After the expansion we obtain the following Lagrangian: 
\be\label{fRexpanded}
 \mathcal{S}_f =\int{}d^4x\sqrt{-g}\f{m_0^2}{2}\left\{\l[1+f_R(R^{(0)})\r]R+f(R^{(0)})-R^{(0)}f_R(R^{(0)})\right\},
\ee
where $f_R\equiv \f{df}{dR}$. In the ADM formalism the above action reads:
\be
 \mathcal{S}_f =\int{}d^4x\sqrt{-g}\f{m_0^2}{2}\left\{\l[1+f_R(R^{(0)})\r]\l[\mathcal{R} +\mathcal{S}-K^2\r]+\f{2}{N}\dot{f}_R K+f(R^{(0)})-R^{(0)}f_R(R^{(0)})\right\},
\ee
where we have used as usual the Gauss Codazzi relation~(\ref{GaussCodazzi}). Using Eqs.~(\ref{Map}), it is easy to calculate that the only non zero EFT functions for $f(R)$ gravity are:
\be \label{fRmapping}
\Omega(t)=f_R(R^{(0)})\,, \qquad \Lambda(t)=\f{m_0^2}{2}f(R^{(0)})-R^{(0)}f_R(R^{(0)})\,.
\ee 
The  public version of  EFTCAMB already contains the designer $f(R)$ models~\cite{Hu:2014oga,Song:2006ej,Pogosian:2007sw}, while the specific Hu-Sawicki model is currently being implemented through the full mapping procedure~\cite{Hu:2016zrh}.

\subsection{The Galileon Lagrangians}\label{galileons}
The Galileon class of theories were derived  in Ref.~\cite{Nicolis:2008in}, by  studying the decoupling limit of the  five dimensional model of modified gravity known as DGP~\cite{Dvali:2000hr}. In this limit, the dynamics of the scalar DoF, corresponding to the longitudinal mode of the massive graviton, decouple from gravity and enjoy a galilean shift symmetry around Minkowski background, as a remnant of the five dimensional Poincare' invariance~\cite{Joyce:2014kja}. Requiring the scalar field to obey this symmetry and to have second order equations of motion allows one to identify a finite amount of terms that can enter the action. These terms are typically organized into a set of Lagrangians which, subsequently, have been covariantized~\cite{Deffayet:2009wt} and the final form is what is known as the Generalized Galileon (GG) model~\cite{Deffayet:2009mn}.   This set of models represent the most general theory of gravity with a scalar DoF and second order field equations in four dimensions and has been shown to coincide with the class of theories derived by Horndeski in Ref.~\cite{Horndeski:1974wa}. It is therefore common to refer to these models with the terms GG and Horndeski gravity, alternatively. 
GG models  have been deeply investigated in the cosmological context, since they display self accelerated solutions which can be used to realize both a single field inflationary scenario  at early times~\cite{Creminelli:2010ba,Kobayashi:2010cm,Burrage:2010cu,Kamada:2010qe,Creminelli:2010qf,Kobayashi:2011nu,Gao:2011qe,DeFelice:2013ar,Takamizu:2013gy,Frusciante:2013haa} and a late time accelerated expansion~\cite{Chow:2009fm,Silva:2009km,DeFelice:2010pv,Deffayet:2010qz,Pujolas:2011he}. Moreover, on small scales these models naturally display  the Vainshtein screening mechanism~\cite{Vainshtein:1972sx,Babichev:2013usa}, which can efficiently hide the extra DoF from local tests of gravity~\cite{Nicolis:2008in,Burrage:2010rs,DeFelice:2011th,Brax:2011sv,Kase:2013uja,Bloomfield:2014zfa,Joyce:2014kja}.

GG models include most of the interesting and viable theories of DE/MG that we aim to test against cosmological data. To this extent, the Einstein-Boltzmann solver EFTCAMB can be readily used to explore these theories both in a model-independent way, through a subset of the EFT functions, and in a model-specific way~\cite{Hu:2013twa,Hu:2014oga}. In the latter case, the first  step consists of mapping a given GG model into the EFT language.  In the following we derive the general mapping between GG and EFT functions, in order to provide an instructive and self-consistent compendium  to easily map any given GG model into the formalism at the basis of EFTCAMB.   

Let us introduce the GG action:
\be
\mathcal{S}_{GG}=\int{}d^4x\sqrt{-g}\l(L_2+L_3+L_4+L_5\r),
\ee
where the Lagrangians have the following structure:
\begin{align}\label{GGlagrangians}
L_2&=\mathcal{K}(\phi,X)\,,\nn \\
L_3&=G_3(\phi,X)\Box\phi\,,\nn\\
L_4&=G_4(\phi,X)R-2G_{4X}(\phi,X)\l[\l(\Box \phi\r)^2-\phi^{;\mu\nu}\phi_{;\mu\nu}\r]\,,\nn\\
L_5&=G_5(\phi,X)G_{\mu\nu}\phi^{;\mu\nu}+\frac{1}{3}G_{5X}(\phi,X)\l[\l(\Box\phi\r)^3-3\Box\phi\phi^{;\mu\nu}\phi_{;\mu\nu}+2\phi_{;\mu\nu}\phi^{;\mu\sigma}\phi^{;\nu}_{;\sigma}\r]\,,
\end{align}
here $G_{\mu\nu}$ is the Einstein tensor, $X\equiv\phi^{;\mu}\phi_{;\mu}$ is the kinetic term and $\{\mathcal{K}, G_i\}$ ($i=3,4,5$) are general functions of the scalar field $\phi$ and $X$, and $G_{iX}\equiv \partial G_i/\partial X$. Moreover, $\Box=\nabla^2$  and ; stand for the covariant derivative w.r.t. the metric $g_{\mu\nu}$.
The mapping of  GG is already present in the literature. For instance in Ref.~\cite{Bloomfield:2013efa} the mapping is obtained directly from the covariant Lagrangians, while in Refs.~\cite{Gleyzes:2013ooa,Gleyzes:2014rba} the authors start from the ADM version of the action.  In this paper we present in details all the steps from the covariant Lagrangians~(\ref{GGlagrangians}) to their expressions in ADM quantities; we then use the  mapping~(\ref{Map}) to obtain the EFT functions corresponding to GG. This allows us to give an instructive presentation of the method, while providing a final result consistent with the EFT conventions at the basis of EFTCAMB. Throughout these steps, we will highlight the differences w.r.t. Refs.~\cite{Bloomfield:2013efa,Gleyzes:2013ooa,Gleyzes:2014rba}  which arise because of different conventions.  Finally, in Appendix~\ref{Conformal} we rewrite the results of this Section with the scale factor as the independent variable and the Hubble parameter defined w.r.t. the  conformal time, making them readily implementable in EFTCAMB.

Since the GG action is formulated in covariant form, we shall use the following relations to rewrite the GG Lagrangians in ADM form:
\be
n_{\mu}=\gamma\phi_{;\mu},\qquad \gamma=\frac{1}{\sqrt{-X}}, \qquad \dot{n}_{\mu}=n^{\nu}n_{\mu;\nu}\,,
\ee
where we have, as usual, assumed that constant time hypersurfaces correspond to uniform field ones.
We notice that the acceleration, $\dot{n}_{\mu}$, and the extrinsic curvature $K^{\mu\nu}$ are orthogonal to the normal vector. This allows us to decompose the covariant derivative of the normal vector as follows:
\begin{equation}
n_{\nu;\mu}=K_{\mu\nu}-n_{\mu}\dot{n}_{\nu}.
\end{equation}
With these definitions it can be easily verified that:
\begin{align}\label{DDphi}
\phi_{;\mu\nu}&=\gamma^{-1}(K_{\mu\nu}-n_{\mu}\dot{n}_{\nu}-n_{\nu}\dot{n}_{\mu})+\frac{\gamma^2}{2}\phi^{;\lambda}X_{;\lambda}n_{\mu}n_{\nu}, \\
\label{BoxPhi}
\Box\phi&=\gamma^{-1}K-\frac{\gamma^2}{2}\phi^{;\lambda}X_{;\lambda}.
\end{align}

\begin{itemize}
\item $\mathbf{L_2}$- \textbf{Lagrangian}
\end{itemize}
Let us start with the simplest of the Lagrangians  which can be Taylor expanded in the kinetic term $X$, around its background value $X_0$, as follows:
\be
\mathcal{K}(\phi,X)=\mathcal{K}(\phi_0, X_0)+\mathcal{K}_X(\phi_0, X_0)(X-X_0)+\f{1}{2}\mathcal{K}_{XX}(X-X_0)^2,
\ee
where in terms of ADM quantities we have:
\be\label{kineticterADM}
X=-\frac{\dot{\phi}_0(t)^2}{N^2}=\frac{X_0}{N^2}\,.
\ee
Now by applying the results in Eqs.~(\ref{Map}), the corresponding EFT functions can be written as:
\be
\Lambda(t)=\mathcal{K}(\phi_0,X_0), \qquad c(t)=\mathcal{K}_X(\phi_0,X_0)X_0 \qquad M^4_2(t)=\mathcal{K}_{XX}(\phi_0,X_0)X_0^2 .
\ee

The differences with previous works in this case are the ones listed in Eq.~(\ref{backgrounddifference}). 

\begin{itemize}
 \item $\mathbf{L_3}$- \textbf{Lagrangian}
\end{itemize}
In order to rewrite this Lagrangian into the desired form, which depends only on ADM quantities, we  introduce an auxiliary function:
\begin{equation}
G_3\equiv F_3+2XF_{3X}\,.
\end{equation}
We proceed to plug this in the $L_3$-Lagrangian~\eqref{GGlagrangians} and using Eq.~\eqref{BoxPhi} we obtain, up to a total derivative:
\be
L_3=-F_{3\phi}X-2(-X)^{3/2}F_{3X}K\,.
\ee
Now going to unitary gauge and considering Eq.~(\ref{kineticterADM}), we can directly use ~(\ref{Map}). Let us start with $c(t)$:
\be
c(t)=\frac{1}{2}(\mathcal{F}-L_N)=-3\dot{\phi}_0^2\ddot{\phi}_0F_{3X}+2\ddot{\phi}_0F_{3XX}\dot{\phi}_0^4-\dot{\phi}_0^4F_{3X\phi}+F_{3\phi}\dot{\phi}_0^2-F_{3\phi X}\dot{\phi}_0^4-6H\dot{\phi}_0^5F_{3XX}+9HF_{3X}\dot{\phi}_0^3\,.
\ee
Now we want to eliminate the dependence on the auxiliary function $F_3$. In order to do this, we need to recombine terms by using the following:
\begin{align}\label{FtoG}
G_3&=F_3+2XF_{3X},\quad G_{3\phi}=F_{3\phi}-2\dot{\phi}_0^2 F_{3X\phi},\quad G_{3X}=3F_{3X}-2\dot{\phi}_0^2F_{3XX}\,, \nn\\
G_{3XX}&=3F_{3XX}-2\dot{\phi}_0^2F_{3XXX}+2F_{3XX},\quad G_{3\phi X}=3F_{3X\phi}-2\dot{\phi}_0^2 F_{3\phi XX}\,,
\end{align}
which gives the final expression:
\begin{equation}
c(t)=\dot{\phi}_0^2G_{3X}(3H\dot{\phi}_0-\ddot{\phi}_0)+G_{3\phi}\dot{\phi}_0^2.
\end{equation}

Now let us move on to the remaining non zero EFT functions corresponding to the $L_3$ Lagrangian:
\ba
&&\Lambda(t)=\bar{L}+\dot{\mathcal{F}}+3H\mathcal{F}\nonumber=G_{3\phi}\dot{\phi}_0^2-2\ddot{\phi}_0\dot{\phi}_0^2G_{3X} \,,\nn\\
&&\bar{M}^3_1(t)=-L_{KN}=-2 G_{3X} \dot{\phi}_0^3 \,,\nn\\
&&M^4_2(t)=\frac{1}{2}\l(L_N+\frac{L_{NN}}{2}\r)-\frac{c}{2}=G_{3X}\frac{\dot{\phi}_0^2}{2}(\ddot{\phi}_0+3H\dot{\phi}_0)-3HG_{3XX}\dot{\phi}_0^5-G_{3\phi X}\frac{\dot{\phi}_0^4}{2}\,,
\ea
where we have used the relations~(\ref{FtoG}). In the definitions of the EFT functions, $G_3$ and its derivatives are evaluated on the background. We suppressed the dependence on $(\phi_0,X_0)$ to simplify the final expressions. Before proceeding to map the remaining GG Lagrangians, let us comment on the differences w.r.t. the results in literature~\cite{Bloomfield:2013efa,Gleyzes:2013ooa,Gleyzes:2014rba}.  The results coincide up to two notable exceptions. The background functions are redefined as presented in Eq.~\eqref{backgrounddifference} and $\bar{M}^3_1=-\bar{m}^3_1$. In the latter term, the minus sign  is not a simple redefinition but rather comes from the fact that our extrinsic curvature has an overall minus sign difference due to the definition of the normal vector.  Therefore, the term  proportional to $\delta K\delta g^{00}$ will always differ by a minus sign.   
\begin{itemize}
\item $\mathbf{L_4}$- \textbf{Lagrangian}
\end{itemize}
Let us now consider the $L_4$ Lagrangian:
\begin{equation}
L_4=G_4 R-2G_{4X}\l[\l(\Box\phi\r)^2-\phi^{;\mu\nu}\phi_{;\mu\nu}\r] .
\end{equation}

After some preliminary manipulations of the Lagrangian, we get:
\be\label{L41}
L_4=G_4 \mathcal{R}+2G_{4X}(K^2-K_{\mu\nu}K^{\mu\nu})+2G_{4X}X_{;\lambda}(Kn^{\lambda}-\dot{n}^{\lambda})\,.
\ee
We proceed by using the relation:
\begin{equation}
\partial_{\mu} G_4=G_{4X} X_{;\mu}+G_{4\phi} \phi_{;\mu}\,,
\end{equation}
which we substitute in the last term of the Lagrangian~\eqref{L41} and, using integration by parts, we get:
\be
\label{L4Fin}
L_4=G_4\mathcal{ R}+(2G_{4X} X-G_4)(K^2-K_{\mu\nu}K^{\mu\nu})+2G_{4\phi}\sqrt{-X}K\,,
\ee
where we have used the Gauss-Codazzi relation~(\ref{GaussCodazzi}). Let us recall that we can relate $\phi_{;\mu}$ to $X$ by using Eq.~(\ref{kineticterADM}). 

Finally, in the same spirit as for $L_3$, we derive from the Lagrangian~(\ref{L4Fin}) the corresponding non zero EFT functions by using the results~(\ref{Map}):
\begin{align}
\Omega(t)&=-1+\f{2}{m_0^2}G_4\,,\nn\\
%%%%%%%%%
c(t)&=-\frac{1}{2}\Big(-\dot{L}_K+2\dot{H}L_{\mathcal{S}}+2H\dot{L}_{\mathcal{S}}\Big)+H\dot{L}_{\mathcal{R}}-\ddot{L}_{\mathcal{R}}-2\dot{H}L_{\mathcal{R}}=G_{4X}(2\ddot{\phi}_0^2+2\dot{\phi}_0\dddot{\phi}_0+4\dot{H}\dot{\phi}_0^2+2H\dot{\phi}_0\ddot{\phi}_0-6H^2 \dot{\phi}_0^2)\nonumber\\
&\,\,+G_{4X\phi}(2\dot{\phi}_0^2\ddot{\phi}_0+10H\dot{\phi}_0^3)+G_{4XX}(12H^2 \dot{\phi}_0^4-8H\dot{\phi}_0^3\ddot{\phi}_0-4\dot{\phi}_0^2\ddot{\phi}_0^2)\,,\nn\\
%%%%%%%%%%%%%%%
\Lambda(t)&=\bar{L}+\dot{\mathcal{F}}+3H\mathcal{F}-(6H^2L_{\mathcal{R}}+2\ddot{L}_{\mathcal{R}}+4H\dot{L}_{\mathcal{R}}+4\dot{H}L_{\mathcal{R}})\nonumber,\\
&\,\,=G_{4X}\l[ 12H^2 \dot{\phi}_0^2+8\dot{H}\dot{\phi}_0^2+16H\dot{\phi}_0\ddot{\phi}_0+4(\ddot{\phi}_0^2+\dot{\phi}_0\dddot{\phi}_0)\r]-G_{4XX}\big(16H\dot{\phi}_0^3\ddot{\phi}_0 +8\dot{\phi}_0^2\ddot{\phi}_0^2\big)+8HG_{4X\phi} \dot{\phi}_0^3\,,\nn\\
%%%%%%%%%%%%%%%%%%%%%
M_2^4(t)&=\frac{1}{2}(L_N+L_{NN}/2)-\frac{c}{2}=G_{4\phi X}\big(4H\dot{\phi}_0^3  -\ddot{\phi}_0\dot{\phi}_0^2  \big)-6H\dot{\phi}_0^5 G_{4\phi XX}-G_{4X}\l(2\dot{H}\dot{\phi}_0^2+H\dot{\phi}_0\ddot{\phi}_0+\dot{\phi}_0\dddot{\phi}_0+\ddot{\phi}_0^2  \r)\nonumber\\
&\,\,+G_{4XX}\big( 18H^2 \dot{\phi}_0^4+2 \dot{\phi}_0^2\ddot{\phi}_0^2  +4H\ddot{\phi}_0\dot{\phi}_0^3\big)-12H^2 G_{4XXX}\dot{\phi}_0^6\,,\nonumber\\
%%%%%%%%%%%%%%
\bar{M}^2_2(t)&=-L_{KK}-2L_{\mathcal{R}}=4G_{4X}\dot{\phi}_0^2\,,\nn\\
%%%%%%%%%%%%%%
\bar{M}^2_3(t)&=-2L_{\mathcal{S}}+2L_{\mathcal{R}}=-4G_{4X}\dot{\phi}_0^2\equiv -\bar{M}^2_2(t)\,,\nn\\
%%%%%%%%%%%%%
\hat{M}^2(t)&=L_{N\mathcal{R}}=2\dot{\phi}_0^2 G_{4X}\,, \nn\\
%%%%%%%%%%%%%
\bar{M}^3_1(t)&=2HL_{\mathcal{S}N}-2\dot{L}_{\mathcal{R}}-L_{KN}=G_{4X}(4\dot{\phi}_0\ddot{\phi}_0+8H\dot{\phi}_0^2)-16H G_{4XX} \dot{\phi}_0^4-4 G_{4\phi X}\dot{\phi}_0^3\,,
\end{align}
where also in this case  $G_4$ and its derivative are evaluated on the background. Let us notice that the above relations satisfy the conditions which define Horndeski/GG theories, i.e.:
\be\label{HorndeskiGG}
\bar{M}_2^2=-\bar{M}^2_3(t)=2\hat{M}^2(t),
\ee 
as found in Refs.~\cite{Bloomfield:2013efa,Gleyzes:2013ooa}. Finally, besides the differences mentioned previously for the $L_2$ and $L_3$ Lagrangians which also apply here, we notice that $\hat{M}^2=\mu_1^2$ when comparing with Ref.~\cite{Gleyzes:2013ooa}. 

\begin{itemize}
\item $\mathbf{L_5}$- \textbf{Lagrangian}
\end{itemize}
Finally, let us conclude with the $L_5$ Lagrangian. This Lagrangian contains cubic terms which makes it more complicated to express it in the ADM form:
\begin{equation} \label{L5lagrangian}
 L_5=G_5(\phi,X)G_{\mu\nu}\phi^{;\mu\nu}+\frac{1}{3}G_{5X}(\phi,X)\l[\l(\Box\phi\r)^3-3\Box\phi\phi^{;\mu\nu}\phi_{;\mu\nu}+2\phi_{;\mu\nu}\phi^{;\mu\sigma}\phi^{;\nu}_{;\sigma}\r]\,.
\end{equation}
In order to rewrite $L_5$, we have to enlist once again  the help of an auxiliary function, $F_5$, which is defined as follows:
\begin{equation}
G_{5X}\equiv F_{5X}+\frac{F_5}{2X}.
\end{equation}
Then, using this definition, we get the following relation:
\begin{align}
G_{5X}X_{;\rho}&=\gamma\nabla_{\rho}(\gamma^{-1}F_5)-F_{5\phi}\gamma^{-1}n_{\rho}.
\end{align}
Let us start with  the first term of the Lagrangian, which can be written as:
\be
G_5 G_{\mu\nu} \phi^{;\mu\nu}=F_5\phi^{;\mu\nu}G_{\mu\nu}-\frac{\gamma}{2}X^{;\nu}n^{\mu}G_{\mu\nu}F_5+(F_{5\phi}-G_{5\phi})\gamma^{-2}n^{\mu}n^{\nu}G_{\mu\nu},
\ee
hence  we need to rewrite $F_5 \phi^{;\mu\nu}G_{\mu\nu}$ in terms of ADM quantities which can be achieved by employing the following relation: 
\be
K^{\mu\nu}G_{\mu\nu}=K K^{\mu\nu}K_{\mu\nu}-K_{\mu\nu}^3+\mathcal{R}_{\mu\nu}K-K^{\mu\nu}n^{\sigma}n^{\rho}R_{\mu\sigma\nu\rho}-\frac{1}{2}K\big(\mathcal{R}-K^2+K_{\mu\nu}K^{\mu\nu}-2R_{\mu\nu}n^{\mu}n^{\nu}\big)\,.
\ee
This leads to the following:
\begin{align}
F_5 \phi^{;\mu\nu}G_{\mu\nu}&=F_5(\gamma^{-1}(-2R_{\mu\nu}n^{\mu}\dot{n}^{\nu})+\frac{\gamma^2}{2}n^{\mu}n^{\nu}\phi^{;\lambda}X_{;\lambda}G_{\mu\nu})\nonumber\\
&+F_5\gamma^{-1}\big[   K K^{\mu\nu}K_{\mu\nu}-K_{\mu\nu}^3+\mathcal{R}_{\mu\nu}K^{\mu\nu}-K^{\mu\nu}n^{\sigma}n^{\rho}R_{\mu\sigma\nu\rho}-\frac{1}{2}K\big(\mathcal{R}-K^2+K_{\mu\nu}K^{\mu\nu}-2R_{\mu\nu}n^{\mu}n^{\nu}\big) \big]\,.
\end{align}
The second term of the Lagrangian can be computed by considering Eqs.~\eqref{DDphi}-\eqref{BoxPhi}, which yields:
\begin{align}\label{defKJ}
&\frac{1}{3}G_{5X}\l[\l( \Box\phi\r)^3-3\Box\phi \phi^{;\mu\nu}\phi_{;\mu\nu}+2\phi_{;\mu\nu}\phi^{;\mu\sigma}\phi_{;\sigma}^{;\nu}\r]=\nonumber\\
&=\frac{G_{5X}}{3}\gamma^{-3}\big(K^3-3K S+2K_{\mu\nu}K^{\mu\sigma}K_{\sigma}^{\nu} \big)+G_{5X}\big(-\frac{1}{2} K^2\phi_{;\lambda}X^{;\lambda}   -2\dot{n}_{\sigma}\dot{n}_{\nu}K^{\nu\sigma}+\frac{S}{2}\phi_{;\lambda}X^{;\lambda}+2\gamma^{-3}K \dot{n}^{\nu}\dot{n}_{\nu}\big)\nonumber\\
&=\frac{G_{5X}}{3}\gamma^{-3}\tilde{\mathcal{K}}+G_{5X}\mathcal{J}\,,
\end{align}
where the definitions of $\tilde{\mathcal{K}}$ and $\mathcal{J}$ come directly from the second line of the above expression. In Appendix~\ref{Jcoefficient} we treat in detail the $G_{5X}\mathcal{J}$ term but for now we simply state the final result:
\begin{equation}
G_{5X}\mathcal{J}=F_5\gamma^{-1}\Big[ \frac{\tilde{\mathcal{K}}}{2}+K^{\mu\nu}n^{\sigma}n^{\rho}R_{\mu\sigma\nu\rho}+\dot{n}^{\sigma}n^{\rho}R_{\sigma\rho}-Kn^{\sigma}n^{\rho}R_{\sigma\rho}  \Big]-\frac{F_{5\phi}}{2}(K^2-S).
\end{equation}
Hence, after collecting all the terms, we get:
\begin{align}
L_5=F_5\sqrt{-X}\Big(  K^{\mu\nu}\mathcal{R}_{\mu\nu}-\frac{1}{2}K\mathcal{R}\Big) +(G_{5\phi}-F_{5\phi})X\frac{\mathcal{R}}{2}+\frac{(-X)^{3/2}}{3}G_{5X}\tilde{\mathcal{K}}+\frac{G_{5\phi}}{2}X(K^2-K_{\mu\nu}K^{\mu\nu})\,.
\end{align}
Now, in order to proceed with the mapping, we need to analyse $\tilde{\mathcal{K}}$ and $\mathcal{U}=K^{\mu\nu}\mathcal{R}_{\mu\nu}$ terms. The latter will be treated as in Appendix \ref{Uperturbations}, while the former can be written up to third order as follows:
\begin{equation}
\tilde{\mathcal{K}}=-6H^3-6H^2K-3HK^2+3H K_{\mu\nu}K^{\mu\nu}+\mathcal{O}(3).
\end{equation}
Finally, the ultimate Lagrangian is:
\begin{align}
L_5=F_5\sqrt{-X}\Big( \mathcal{U}-\frac{1}{2}K\mathcal{R}\Big) +(G_{5\phi}-F_{5\phi})X\frac{\mathcal{R}}{2}+\frac{(-X)^{3/2}}{3}G_{5X}(-6H^3-6H^2K-3HK^2+3H \mathcal{S})+\frac{G_{5\phi}}{2}X(K^2-\mathcal{S})\,.
\end{align}
 Although $F_5$ is present in the above Lagrangian, it will disappear when computing the EFT functions as was the case for $L_3$. 
At this point we can write down the non zero EFT functions as follows: 
\begin{align}
\Omega(t)&= \frac{2}{m_0^2}\l(G_{5X}\ddot{\phi}_0\dot{\phi}_0^2-G_{5\phi}\frac{\dot{\phi}_0^2}{2}\r)-1\,,\nn\\
c(t)&=\frac{1}{2}\dot{\tilde{\mathcal{F}}}+\frac{3}{2}Hm_0^2\dot{\Omega} -3H^2\dot{\phi}^2_0G_{5\phi}+3H^2\dot{\phi}_0^4G_{5\phi X}-3H^3\dot{\phi}_0^3G_{5X}+2H^3 \dot{\phi}_0^5G_{5XX}\,,\nn\\
%%%%%%%%
\Lambda(t)&=\tilde{\mathcal{F}}-3m_0^2H^2(1+\Omega)+4G_{5X}H^3\dot{\phi}_0^3+3HG_{5\phi}\dot{\phi}_0^2\,,\nn\\
%%%%%%%%%%%%%%%%
M^4_2(t)&=-\frac{\tilde{\mathcal{F}}}{4}-\frac{3}{4}Hm_0^2\dot{\Omega}-2H^3G_{5XXX}\dot{\phi}_0^7-3H^2\dot{\phi}_0^6 G_{5\phi XX}+6G_{5XX}H^3\dot{\phi}_0^5 +6H^2G_{5\phi X}\dot{\phi}_0^4-\frac{3}{2}H^3G_{5X} \dot{\phi}_0^3\,,\nonumber\\
%%%%%%%%%%%%%%%%%%%%
\hat{M}^2(t)&=-G_{5X}\dot{\phi}_0^2\ddot{\phi}_0+HG_{5X}\dot{\phi}_0^3+G_{5\phi}\dot{\phi}_0^2\,,\nonumber\\
%%%%%%%%%%%%%%%%%%%%%%%%%%%%
%%%%%%%%%%%%%%%%%%%%%%
\bar{M}^2_2(t)&=-\bar{M}_3^2(t)=2\hat{M}^2(t)\,,\nonumber\\
%%%%%%%%%%%%%%%%%
\bar{M}_1^3(t)&=-m_0^2\dot{\Omega}+4H\dot{\phi}_0^2G_{5\phi}-4H\dot{\phi}_0^4G_{5\phi X}-4H^2 \dot{\phi}_0^5 G_{5XX}+6H^2\dot{\phi}_0^3 G_{5X}\,,
\end{align}
with $\tilde{\mathcal{F}}=\mathcal{F}-m_0^2\dot{\Omega}-2Hm_0^2 (1+\Omega)=2H^2 G_{5X}\dot{\phi}_0^3+2HG_{5\phi}\dot{\phi}_0^2-m_0^2\dot{\Omega}-2Hm_0^2 (1+\Omega)$.
 We have omitted, in the EFT functions, the dependence on the background quantities $\phi_0$ and $X_0$ of $G_5$ and its derivatives.  Finally we recover, as expected, the relation~(\ref{HorndeskiGG}). 
 
\subsection{GLPV Lagrangians}\label{Sec:GLPVmapping}

We shall now move on to  the beyond Hordenski models derived by Gleyzes \emph{et al.}~\cite{Gleyzes:2014dya,Gleyzes:2014qga}, known as GLPV.
These build on the premises of the Galileon models and include some extra terms in the Lagrangians that, while contributing higher order spatial derivatives in the field equations, maintain second order equations of motion for the true propagating DoF. Specifically, the GLPV action assumes the following form:
\be\label{GLPVaction}
\mathcal{S}_{\rm GLPV}=\int{}d^4x\sqrt{-g}\l[L_2^{GG}+L_3^{GG}+L_4^{GG}+L_5^{GG}+ L_4^{\rm GLPV}+L_5^{\rm GLPV}\r]\,,
\ee
where $L_i^{GG}$(i=2,3,4,5) are the GG Lagrangians listed in Eq.(\ref{GGlagrangians}) and the new terms to be added to the GG Lagrangians are the following:
 \begin{align}\label{GLPVlagrangians}
 L_4^{\rm GLPV}&=\tilde{F}_4(\phi,X)\epsilon^{\mu\nu\rho}_{\phantom{\mu\nu\rho}\sigma}
 \epsilon^{\mu'\nu'\rho'\sigma}\phi_{;\mu}\phi_{;\mu'}\phi_{;\nu\nu'}
 \phi_{\rho\rho'}\,,\nn\\
 L_5^{\rm{GLPV}}&=\tilde{F}_5(\phi,X)\epsilon^{\mu\nu\rho\sigma}
 \epsilon^{\mu'\nu'\rho'\sigma'}\phi_{;\mu}\phi_{;\mu'}\phi_{;\nu\nu'}
 \phi_{;\rho\rho'}\phi_{;\sigma\sigma'}\,,
\end{align}  
where $\epsilon^{\mu\nu\rho\sigma}$  is the totally antisymmetric Levi-Civita tensor and $\tilde{F}_4, \tilde{F}_5$ are two new arbitrary functions of $(\phi,X)$.
 
As usual, we will first express the new Lagrangians in terms of ADM quantities using, among others, relations~\eqref{DDphi}-\eqref{BoxPhi}, and we get:
\begin{align}\label{ADMGLPV}
 L_4^{\rm GLPV}&=-X^2\tilde{F}_4(\phi,X)(K^2-K_{ij}K^{ij})\,, \nn\\
 L_5^{\rm GLPV}&=\tilde{F}_5(\phi,X)(-X)^{5/2}\tilde{\mathcal{K}}=\tilde{F}_5(\phi,X)(-X)^{5/2}(-6H^3-6H^2 K-3H K^2+3HK_{\mu\nu}K^{\mu\nu})\,.
\end{align}
The last equality holds up to second order in perturbations. It is now easy to apply the familiar procedure. Moreover, since  different Lagrangians contribute separately to the EFT functions, we can simply calculate  the EFT functions corresponding to the new Lagrangians~(\ref{ADMGLPV}) and  add those to the results previously derived  for the GG Lagrangians. 

\begin{itemize} 
\item ${\mathbf L_4^{\rm GLPV}}$-\textbf{ Lagrangian }
\end{itemize}
Let us start with the operators included in the $L_4^{\rm GLPV}$ Lagrangian:
\begin{equation}
L_4^{\rm GLPV}=-X^2\tilde{F}_4(K^2-\mathcal{S}).
\end{equation}
We can easily derive the following quantities that are useful for the mapping:
\begin{align}
L_K&=6H\dot{\phi}_0^4\tilde{F}_4,\quad L_{\mathcal{S}}=\dot{\phi}_0^4 \tilde{F}_4,\quad L_{KK}=-2\dot{\phi}_0^4\tilde{F}_4, \quad L_N=4\frac{\dot{\phi}_0^4}{N^5}\tilde{F}_{4}(K^2-\mathcal{S})=24H^2\dot{\phi}_0^4\tilde{F}_4\,,\nn\\
L_{NN}&=-120\dot{\phi}_0^4\tilde{F}_4 H^2,\quad L_{NK}=-24H\dot{\phi}_0^4\tilde{F}_4,\quad L_{N\mathcal{S}}=-4\dot{\phi}_0^4 \tilde{F}_4,\quad \mathcal{F}=4H\dot{\phi}_0^4 \tilde{F}_4\,,\nn\\
\dot{\mathcal{F}}&=4\dot{H}\dot{\phi}_0^4 \tilde{F}_{4}+16H\tilde{F}_4\dot{\phi}_0^3 \ddot{\phi}_0-8H\dot{\phi}_0^5 \ddot{\phi}_0 \tilde{F}_{4X}+4H\dot{\phi}_0^5 \tilde{F}_{4\phi}\,.
\end{align}
Using the relations~(\ref{Map}), we obtain the non-zero EFT functions corresponding to $L_4^{\rm GLPV}$:
\begin{align}
c(t)&=2\dot{H}\dot{\phi}_0^4\tilde{F}_4+8H\dot{\phi}_0^3\ddot{\phi}_0\tilde{F}_4-4H\dot{\phi}_0^5\ddot{\phi}_0\tilde{F}_{4X}+2H\tilde{F}_{4\phi}\dot{\phi}_0^5-12H^2\dot{\phi}_0^4\tilde{F}_4\,,\nn\\
%%%%%%%%%%%%%%%%%%%%
\Lambda(t)&=6H^2\dot{\phi}_0^4\tilde{F}_4+4\dot{H}\dot{\phi}_0^4\tilde{F}_4+
16H\dot{\phi}_0^3\ddot{\phi}_0\tilde{F}_4 +4H\dot{\phi}_0^5\tilde{F}_{4\phi} -8H\dot{\phi}_0^5\ddot{\phi}_0\tilde{F}_{4X}\,,\nn\\
%%%%%%%%%%%%%%%%%%%%%
M^4_2(t)&=-18\dot{\phi}_0^4\tilde{F}_4H^2-\dot{H}\dot{\phi}_0^4\tilde{F}_4
-4H\dot{\phi}_0^3\ddot{\phi}_0\tilde{F}_4+2H\dot{\phi}_0^5\ddot{\phi}_0\tilde{F}_{4X}-H\tilde{F}_{4\phi}\dot{\phi}_0^5+6H^2\dot{\phi}_0^4\tilde{F}_4\,,\nn\\
%%%%%%%%%%%%%%%
\bar{M}^2_2(t)&=2\dot{\phi}_0^4\tilde{F}_4, \nn\\
\bar{M}_1^3(t)&=16H\dot{\phi}_0^4\tilde{F}_4, \nn\\
\bar{M}_3^2(t)&=-\bar{M}_2^2(t)\,.
\end{align}
As before, $\tilde{F}_4$ and its derivatives are evaluated on the background, therefore they only depend on time.

\begin{itemize}
\item ${\mathbf L_5^{\rm GLPV}}$-\textbf{ Lagrangian }
\end{itemize}
Let us now consider the last Lagrangian:
\begin{equation}
L_5^{\rm GLPV}=-(-X)^{5/2}\tilde{F}_5(-6H^3 -6H^2K-3HK^2+3H\mathcal{S})\,,
\end{equation}
which gives the derivatives, w.r.t. ADM quantities, one needs to obtain the mapping:
\begin{align}
L_K&=-12H^2\dot{\phi}_0^5\tilde{F}_5,\quad L_{\mathcal{S}}=-3H\dot{\phi}_0^5\tilde{F}_5,\quad L_{KK}=6H\dot{\phi}_0^5\tilde{F}_5,\quad L_N=5\frac{\dot{\phi}_0^5}{N^6}\tilde{F}_5\tilde{\mathcal{K}}=-30\dot{\phi}_0^5H^3\tilde{F}_5\,,\nn\\
L_{NN}&=180H^3\dot{\phi}_0^5 \tilde{F}_5,\quad L_{NK}=60\dot{\phi}_0^5 \tilde{F}_5 H^2,\quad L_{N\mathcal{S}}=15H\dot{\phi}_0^5 \tilde{F}_5, \quad \mathcal{F}=-6H^2\dot{\phi}_0^5\tilde{F}_5\,,\nn\\
\dot{\mathcal{F}}&=12H^2\dot{\phi}_0^6 \tilde{F}_{5X}\ddot{\phi}_0-12H \dot{H}\dot{\phi}_0^5 \tilde{F}_5-30H^2\dot{\phi}_0^4\tilde{F}_5\ddot{\phi}_0
-6H^2\dot{\phi}_0^6\tilde{F}_{5\phi}\,.
\end{align}
Employing these, allows us to obtain the non-zero EFT functions:
\begin{align}
\Lambda(t)&=-3H^3\dot{\phi}_0^5\tilde{F}_5-12H\dot{H}\dot{\phi}_0^5\tilde{F}_5
-30H^2\dot{\phi}_0^4\tilde{F}_{5}\ddot{\phi}_0
+12H^2\dot{\phi}_0^6\tilde{F}_{5X}\ddot{\phi}_0
-6H^2\dot{\phi}_0^6\tilde{F}_{5\phi}\,,\nn\\
c(t)&=6H^2\dot{\phi}_0^6\ddot{\phi}_0\tilde{F}_{5X}
-6H\dot{H}\dot{\phi}_0^5\tilde{F}_5
-15H^2\dot{\phi}_0^4\tilde{F}_{5}\ddot{\phi}_0
-3H^2\dot{\phi}_0^6\tilde{F}_{5\phi}+15\dot{\phi}_0^5H^3\tilde{F}_5\,,\nn\\
M^4_2(t)&=\frac{45}{2}\dot{\phi}_0^5H^3\tilde{F}_5
+3H\dot{H}\dot{\phi}_0^5\tilde{F}_5+\frac{15}{2}H^2\dot{\phi}_0^4\ddot{\phi}_0\tilde{F}_5-3H^2\dot{\phi}_0^6 \ddot{\phi}_0\tilde{F}_{5X}+\frac{3}{2}H^2\dot{\phi}_0^6 \tilde{F}_{5\phi}\,,\nn\\
\bar{M}_2^2(t)&=-6H\dot{\phi}_0^5 \tilde{F}_5,\nn\\
\bar{M}^3_1(t)&=-30H^2\dot{\phi}_0^5\tilde{F}_5,\nn \\
\bar{M}^2_3(t)&=-\bar{M}_2^2(t)\,.
\end{align}
As usual the functions $\tilde{F}_5$ and its derivatives are functions of time. Their expressions in terms of the scale factor and the Hubble parameter w.r.t. conformal time can be found in Appendix~\ref{Conformal}. Let us notice that GLPV models correspond to:
\be\label{GLPVcondition}
\bar{M}^2_2=-\bar{M}_3^2,
\ee
which is a less restrictive condition than the one defining GG theories~(\ref{HorndeskiGG}); indeed $\bar{M}^2_2\neq 2\hat{M}^2$ for GLPV. 
 
\vspace{0.2cm}
Let us conclude this Section by working out the mapping between the EFT functions and a common way to write the GLPV action. This action is built directly in terms of geometrical quantities, hence guaranteeing the unitary gauge since the scalar DoF has been eaten by the metric~\cite{Gleyzes:2014dya}. Therefore now we will consider the following GLPV Lagrangian instead of the one defined previously:
\ba
L_{\rm GLPV}&=&A_2(t,N)+A_3(t,N)K+A_4(t,N)(K^2-K_{ij}K^{ij})+B_4(t,N)\mathcal{R}\nn\\
&+&A_5(t,N)\l(K^3-3KK_{ij}K^{ij}+2K_{ij}K^{ik}K^{j}_k\r)+B_5(t,N)K^{ij}\l(\mathcal{R}_{ij}-h_{ij}\f{\mathcal{R}}{2}\r)\,,
\ea
where $A_{i}, B_{i}$ are general functions of $t$ and $N$, and can be expressed in terms of the scalar field, $\phi$, , as shown in Ref.~\cite{Gleyzes:2014dya}, effectively creating the equivalence between the above Lagrangian and the one introduced in Eq.~(\ref{GLPVaction}). 

It is very easy to write the above Lagrangian in terms of the quantities introduced in Section~\ref{generallagrangian}, indeed we get:
\ba \label{GLPVADM}
L_{\rm GLPV}&=&A_2(t,N)+A_3(t,N)K+A_4(t,N)(K^2-\mathcal{S})+B_4(t,N)\mathcal{R}\nn\\
&+&A_5(t,N)\l(-6H^3-6H^2K-3HK^2+3H\mathcal{S}\r)+B_5(t,N)\l(\mathcal{U}-\f{\mathcal{R}K}{2}\r)\,.
\ea
Now, we can compute the quantities that we need for the mapping~(\ref{Map}):
\ba
&&\bar{L}=\bar{A}_2-3H\bar{A}_3+6H^2\bar{A}_4-6H^3\bar{A}_5\,, \qquad \mathcal{E}=\bar{B}_4-\f{1}{2}\dot{\bar{B}}_5\,, \qquad \mathcal{F}=\bar{A}_3-4H\bar{A}_4+6H^2\bar{A}_5\,, \qquad L_{\mathcal{S}}=-\bar{A}_4+3H\bar{A}_5\,,\nn\\
&&L_K=\bar{A}_3-6H\bar{A}_4+12H^2\bar{A}_5\,,\qquad L_N=\bar{A}_{2N}-3H\bar{A}_{3N}+6H^2\bar{A}_{4N}-6H^3\bar{A}_{5N}\,, \qquad L_{\mathcal{U}}=\bar{B}_5\,,\nn\\
&&L_{NN}=\bar{A}_{2NN}-3H\bar{A}_{3N}+6H^2\bar{A}_{4NN}-6H^3\bar{A}_{5NN}\,,\qquad L_{KK}=2\bar{A}_4-6H\bar{A}_5\,,\qquad L_{\mathcal{S}N}=-\bar{A}_{4N}+3H\bar{A}_{5N}\,,\nn\\
&&L_{KN}=\bar{A}_{3N}-6H\bar{A}_{4N}+12H^2\bar{A}_{5N}\,,\qquad L_{K\mathcal{R}}=-\f{1}{2}\bar{B}_5\,,\qquad L_{N\mathcal{U}}=\bar{B}_{5N}\,, \qquad L_{N\mathcal{R}}=\bar{B}_{4N}+\f{3}{2}H\bar{B}_{5N}\,,
\ea
where the quantities with the bar are evaluated in the background and $A_{iY}$ means derivative of $A_i$ w.r.t. $Y$. Then  the EFT functions follow from Eq.~(\ref{Map}):
\ba
\Omega(t)&=&\f{2}{m_0^2}\l(\bar{B}_4-\f{1}{2}\dot{\bar{B}}_5\r)-1\,,\nn\\
\Lambda(t)&=&\bar{A}_2-6H^2\bar{A}_4+12H^3\bar{A}_5+\dot{\bar{A}}_3-4\dot{H}\bar{A}_4-4H\dot{\bar{A}}_4+6H^2\dot{\bar{A}}_5+12H\dot{H}\bar{A}_5\nn\\
&&-\l[2(3H^2+2\dot{H})\l(\bar{B}_4-\f{1}{2}\dot{\bar{B}}_5\r)+2\ddot{\bar{B}}_4-\bar{B}^{(3)}_5+4H\l(\dot{\bar{B}}_4-\f{1}{2}\ddot{\bar{B}}_5\r)\r] \,,\nn\\
c(t)&=& \f{1}{2}\l(\dot{\bar{A}}_3-4\dot{H}\bar{A}_4-4H\dot{\bar{A}}_4+6H^2\dot{\bar{A}}_5+12H\dot{H}\bar{A}_5-\bar{A}_{2N}+3H\bar{A}_3N-6H^2\bar{A}_{4N}+6H^3\bar{A}_{5N}\r)\nn\\
&&+H\l(\dot{\bar{B}}_4-\f{1}{2}\ddot{\bar{B}}_5\r)-\ddot{\bar{B}}_4+\f{1}{2}\bar{B}_5^{(3)}-2\dot{H}\l(\bar{B}_4-\f{1}{2}\dot{\bar{B}}_5\r) \,,\nn\\
\bar{M}^2_2(t)&=&=-2\bar{A}_4+6H\bar{A}_5-2\bar{B}_4+\dot{\bar{B}}_5\,,\nn\\
\bar{M}^3_1(t)&=&-\bar{A}_{3N}+4H\bar{A}_{4N}-6H^2\bar{A}_{5N}-2\dot{\bar{B}}_4+\ddot{\bar{B}}_5\,,\nn\\
\bar{M}^2_3(t)&\equiv& -\bar{M}^2_2(t)\,,\nn\\
M^4_2(t)&=& \f{1}{4}\l(\bar{A}_{2NN}-3H\bar{A}_{3NN}+6H^2\bar{A}_{4NN}-6H^3\bar{A}_{5NN}\r)-\f{1}{4}\l(\dot{\bar{A}}_3-4\dot{H}\bar{A}_4-4H\dot{\bar{A}}_4+6H^2\dot{\bar{A}}_5+12H\dot{H}\bar{A}_5\r)\nn\\
&&+\f{3}{4}\l(\bar{A}_{2N}-3H\bar{A}_{3N}+6H^2\bar{A}_{4N}-6H^3\bar{A}_{5N}\r)-\f{1}{2}\l[H\l(\dot{\bar{B}}_4-\f{1}{2}\ddot{\bar{B}}_5\r)-\ddot{\bar{B}}_4+\f{1}{2}\bar{B}_5^{(3)}-2\dot{H}\l(\bar{B}_4-\f{1}{2}\dot{\bar{B}}_5\r)\r]\,,\nn\\
\hat{M}^2(t)&=&\bar{B}_{4N}+\f{1}{2}H\bar{B}_{5N}+\f{1}{2}\dot{\bar{B}}_5\,.
\ea
The condition (\ref{GLPVcondition}) is satisfied as desired and one can focus on the GG subset of theories by enforcing the condition $\bar{M}^2_2(t)=2\hat{M}^2(t)$ .

\subsection{Ho\v rava Gravity}\label{Sec:Horava}

One of the main aspects of our paper is the inclusion of  operators with higher order spatial derivatives in the EFT action. Thus, it is natural to proceed with the mapping of the most popular theory containing such operators, i.e. Ho\v rava gravity~\cite{Horava:2008ih,Horava:2009uw}. This theory is a recent proposed candidate to describe the gravitational interaction in the ultra-violet regime (UV). This is done by breaking the Lorentz symmetry resulting in a modification of the graviton propagator. Practically, this amounts to adding higher-order spatial derivatives to the action while keeping the time  derivatives at most second order, in order to avoid Ostrogradski instabilities~\cite{Ostrogradski}. As a result, time and space are treated on a different footing, therefore the natural formulation in which to construct the action is the ADM one. It has been shown that, in order to obtain a power-counting renormalizable theory,  the action needs to contain terms with up to sixth-order spatial derivatives~\cite{Visser:2009fg,Visser:2009ys,Blas:2009qj}. The resulting action  does not demonstrate full diffeomorphism invariance but is rather invariant under a restricted symmetry, the foliation preserving diffeomorphisms (for a review see~\cite{Sotiriou:2010wn,Mukohyama:2010xz} and references therein). Besides the UV regime, Ho\v rava gravity  has taken hold on the cosmological side as well as it exhibits a rich phenomenology~\cite{Calcagni:2009ar,Kiritsis:2009sh,Brandenberger:2009yt,Mukohyama:2009gg,Cai:2009dx,Chen:2009jr,Cai:2010hi,Carroll:2004ai,Gao:2009ht,Wang:2009yz,Kobayashi:2009hh,Kobayashi:2010eh} and very recently it has started to be constrained in that context~\cite{Zuntz:2008zz,Dutta:2009jn,Dutta:2010jh,Audren:2013dwa,Blas:2012vn,Audren:2014hza,Frusciante:2015maa}.

Here, we will consider the following action which contains up to six order spatial derivatives, (and is  therefore  included in the extended EFT action): 
\begin{eqnarray}\label{actionhorava}
\mathcal{S}_{H}&=&\f{1}{16\pi G_H}\int{}d^4x\sqrt{-g}\left[K_{ij}K^{ij}-\lambda K^2 -2 \xi\bar{\Lambda}+\xi \mathcal{R}+\eta a_i a^i + g_1\mathcal{R}^2+g_2\mathcal{R}_{ij}\mathcal{R}^{ij}+g_3\mathcal{R}\nabla_i a^i\right. \nn \\ &&\,\,\,\,\left.+g_4 a_i \Delta a^i +g_5\mathcal{R} \Delta \mathcal{R}+g_6\nabla_i\mathcal{R}_{jk}\nabla^i\mathcal{R}^{jk}+g_7a_i\Delta^2 a^i+g_8\Delta\mathcal{R}\nabla_i a^i \right],
\end{eqnarray}
where  the coefficients $\lambda$, $\eta$, $\xi$ and $g_i$ are running coupling constants,  $\bar{\Lambda}$ is the ''bare'' cosmological constant and $G_H$ is the coupling constant~\cite{Blas:2009qj,Frusciante:2015maa}:
\be
\f{1}{16\pi G_H}=\f{m_0^2}{(2\xi-\eta)}\,.
\ee
The above action is already in unitary gauge and ADM form, then we just need few steps to write it in terms of the quantities introduced in Section~\ref{generallagrangian}:
\begin{eqnarray}
\mathcal{S}_{H}&=&\f{1}{16\pi G_H}\int{}d^4x\sqrt{-g}\left[S-\lambda K^2 -2 \xi\bar{\Lambda}+\xi \mathcal{R}+\eta \alpha_1 + g_1\mathcal{R}^2+g_2\mathcal{Z}+g_3\alpha_3+g_4 \alpha_2 -g_5\mathcal{Z}_1+g_6\mathcal{Z}_2+g_7\alpha_4+g_8\alpha_5 \right],\nn\\
&&
\end{eqnarray}
 then by using the results~(\ref{Map}) it is easy to show that the EFT functions read:
\begin{align}\label{horavamapping}
&m_0^2(1+\Omega)=\f{2m_0^2 \xi}{(2\xi-\eta)}, \qquad  c(t)= -\f{m_0^2}{(2\xi-\eta)}(1+2\xi-3\lambda)\dot{H},\nn \\
&\Lambda(t)=\f{2m_0^2}{(2\xi-\eta)}\l[-\xi\bar{\Lambda}-(1-3\lambda+2\xi)\l(\f{3}{2}H^2+\dot{H}\r)\r], \nn\\
&\bar{M}_3^2= -\f{2m_0^2}{(2\xi-\eta)}(1-\xi),\qquad \bar{M}_2^2 =-2\f{m_0^2}{(2\xi-\eta)}(\xi-\lambda), \qquad m^2_2=\f{m_0^2}{4(2\xi-\eta)}\eta,\nn \\
&M_2^4(t)=\f{m_0^2}{2(2\xi-\eta)}(1+2\xi-3\lambda)\dot{H}, \qquad \lambda_1=g_1\f{m_0^2}{(2\xi-\eta)},\qquad \lambda_2=g_2\f{m_0^2}{(2\xi-\eta)},\nn \\
&\lambda_3=g_3\f{m_0^2}{2(2\xi-\eta)},\qquad \lambda_4=g_4\f{m_0^2}{4(2\xi-\eta)},\qquad \lambda_5=-g_5\f{m_0^2}{(2\xi-\eta)} \nn \\
&\lambda_6=g_6\f{m_0^2}{(2\xi-\eta)},\qquad \lambda_7=g_7\f{m_0^2}{4(2\xi-\eta)},\qquad \lambda_8=g_8\f{m_0^2}{2(2\xi-\eta)},
\end{align}
and the remaining EFT functions are zero. The mapping of Ho\v rava gravity has been worked out in details in Ref.~\cite{Frusciante:2015maa}, by some of the authors of this paper. 
Subsequently, the low-energy part of Ho\v rava action, which is described by  $\{\Omega, c, \Lambda, \bar{M}_3^2, \bar{M}_2^2,M_2^4, m^2_2 \}$, has been implemented in EFTCAMB~\cite{Hu:2014oga} and constraints on the low-energy parameters $\{\xi,\eta,\lambda\}$ have been obtained in Ref.~\cite{Frusciante:2015maa}.

\section{Stability}\label{Sec:stability}

Along with its unifying aspect, a very important advantage of the EFT formalism, which we already mentioned, is that of being formulated at the level of the action. This in fact offers a  powerful, model-independent handle on the theoretical viability of the theories explored within this framework. Indeed, by inspecting the EFT action expanded to quadratic order in the perturbations, it is possible to impose conditions on the EFT functions to ensure that unphysical behaviours do not develop. This is done at the level of the action, before making any choice for the functional form of the EFT functions, hence the resulting conditions are very general. As it has been preliminary shown in Ref.~\cite{Raveri:2014cka}, the impact of such  conditions can be quite significant as they can efficiently reduce the parameter space that one needs to explore when performing a fit to data. In some cases they have been shown to dominate over the constraining power of current data~\cite{Raveri:2014cka}.

The study of the theoretical viability of the EFT action has already been performed to some extent in the literature~\cite{Gubitosi:2012hu,Bloomfield:2012ff,Gleyzes:2013ooa,Piazza:2013pua,Gleyzes:2014qga}, however here we will include in the analysis, for the first time,  higher order operators and consider also the instabilities related to a negative squared mass of the scalar DoF. Specifically, we will consider three possible instabilities:  ghost and gradient instabilities both in the scalar and tensor sector, and tachyonic scalar modes (for a review see Ref.~\cite{Sbisa:2014pzo}). Starting from the general action~(\ref{EFTADM}), we expand it up to quadratic order in tensor and scalar perturbations of the metric around a flat FLRW background. Our focus is on  the stability of the gravity sector, hence we will  not consider matter fluids. The complete analysis of the stability of the general action~(\ref{EFTADM}) in the presence of a matter sector is  work in progress~\cite{generalstability}. 

Let us consider the following metric perturbations for the scalar components:
\be
ds^2=-(1+2\delta N)dt^2+2\partial_i\psi dtdx^i+a^2(1+2\zeta)\delta_{ij}dx^idx^j \,,
\ee
where as usual $\delta N (t,x^i)$ is the perturbation of the lapse function, $\partial_i\psi(t,x^i)$ and $\zeta(t,x^i)$ are the scalar perturbations respectively of the shift function and the three dimensional metric. Then, the scalar perturbations of the quantities involved in the  action~(\ref{EFTADM}) are: 
\ba\label{scalarperturbations}
&&\delta K=-3\dot{\zeta}+3H\delta N+\f{1}{a^2}\partial^2\psi\,, \nn\\
&&\delta K_{ij}=a^2\delta_{ij}(H\delta N-2H\zeta-\dot{\zeta})+\partial_i\partial_j\psi\,, \nn\\
&&\delta K^i_j=(H\delta N-\dot{\zeta})\delta^i_j+\f{1}{a^2}\partial^i\partial_j\psi\,, \nn\\
&&\delta \mathcal{R}_{ij}=-(\delta_{ij}\partial^2\zeta+\partial_i\partial_j\zeta)\,, \nn\\
&&\delta_1 \mathcal{R}=-\f{4}{a^2}\partial^2\zeta\,, \nn\\
&&\delta_2 \mathcal{R}=-\f{2}{a^2}[(\partial_i\zeta)^2-4\zeta\partial^2\zeta].
\ea
Now, we can expand action~(\ref{EFTADM}) to quadratic order in metric perturbations. In the following we will Fourier transform the spatial part \footnote{More properly, in Fourier space we should write $(\zeta(t,k))^2 \rightarrow \zeta_{\textbf{k}}\zeta_{-\textbf{k}}$,  however in the following we prefer to drop the indices in order to simplify the notation. } and after regrouping  terms, we obtain:
\ba\label{full}
\mathcal{S}_{EFT}^{(2)}&=&\f{1}{(2\pi)^3}\int{}d^3kdt\,a^3\l\{-\l(\mathcal{W}_0+\mathcal{W}_{3}k^2 +\mathcal{W}_{2}k^4 \r)k^2\zeta^2-3a^2\mathcal{W}_4\dot{\zeta}\delta N-\f{3}{2}a^2\mathcal{W}_5(\dot{\zeta})^2\r.\nn\\
&&\l.-\l(\mathcal{W}_4\delta N+\mathcal{W}_5\dot{\zeta}-\mathcal{W}_7 k^2\psi+\f{2}{a^4}\bar{m}_5k^2\zeta \r)k^2\psi
 +\l(\mathcal{W}_1+4m_2^2\f{k^2}{a^2}-4\f{\lambda_4}{a^4}k^4+4\f{\lambda_7}{a^6}k^6\r)(\delta N)^2\r.\nn\\
&&\l.-\l(\mathcal{W}_6+8\lambda_3\f{k^2}{a^4}+8\f{\lambda_8}{a^6}k^4\r) \delta N k^2\zeta \r\}\,,
\ea
where: 
\ba
&&\mathcal{W}_0=-\f{1}{a^2}\l[m_0^2(1+\Omega)+3H\bar{m}_5+3\dot{\bar{m}}_5\r]\,,\nn \\
&&\mathcal{W}_1= c+2M^4_2-3m_0^2H^2(1+\Omega)-3m_0^2H\dot{\Omega}-\f{3}{2}H^2\bar{M}^2_3-\f{9}{2}H^2\bar{M}^2_2-3H\bar{M}^3_1\,, \nn \\
&&\mathcal{W}_{2}=-16\frac{\lambda_5}{a^6}-6\frac{\lambda_6}{a^6}  \,,\nn\\
&&\mathcal{W}_{3}=-16\frac{\lambda_1}{a^4}-6\frac{\lambda_2}{a^4} \,,\nn\\
&&\mathcal{W}_4= \f{1}{a^2}\l(-2m_0^2H(1+\Omega)-m_0^2\dot{\Omega}-H\bar{M}^2_3-\bar{M}^3_1-3H\bar{M}^2_2 \r)\,,\nn\\ 
&&\mathcal{W}_5= \f{1}{a^2}\l( 2m_0^2(1+\Omega) +\bar{M}^2_3+3\bar{M}^2_2\r)\,,\nn\\
&&\mathcal{W}_6= -\f{4}{a^2}\l(\f{1}{2}m_0^2(1+\Omega)+\hat{M}^2\r)-6H\f{\bar{m}_5}{a^2}\,,\nn\\
&&\mathcal{W}_7= -\f{1}{2a^4}\l(\bar{M}^2_3+\bar{M}^2_2\r)\,.
\ea

In this action we have three DoFs $\{\zeta,\delta N, \psi\}$, but in reality only one, $\zeta$, is dynamical, while the other two, $\{\delta N, \psi\}$, are auxiliary fields. This implies that they can be eliminated through the constraint equations obtained by varying the above action  w.r.t. them. We will leave for the next Sections the details of such a calculation, here we want to outline the general procedure we are adopting.  After replacing back in the action the general expression for $\delta N$ and $\psi$ , we end up with an action of the form:
\be\label{action1}
\mathcal{S}_{EFT}^{(2)}=\f{1}{(2\pi)^3}\int{}d^3kdt\,a^3\l\{\mathcal{L}_{\dot{\zeta}\dot{\zeta}}(t,k)\dot{\zeta}^2-\l[k^2G(t,k)+\bar{M}(t,k)\r] \zeta^2\r\}.
\ee
where $\bar{M}(t,k)$  depends on inverse powers of k. $\mathcal{L}_{\dot{\zeta}\dot{\zeta}}(t,k) $ is usually called the kinetic term and its positivity guarantees that the theory is free from ghost in the scalar sector. The variation of the above action w.r.t. $\zeta$ gives:
\be
\ddot{\zeta}+\l(3H+\f{\dot{\mathcal{L}}_{\dot{\zeta}\dot{\zeta}}}{\mathcal{L}_{\dot{\zeta}\dot{\zeta}}}\r)\dot{\zeta}+\l(k^2\f{G}{\mathcal{L}_{\dot{\zeta}\dot{\zeta}}}+\f{\bar{M}}{\mathcal{L}_{\dot{\zeta}\dot{\zeta}}}\r)\zeta=0\,,
\ee
where the coefficient of $\dot{\zeta}$ is called the friction term and its sign will damp  or enhance the amplitude of the field fluctuations.   $\bar{M}/\mathcal{L}_{\dot{\zeta}\dot{\zeta}}$ is called the dispersion coefficient which, in principle, can be both negative and positive. Finally, we define the propagation speed as:
\be\label{speeddef}
c_s^2\equiv \f{G}{\mathcal{L}_{\dot{\zeta}\dot{\zeta}}}.
\ee 
%which must be positive definite in order to avoid gradient instabilities. 
Let us note that the  speed of propagation and the dispersion coefficient (or ''mass'' term)  and their effective counterparts have non-local expressions. Therefore, their interpretation as the actual physical entities might be ambiguous at first glance because usually these quantities are defined in some specific limit, where they assume local expressions. In this work, we still retain the labeling of speed of propagation and mass term for the non-local expressions, because they reduce to the corresponding  local and physical quantity when the proper limit is considered. Moreover, the non-local definitions are the ones which serve to our purpose, since they represent the proper quantities on which the stability conditions have to be imposed in order to guarantee a viable theory at all times and scales. 

Now, let us  perform a field redefinition in order to have a canonical action. This step is important in order to identify the correct  conditions to avoid the gradient and tachyonic instabilities, in particular the last one which is related to the  condition of boundedness  from below of the corresponding canonical Hamiltonian. 
 We will show that not only the mass is sensitive to this normalization, as it is known, but that in the general case in which the kinetic term is scale-dependent also the speed of propagation, is affected by the field redefinition. In general, we can use:
\be\label{redefinition}
\zeta(t,k)=\f{\phi(t,k)}{\sqrt{2\mathcal{L}_{\dot{\zeta}\dot{\zeta}}(t,k)}},
\ee
which, once applied to the action~(\ref{action1}), gives: 
\be\label{afternormalization}
\mathcal{S}_{EFT}^{(2)}=\f{1}{(2\pi)^3}\int{}d^3kdt\,a^3\l[\f{1}{2}\dot{\phi}^2-c_{s,{\rm eff}}^2(t,k)\f{k^2}{2}\phi^2-m_{{\rm eff}}^2(t,k)\phi^2\r]\,,
\ee
where $m_{{\rm eff}}(t,k)$ is an effective mass and depends on inverse powers of k, while $c_{s,{\rm eff}}^2(t,k)$ is the effective speed of propagation. 

When $\mathcal{L}_{\dot{\zeta}\dot{\zeta}}$ is only a function of time, the field redefinition~(\ref{redefinition}) will give time-dependent contributions only to $\bar{M}$ thus generating $m_{{\rm eff}}^2$ and leaving $G$ unaffected. In this case we have:
\ba \label{effective}
&&c_{s,{\rm eff}}^2(t,k)= c^2_s(t,k)\,, \nn\\
&&m_{\rm eff}^2(t)= \frac{ \mathcal{L}_{\dot{\zeta}\dot{\zeta}} \left(4 \bar{M}(t)-2 \ddot{\mathcal{L}}_{\dot{\zeta}\dot{\zeta}}\right)+\dot{\mathcal{L}}_{\dot{\zeta}\dot{\zeta}}^2-6 H\mathcal{L}_{\dot{\zeta}\dot{\zeta}} \dot{\mathcal{L}}_{\dot{\zeta}\dot{\zeta}}}{8  \mathcal{L}_{\dot{\zeta}\dot{\zeta}}^2}\,.
\ea
Let us notice that in case in which the kinetic term depends only on time, the term $\bar{M}$ usually turns out to be zero or at most a function of time. 

On the contrary, when $\mathcal{L}_{\dot{\zeta}\dot{\zeta}}$ exhibits a k-dependence, the field redefinition will affect both $\bar{M}$ and $G$ and in general $c_{s,{\rm eff}}^2\neq c^2_s$ and the above expression for the effective mass does not hold anymore. In Section~\ref{Sec:bGLPV} we will discuss the general expressions for these two quantities. In general, the GLPV class of theories belongs to the case in which $\mathcal{L}_{\dot{\zeta}\dot{\zeta}}$ is only a function of time. When one starts including operators like $\{m^2_2, \bar{m}_5,\lambda_i,\bar{M}^2_3\neq-\bar{M}^2_2\}$,  k-dependence will be generated in the kinetic term. In the following sections we will analyse these cases in details. 

Finally, in order to study the stability, one has to analyse the evolution of the field equation obtained by varying the action~(\ref{afternormalization}) w.r.t. $\phi$, i.e.:
\be
\ddot{\phi}+ 3H\dot{\phi}+\l(k^2c_{s,{\rm eff}}^2+m_{{\rm eff}}^2\r)\phi =0,
\ee
In this case $H$ represents a friction term, which is always positive, and $m_{{\rm eff}}^2$ is the  dispersion coefficient. A negative value of the effective mass squared generates tachyonic instability, however  requiring $m_{{\rm eff}}^2$ to be positive is a stringent condition, indeed to guarantee stability it is necessary to ensure that the time scale on which the instability occurs is longer then the time evolution of the system~\cite{Sbisa:2014pzo}. Therefore, we can require that it is longer than the Hubble time, $H_0$. Moreover, one has to consider also the condition to avoid gradient instabilities which is obtained by enforcing a positive value of the effective speed of propagation.  In the simpler cases in which the kinetic term depends only on time (e.g. Horndeski and GLPV theories),  the normalization of the field leaves the speed of sound unchanged, i.e. $c_s^2=c_{\rm eff}^2$, thus the condition to impose is $c^2_{\rm eff}=c_s^2>0$.  For the more general case in which the kinetic term depends also on scale,  $c_{\rm eff}^2 = c_s^2+f(t,k)$ (see Section~\ref{Sec:bGLPV} for the full expression of $f(t,k)$); however, in the high k-limit, where the gradient instability shows up, $f(t,k)$ is maximally of order $\mathcal{O}(1/k^2)$ which can be neglected in this limit. Therefore, the condition on the effective speed of propagation reduces indeed to the original condition on the speed of propagation, i.e. $c_s^2>0$. In summary, in order to guarantee the stability of the scalar sector  the combination of $c_{\rm eff}^2>0$ and $m^2_{\rm eff}>0$, along with the no-ghost condition, i.e. $\mathcal{L}_{\dot{\zeta}\dot{\zeta}}>0$, provides the full set of stability conditions.

We conclude with the stability analysis on the tensor modes. The perturbed metric components which contribute to tensor modes are:
\be
g_{ij}^T(t,x^i)=a^2h_{ij}^T(t,x^i)\,,
\ee
therefore, the terms containing tensor perturbations in~(\ref{EFTADM}), are the following: 
\ba \label{tensorperturbations}
\delta K^i_j=-\f{\dot{h}^{i\,T}_j}{2} \qquad \delta \mathcal{R}_{ij}=-\f{\delta^{lk}}{a^2}\partial_l\partial_k h_{ij}^T \qquad \delta_2\mathcal{R}= \f{1}{a^2}\Big(\f{3}{4}\partial_k h_{ij}^T \partial^k h^{ij\,T}+h_{ij}^T\partial^2 h^{ij\, T}-\f{1}{2}\partial_k h_{ij}^T\partial^j h^{ik \,T}\Big)\,, 
\ea
where $\delta_2\mathcal{R}$ is the second order perturbation of the Ricci scalar, $\mathcal{R}$. Then, the EFT action for tensor perturbations  up to second order reads:
\be
\mathcal{S}_{EFT}^{T\,(2)}=\int{}d^4x\,a^3\l\{\f{m_0^2}{2}(1+\Omega)\delta_2\mathcal{R}+\l(\f{m_0^2}{2}(1+\Omega)-\f{\bar{M}^2_3}{2}\r)\delta K^i_j\delta K^j_i+\lambda_2\delta \mathcal{R}_{ij}\delta \mathcal{R}^{ij}+\lambda_6\f{\tilde{g}^{kl}}{a^2}\partial_k \mathcal{R}_{ij}  \partial_l \mathcal{R}^{ij}  \r\},
\ee
from which we can notice that only four EFT functions describe the dynamics of tensors, i.e. $\{\Omega, \bar{M}^2_3, \lambda_2, \lambda_8\}$. Among the extra operators  that we added in action~(\ref{EFTADM}), only two  contribute to tensor modes $\{\lambda_2, \lambda_8\}$. Now, using~(\ref{tensorperturbations}), the action becomes:
\be\label{actiontensor}
\mathcal{S}_{EFT}^{T\,(2)}=\int{}d^4x\,a^3\l\{-\f{m_0^2}{2}(1+\Omega)\f{1}{4a^2}(\partial_k h_{ij}^T)^2 +\l(\f{m_0^2}{2}(1+\Omega)-\f{\bar{M}^2_3}{2}\r)\f{(\dot{h}_{ij}^T)^2}{4}+\lambda_2\l(\f{\delta^{lk}}{a^2} \partial_l\partial_k h_{ij}^T\r)^2+\lambda_6\f{1}{a^6}(\partial_k \partial_l\partial^l h_{ij}^T)^2  \r\}.
\ee
It is clear that the additional operators associated to higher spatial derivatives do not affect the kinetic term. However, they  affect the speed of propagation of the tensor modes, as we will show in the following. Indeed, action~(\ref{actiontensor}) can be written in the compact form: 
\be
\mathcal{S}_{EFT}^{T\,(2)}=\f{1}{(2\pi)^3}\int{}d^3k dt\,a^3\f{A_T(t)}{8}\l[(\dot{h}_{ij}^T)^2-\f{c^2_T(t,k)}{a^2}k^2(h_{ij}^T)^2\r]\,,
\ee
with
\ba\label{tensordefinition}
&&A_T(t)=m_0^2(1+\Omega)-\bar{M}^2_3 \,, \nn\\
&&c^2_T(t,k)=\bar{c}^2_T(t)-8\f{\lambda_2 \f{k^2}{a^2}+\lambda_6 \f{k^4}{a^4}}{m_0^2(1+\Omega)-\bar{M}^2_3}\,, \nn\\
&&\bar{c}_T^2(t)=\f{m_0^2(1+\Omega)}{m_0^2(1+\Omega)-\bar{M}^2_3},
\ea
where we have Fourier transformed the spatial part.  $\bar{c}_T^2$  is the tensor speed of propagation for  all the theories belonging to the GLPV class, as shown in Refs.~\cite{Bellini:2014fua,Gleyzes:2014qga}. However, GLPV theories are characterized by the condition $\bar{M}^2_3(t)= -\bar{M}^2_2(t)$, while the present definition of the tensor speed does not rely on this constraint as it holds for a wider class of theories.   In order to avoid the development of instabilities in the tensorial sector, one generally demands the kinetic term to be positive, i.e. $A_T>0$,  and to have a positive speed of propagation $c_T^2>0$. From Eqs.~(\ref{tensordefinition}) it is easy to identify the corresponding conditions on the EFT functions.

\subsection{Stability conditions for the GLPV class of theories}

Let us focus on the GLPV class of theories by considering the appropriate set of operators:
\be\label{stabGLPV}
\mathcal{S}_{\rm{GLPV}}^{(2)}=\f{1}{(2\pi)^3}\int{}d^3kdt\,a^3\l[-\mathcal{W}_6\delta N k^2 \zeta - \mathcal{W}_4\delta N k^2 \psi -\mathcal{W}_5k^2 \psi \dot{\zeta}- \mathcal{W}_0k^2 \zeta^2 +\mathcal{W}_1(\delta N)^2 -3a^2\mathcal{W}_4\delta N  \dot{\zeta}-\f{3}{2}a^2\mathcal{W}_5 \dot{\zeta}^2\r]\,,
\ee
which is obtained from action~(\ref{full}) by imposing the following constraints:
\be
\mathcal{W}_7=0\,, \qquad \l\{m_2^2,\bar{m}_5,\lambda_i \r\}=0.
\ee
By varying the above action w.r.t. $\delta N$ and $\psi$ we get, respectively,:
\ba
&&-\mathcal{W}_6k^2 \zeta -\mathcal{W}_4k^2 \psi +2 \mathcal{W}_1\delta N -3a^2\mathcal{W}_4 \dot{\zeta}=0\,,\nn \\
&&-\mathcal{W}_4\delta N -\mathcal{W}_5 \dot{\zeta}=0.
\ea
Inverting these relations gives:
\ba
&&\delta N=-\frac{\mathcal{W}_5 }{\mathcal{W}_4}\dot{\zeta}\,,\nn\\
&&k^2\psi=-\frac{1}{\mathcal{W}_4^2}\l[\left(3 a^2 \mathcal{W}_4^2+2 \mathcal{W}_1 \mathcal{W}_5\right) \dot{\zeta}+  \mathcal{W}_4 \mathcal{W}_6k^2 \zeta\r]\,,
\ea
which, once substituted back  in the action~(\ref{stabGLPV}), yields: 
\be \label{GLPVonefield}
\mathcal{S}_{\rm{GLPV}}^{(2)}=\f{1}{(2\pi)^3}\int{}d^3kdt\,a^3 \l\{\l(\frac{3}{2} a^2 \mathcal{W}_5 +\frac{\mathcal{W}_1 \mathcal{W}_5^2 }{\mathcal{W}_4^2}\r)\dot{\zeta}^2-k^2\l[\f{3}{2}H \frac{\mathcal{W}_5 \mathcal{W}_6 }{\mathcal{W}_4}+ \f{1}{2}\f{d}{dt}\l(\frac{\mathcal{W}_5 \mathcal{W}_6 }{\mathcal{W}_4}\r)+ \mathcal{W}_0\r] \zeta^2\r\}\,.
\ee
This particular form has been obtained after integrating by parts the term containing $\dot{\zeta}\zeta$. The above action has the same form of~(\ref{action1}), where $\bar{M}=0$. Therefore, it is easy to read  the no-ghost condition: 
\be\label{ghostGLPV}
\mathcal{L}_{\dot{\zeta}\dot{\zeta}}(t)\equiv\frac{3}{2} a^2 \mathcal{W}_5+\frac{\mathcal{W}_1 \mathcal{W}_5^2}{\mathcal{W}_4^2}>0\,,
\ee
and the condition on the speed of propagation ($c_s^2>0$): 
\be
c_s^2(t)=\frac{3 H \mathcal{W}_5 \mathcal{W}_6 \mathcal{W}_4+\mathcal{W}_6 \mathcal{W}_4 \dot{\mathcal{W}}_5+\mathcal{W}_5 \mathcal{W}_4 \dot{\mathcal{W}}_6-\mathcal{W}_5 \mathcal{W}_6 \dot{\mathcal{W}}_4+2 \mathcal{W}_0 \mathcal{W}_4^2}{3 a^2 \mathcal{W}_5 \mathcal{W}_4^2+2 \mathcal{W}_1 \mathcal{W}_5^2}\,.
\ee
The speed of propagation coincides with the phase velocity due to the lack of k-dependence in the kinetic term, as discussed at  earlier stage. Additionally, this implies that only the mass term will be sensitive to the field redefinition which, in this case, reads: 
\be
\zeta(t,k)=\frac{\phi(t,k) }{\sqrt{2 \left(\frac{3}{2} a^2 \mathcal{W}_5+\frac{\mathcal{W}_1 \mathcal{W}_5^2}{\mathcal{W}_4^2}\right)}}\,.
\ee
After this transformation the effective mass follows directly form Eq.~(\ref{effective}), i.e.:
\be
m_{\rm{eff}}^2(t)=\frac{ -2\mathcal{L}_{\dot{\zeta}\dot{\zeta}}\ddot{\mathcal{L}}_{\dot{\zeta}\dot{\zeta}}+\dot{\mathcal{L}}_{\dot{\zeta}\dot{\zeta}}^2-6 H\mathcal{L}_{\dot{\zeta}\dot{\zeta}} \dot{\mathcal{L}}_{\dot{\zeta}\dot{\zeta}}}{8  \mathcal{L}_{\dot{\zeta}\dot{\zeta}}^2}\,,
 \ee
where the kinetic term is given by Eq.~(\ref{ghostGLPV}).

\subsection{Stability conditions for the class of theories beyond GLPV}\label{Sec:bGLPV}

To go beyond the GLPV class of theories we start by naively considering the general action~(\ref{full}) with all the higher order operators. We proceed to integrate out the auxiliary fields $\delta N$ and $\psi$ by solving the following field equations:
\ba\label{fieldeqsauxiliaryfields}
&&-2\bar{m}_5 \frac{k^2}{a^4}\zeta +2\mathcal{W}_7 k^2 \psi -\mathcal{W}_4\delta N -\mathcal{W}_5 \dot{\zeta}=0\,,\nn\\
&&8\l(m_{2}^2-\f{\lambda_4}{a^2}k^2+\f{\lambda_7}{a^4}k^4\r)\frac{k^2}{a^2} \delta N -\l(\mathcal{W}_6+8\lambda_3\f{k^2}{a^4}+8\f{\lambda_8}{a^6}k^4\r)k^2 \zeta -  \mathcal{W}_4k^2 \psi+2  \mathcal{W}_1\delta N-3a^2 \mathcal{W}_4 \dot{\zeta}=0\,,
\ea
and we finally end up with an action of the form:
\be\label{actiondef2}
\mathcal{S}_{EFT}^{(2)}=\f{1}{(2\pi)^3}\int{}d^3kdt\,a^3 \left\{\mathcal{L}_{\dot{\zeta}\dot{\zeta}}(t,k) \dot{\zeta}^2-k^2\bar{\mathcal{B}}(t,k)\zeta^2-k^2\bar{\mathcal{V}}(t,k) \dot{\zeta}\zeta\right\}\,,
\ee
where: 
\ba\label{definition1}
&&\mathcal{L}_{\dot{\zeta}\dot{\zeta}}(t,k)=\frac{\left(6 a^2 \mathcal{W}_7+\mathcal{W}_5\right) \left[3 a^4 \mathcal{W}_4{}^2+2 a^2 \mathcal{W}_1 \mathcal{W}_5+8 k^2 \mathcal{W}_5\l(m_{2}^2-\f{\lambda_4}{a^2}k^2+\f{\lambda_7}{a^4}k^4\r)\right]}{2 a^2 \left(\mathcal{W}_4{}^2-4 \mathcal{W}_1 \mathcal{W}_7\right)-32 k^2 \mathcal{W}_7 \l(m_{2}^2-\f{\lambda_4}{a^2}k^2+\f{\lambda_7}{a^4}k^4\r)}\,,\nn \\
%%%%
%%%%
&&\bar{\mathcal{B}}(t,k)=\l\{ a^{2} \mathcal{W}_0\left(\mathcal{W}_4^2-4 \mathcal{W}_1 \mathcal{W}_7\right)+k^2\l[\f{1}{a^6} \left(-a^6 \mathcal{W}_7 \left(a^2 \mathcal{W}_6^2+16 m_{2}^2 \mathcal{W}_0\right)-2 a^4 \bar{m}_5 \mathcal{W}_4 \mathcal{W}_6+a^8 \left(\mathcal{W}_4^2-4 \mathcal{W}_1 \mathcal{W}_7\right) \mathcal{W}_{3}\r.\r.\r.\nn\\
&&\l.\l.\l.-4 \bar{m}_5^2 \mathcal{W}_1\right)+k^2 \l(\f{1}{a^8} \left(a^{10} \left(\mathcal{W}_4^2-4 \mathcal{W}_1 \mathcal{W}_7\right) \mathcal{W}_{2}-16 \left(a^6 \mathcal{W}_7 \left(a^2 m_{2}^2 \mathcal{W}_{3}+\lambda _3 \mathcal{W}_6-\lambda _4 \mathcal{W}_0\right)+a^2 \bar{m}_5 \lambda _3 \mathcal{W}_4+\bar{m}_5^2 m_{2}^2\right)\right)\r)\r.\r.\nn\\
&&\l.\l.
+k^4\l(-\f{16}{a^{10}} \left(a^4 \mathcal{W}_7 \left(a^6 m_2^2 \mathcal{W}_{2}-a^4 \lambda _4 \mathcal{W}_{3}+a^2 \lambda _7 \mathcal{W}_0+4 \lambda _3{}^2\right)+a^2 \lambda _8 \left(a^4 \mathcal{W}_6 \mathcal{W}_7+\bar{m}_5 \mathcal{W}_4\right)-\bar{m}_5{}^2 \lambda _4\right)\r)
\r.\r.\nn\\
&&\l.\l.+k^6\l(\f{16}{a^{12}} \left(a^4 \mathcal{W}_7 \left(a^6 \lambda _4 \mathcal{W}_{2}-a^4 \lambda _7 \mathcal{W}_{3}-8 \lambda _3 \lambda _8\right)-\bar{m}_5{}^2 \lambda _7\right)\r) +k^8\l(-\f{16}{a^{10}} \mathcal{W}_7 \left(a^6 \lambda _7 \mathcal{W}_{2}+4 \lambda _8{}^2\right)\r) \r] \r\}/\Big\{ a^2 \left(\mathcal{W}_4^2\r.\nn\\
&&\l.-4 \mathcal{W}_1 \mathcal{W}_7\right)-16 k^2 \mathcal{W}_7 \l(m_{2}^2-\f{\lambda_4}{a^2}k^2+\f{\lambda_7}{a^4}k^4\r)\Big\}\,,\nn\\
%%%
%%%%
&&\bar{\mathcal{V}}(t,k)=-\l\{\f{k^2}{a^2} \l[ 8 \mathcal{W}_4\l(6 a^2 \mathcal{W}_7+ \mathcal{W}_5\r) \l(\lambda_3+\lambda_8\f{k^2}{a^2}\r)+16 \f{\bar{m}_5 \mathcal{W}_5}{a^2} \l(m_{2}^2-\f{\lambda_4}{a^2}k^2+\f{\lambda_7}{a^4}k^4\r)\r]+6 a^4 \mathcal{W}_4 \mathcal{W}_7\mathcal{W}_6\r.\nn\\&&\hspace{1cm}\l.
+a^2 \mathcal{W}_4 \mathcal{W}_5\mathcal{W}_6+6 \bar{m}_5 \mathcal{W}_4^2+4 \f{\bar{m}_5}{a^2} \mathcal{W}_1 \mathcal{W}_5\r\}/\l\{ a^2 \left(\mathcal{W}_4^2-4 \mathcal{W}_1 \mathcal{W}_7\right)-16 k^2 \mathcal{W}_7 \l(m_{2}^2-\f{\lambda_4}{a^2}k^2+\f{\lambda_7}{a^4}k^4\r)\r\}.
\ea
It is easy to notice that the above expressions can be  written in a more compact form as: 
\ba\label{definitions}
&&\mathcal{L}_{\dot{\zeta}\dot{\zeta}}(t,k)=\frac{k^2 \mathcal{A}_4(t,k)+\mathcal{A}_1(t)}{k^2 \mathcal{A}_2(t,k)+\mathcal{A}_3(t)}\,,\nn\\
&&\bar{\mathcal{B}}(t,k)=\frac{k^2 \mathcal{B}_2(t,k)+\mathcal{B}_1(t)}{k^2 \mathcal{A}_2(t,k)+\mathcal{A}_3(t)}\,,\nn\\
&&\bar{\mathcal{V}}(t,k)=\frac{k^2 \mathcal{V}_2(t,k)+\mathcal{V}_1(t)}{k^2 \mathcal{A}_2(t,k)+\mathcal{A}_3(t)}.
\ea
By considering the above definitions the action can be written in the same form  of (\ref{action1}), i.e.:
\be\label{bglpvaction}
\mathcal{S}_{EFT}^{(2)}=\f{1}{(2\pi)^3}\int{}d^3kdt\,a^3\l\{\mathcal{L}_{\dot{\zeta}\dot{\zeta}}(t,k)\dot{\zeta}^2-k^2G(t,k) \zeta^2\r\}\,,
\ee
where we have identified the ``gradient'' term as:
\ba
G(t,k)&=&\l\{ k^2\left[\mathcal{V}_2\left(k^2 \dot{\mathcal{A}}_2+\dot{\mathcal{A}}_3-3 H \left(k^2 \mathcal{A}_2+\mathcal{A}_3\right)\right)+ \mathcal{A}_2\mathcal{A}_3\left(2 \mathcal{B}_1-\dot{\mathcal{V}}_1-k^2 \dot{\mathcal{V}}_2+2 k^2 \mathcal{B}_2\right) +\mathcal{V}_1\left( \dot{\mathcal{A}}_2-3 H \mathcal{A}_2\right)\r]\r.\nn\\
&+&\l.\mathcal{V}_1 \left(\dot{\mathcal{A}}_3-3 H\mathcal{A}_3\right)\r\}/\l\{2  \left(k^2 \mathcal{A}_2+\mathcal{A}_3\right){}^2\r\}\,\nn\\
&&\equiv \frac{k^2 \mathcal{G}_2(t,k)+\mathcal{G}_1(t)}{\l(k^2 \mathcal{A}_2(t,k)+\mathcal{A}_3(t)\r)^2}\,.
\ea
Then the speed of propagation is $c_s^2(t,k)=G/\mathcal{L}_{\dot{\zeta}\dot{\zeta}}$ and the friction term in the field equation of $\zeta$ turn out to be  a function of both $t$ and $k$.
Let us notice that when considering the most general case, at least one of the functions $\{m^2_2, \lambda_i\}$ is not zero and none of the $\mathcal{A}_i$ functions are nil. Additionally the action  does not contain the term $\bar{M}$. We will show in the next Section some particular cases of the action~(\ref{full}) for which such a term is present.

Let us now normalize the field by means of~(\ref{redefinition}) with the kinetic term given by Eq.~(\ref{definition1}). Since the kinetic term is a function of $k$, the normalization will affect both the effective mass and  speed of propagation. Thus we have:
\ba
&&m_{\rm{eff}}^2(t,k)=\frac{\mathcal{A}_1{}^2 \left[2 \mathcal{A}_3 \left(3 H \dot{\mathcal{A}}_3+ \ddot{\mathcal{A}}_3\right)-3  \dot{\mathcal{A}}_3^2\right]-2 \mathcal{A}_3 \mathcal{A}_1 \left[\mathcal{A}_3 \left(3 H \dot{\mathcal{A}}_1+ \ddot{\mathcal{A}}_1\right)- \dot{\mathcal{A}}_1 \dot{\mathcal{A}}_3\right]+ \mathcal{A}_3^2 \dot{\mathcal{A}}_1^2}{8  \left(k^2 \mathcal{A}_4+\mathcal{A}_1\right)^2 \left(k^2 \mathcal{A}_2+\mathcal{A}_3\right)^2} \,,\nn \\
%%%
%%%
&&c_{s,\rm{eff}}^2(t,k)=\left\{6 H \left[\left[k^2 \left(\dot{\mathcal{A}}_4 k^2+\dot{\mathcal{A}}_1\right) \mathcal{A}_2^2+2 \left[\mathcal{A}_3 \left(\dot{\mathcal{A}}_4 k^2+\dot{\mathcal{A}}_1\right)-k^2 \mathcal{A}_4 \left(\dot{\mathcal{A}}_2 k^2+\dot{\mathcal{A}}_3\right)\right] \mathcal{A}_2+\mathcal{A}_3 \left(\mathcal{A}_3 \dot{\mathcal{A}}_4\right.\right.\right.\right.\nonumber\\
&&\hspace{1cm}\left.\left.\left.\left.-2 \mathcal{A}_4 \left(\dot{\mathcal{A}}_2 k^2+\dot{\mathcal{A}}_3\right)\right)\right] \mathcal{A}_1-\mathcal{A}_1^2 \left(\mathcal{A}_3 \dot{\mathcal{A}}_2+\mathcal{A}_2 \left(\dot{\mathcal{A}}_2 k^2+\dot{\mathcal{A}}_3\right)\right)+\left(\mathcal{A}_2 k^2+\mathcal{A}_3\right) \mathcal{A}_4 \left[\mathcal{A}_2 \left(\dot{\mathcal{A}}_4 k^2+\dot{\mathcal{A}}_1\right) k^2\right.\right.\right.\nonumber\\
&&\hspace{1cm}\left.\left.\left.-k^2 \mathcal{A}_4 \left(\dot{\mathcal{A}}_2 k^2+\dot{\mathcal{A}}_3\right)+\mathcal{A}_3 \left(\dot{\mathcal{A}}_4 k^2+\dot{\mathcal{A}}_1\right)\right]\right]+ \left[3 \mathcal{A}_4^2 \dot{\mathcal{A}}_2^2 k^6-4 \mathcal{A}_3 \mathcal{A}_4 \mathcal{G}_2 k^4+6 \mathcal{A}_4^2 \dot{\mathcal{A}}_2 \dot{\mathcal{A}}_3 k^4-2 \mathcal{A}_3 \mathcal{A}_4 \dot{\mathcal{A}}_2 \dot{\mathcal{A}}_4 k^4\right.\right.\nonumber\\
&&\hspace{1cm}
\left.\left.-2 \mathcal{A}_3 \mathcal{A}_4^2 \ddot{\mathcal{A}}_2 k^4+3 \mathcal{A}_4^2 \dot{\mathcal{A}}_3^2 k^2-\mathcal{A}_3^2 \dot{\mathcal{A}}_4^2 k^2-4 \mathcal{A}_3 \mathcal{A}_4 \mathcal{G}_1 k^2-2 \mathcal{A}_3 \mathcal{A}_4 \dot{\mathcal{A}}_1 \dot{\mathcal{A}}_2 k^2-2 \mathcal{A}_3 \mathcal{A}_4 \dot{\mathcal{A}}_3 \dot{\mathcal{A}}_4 k^2\right.\right.\nonumber\\
&&\hspace{1cm}\left.\left.-2 \mathcal{A}_3 \mathcal{A}_4^2 \ddot{\mathcal{A}}_3 k^2+2 \mathcal{A}_3^2 \mathcal{A}_4 \dot{\mathcal{A}}_4 k^2-\mathcal{A}_2^2 \left(\dot{\mathcal{A}}_4^2 k^4+2 \dot{\mathcal{A}}_1 \dot{\mathcal{A}}_4 k^2-2 \mathcal{A}_4 \left(\ddot{\mathcal{A}}_4 k^2+\ddot{\mathcal{A}}_1\right) k^2+\dot{\mathcal{A}}_1^2\right) k^2\right.\right.\nonumber\\
&&\hspace{1cm}\left.\left.-2 \mathcal{A}_3 \mathcal{A}_4 \dot{\mathcal{A}}_1 \dot{\mathcal{A}}_3-2 \mathcal{A}_3{}^2 \dot{\mathcal{A}}_1 \dot{\mathcal{A}}_4+2 \mathcal{A}_3^2 \mathcal{A}_4 \ddot{\mathcal{A}}_1+\mathcal{A}_1^2 \left[3 k^2 \dot{\mathcal{A}}_2{}^2+6 \dot{\mathcal{A}}_3 \dot{\mathcal{A}}_2-2 \left(\mathcal{A}_3 \ddot{\mathcal{A}}_2+\mathcal{A}_2 \left(\ddot{\mathcal{A}}_2 k^2+\ddot{\mathcal{A}}_3\right)\right)\right]\right.\right.\nonumber\\
&&\hspace{1cm}\left.\left.-2 \mathcal{A}_2 \left[\mathcal{A}_4{}^2 \left(\ddot{\mathcal{A}}_2 k^2+\ddot{\mathcal{A}}_3\right) k^4+\mathcal{A}_4 \left(2 \mathcal{G}_2 k^4+\dot{\mathcal{A}}_2 \dot{\mathcal{A}}_4 k^4+2 \mathcal{G}_1 k^2+\dot{\mathcal{A}}_1 \dot{\mathcal{A}}_2 k^2+\dot{\mathcal{A}}_3 \dot{\mathcal{A}}_4 k^2-2 \mathcal{A}_3 \ddot{\mathcal{A}}_4 k^2+\dot{\mathcal{A}}_1 \dot{\mathcal{A}}_3\right.\right.\right.\right.\nonumber\\
&&\hspace{1cm}\left.\left.\left.\left.-2 \mathcal{A}_3 \ddot{\mathcal{A}}_1\right) k^2+\mathcal{A}_3 \left(\dot{\mathcal{A}}_4 k^2+\dot{\mathcal{A}}_1\right)^2\right]+2 \mathcal{A}_1 \left[k^2 \left(\ddot{\mathcal{A}}_4 k^2+\ddot{\mathcal{A}}_1\right) \mathcal{A}_2^2-\left(2 \mathcal{G}_2 k^4+\dot{\mathcal{A}}_2 \dot{\mathcal{A}}_4 k^4+2 \mathcal{A}_4 \ddot{\mathcal{A}}_2 k^4+2 \mathcal{G}_1 k^2\right.\right.\right.\right.\nonumber\\
&&\hspace{1cm}.\left.\left.\left.\left.+\dot{\mathcal{A}}_1 \dot{\mathcal{A}}_2 k^2+\dot{\mathcal{A}}_3 \dot{\mathcal{A}}_4 k^2+2 \mathcal{A}_4 \ddot{\mathcal{A}}_3 k^2-2 \mathcal{A}_3 \ddot{\mathcal{A}}_4 k^2+\dot{\mathcal{A}}_1 \dot{\mathcal{A}}_3-2 \mathcal{A}_3 \dot{\mathcal{A}}_1\right) \mathcal{A}_2+3 \mathcal{A}_4 \left(\dot{\mathcal{A}}_2 k^2+\dot{\mathcal{A}}_3\right)^2-\mathcal{A}_3 \left(2 \mathcal{G}_2 k^2\right.\right.\right.\right.\nonumber\\
&&\hspace{1cm}\left.\left.\left.\left.+\dot{\mathcal{A}}_2 \dot{\mathcal{A}}_4 k^2+2 \mathcal{A}_4 \ddot{\mathcal{A}}_2 k^2+2 \mathcal{G}_1+\dot{\mathcal{A}}_1 \dot{\mathcal{A}}_2+\dot{\mathcal{A}}_3 \dot{\mathcal{A}}_4+2 \mathcal{A}_4 \ddot{\mathcal{A}}_3\right)+\mathcal{A}_3^2 \ddot{\mathcal{A}}_4\right]\right]\right\}/[8  \left(\mathcal{A}_2 k^2+\mathcal{A}_3\right)^2 \left(\mathcal{A}_4 k^2+\mathcal{A}_1\right)^2] \nonumber \\
&&\equiv c_s^2+f(t,k).\nn\\
\ea
As said before the effective mass is a function of  inverse powers of k. For sufficiently high k, the effective mass is negligible while in the low k limit, which is the one of interest in linear cosmology, it is solely a function of time.   Let us notice that the effective mass in this case has been obtained  directly from action~(\ref{bglpvaction}), not from Eq.~(\ref{effective}) which is valid only for cases when  the kinetic term   does not depend on k.

\subsection{Special cases}\label{Sec:cSpecial}

Although the subset of theories with higher than second order spatial derivatives treated in the previous Section is very general, there are some special cases for which the action assumes some particular forms due to specific combinations of the EFT functions in the kinetic term. In order to illustrate said cases, we will consider the following action for practical examples: 
\ba
\mathcal{S}_{EFT}^{(2)}&=&\f{1}{(2\pi)^3}\int{}d^3kdt\,a^3\l[4 m_{2}^2 \frac{k^2 }{a^2}(\delta N)^2-\mathcal{W}_6\delta N k^2 \zeta -  \mathcal{W}_4\delta N k^2\psi -\mathcal{W}_5k^2 \psi  \dot{\zeta}-\mathcal{W}_0k^2 \zeta ^2 + \mathcal{W}_7(k^2 \psi)^2\r.\nn\\&&\l.+\mathcal{W}_1(\delta N)^2 -3a^2 \mathcal{W}_4 \delta N\dot{\zeta}-\f{3}{2}a^2 \mathcal{W}_5 \dot{\zeta}^2\r]\,,
\ea
for which the following conditions hold:
\ba
\mathcal{W}_7\neq 0\, \qquad  \l\{\bar{m}_5,\lambda_i \r\}=0\,.
\ea
By solving the Eqs.~(\ref{fieldeqsauxiliaryfields}) for $\delta N$ and $\psi$ we get:  
\ba
&&\delta N=\frac{\mathcal{W}_4 \left(6 a^2 \mathcal{W}_7+\mathcal{W}_5\right) \dot{\zeta}+2 \mathcal{W}_6 \mathcal{W}_7k^2 \zeta}{16  m_{2}^2 \mathcal{W}_7\f{k^2}{a^2}- \mathcal{W}_4^2+4 \mathcal{W}_1 \mathcal{W}_7} \,, \nn \\
&&k^2\psi=\frac{\mathcal{W}_4 \mathcal{W}_6k^2 \zeta +\l(2 \mathcal{W}_1 \mathcal{W}_5 +3 a^2 \mathcal{W}_4^2 +8  m_{2}^2 \mathcal{W}_5\f{k^2}{a^2}\r) \dot{\zeta}}{ 16  m_{2}^2 \mathcal{W}_7\f{k^2}{a^2}-\mathcal{W}_4^2+4 \mathcal{W}_1 \mathcal{W}_7}\,,
\ea
which allow us to eliminate the two auxiliary fields in the action. Substituting back in the action we get:
\ba \label{bGLPVonefield}
\mathcal{S}_{EFT}^{(2)}&=&\f{1}{(2\pi)^3}\int{}d^3kdt\,a^3\l\{\l[\frac{\left(6 a^2 \mathcal{W}_7+\mathcal{W}_5\right) \left(3a^4 \mathcal{W}_4^2+2a^2 \mathcal{W}_1 \mathcal{W}_5+8m_{2}^2 \mathcal{W}_5k^2\right)}{2 a^2 \left(\mathcal{W}_4^2-4 \mathcal{W}_1 \mathcal{W}_7\right)-32  m_{2}^2 \mathcal{W}_7k^2}\r]\dot{\zeta}^2\nn \r. \\ 
&+&\l.k^2\l[\frac{\l(a^2 \left(\mathcal{W}_0\left(\mathcal{W}_4^2-4 \mathcal{W}_1 \mathcal{W}_7\right)-k^2 \mathcal{W}_6^2 \mathcal{W}_7\right)-16  m_{2}^2 \mathcal{W}_0 \mathcal{W}_7k^2\r)\zeta^2-\l( a^2 \mathcal{W}_4 \mathcal{W}_6 \left(6 a^2 \mathcal{W}_7+\mathcal{W}_5\right)\r)\dot{\zeta}\zeta}{16  m_{2}^2 \mathcal{W}_7k^2-a^2 \left(\mathcal{W}_4^2-4 \mathcal{W}_1 \mathcal{W}_7\right)}\r]\r\}\,,\nn\\
\ea
where the kinetic term reads: 
\be\label{ghostbGLPV}
\mathcal{L}_{\dot{\zeta}\dot{\zeta}}(t,k)\equiv\frac{\left(6 a^2 \mathcal{W}_7+\mathcal{W}_5\right) \left(3 a^4 \mathcal{W}_4^2+2 a^2 \mathcal{W}_1 \mathcal{W}_5+8 k^2 m_{2}^2 \mathcal{W}_5\right)}{2 a^2 \left(\mathcal{W}_4^2-4 \mathcal{W}_1 \mathcal{W}_7\right)-32 k^2 m_{2}^2 \mathcal{W}_7}\,.
\ee
In the following we will consider two special cases in which 1) the kinetic term depends only on time; 2) the kinetic term has a particular k-dependence, which needs to be studied carefully in order to correctly identify the speed of propagation.

\begin{itemize}
\item First case: $3 a^2 \mathcal{W}_4^2+2 \mathcal{W}_1 \mathcal{W}_5 \neq 0$ and $m_2^2 = 0$. The kinetic term is only a function of time:
\be\label{ghost1case}
\mathcal{L}_{\dot{\zeta}\dot{\zeta}}(t)=\frac{\left(6 a^2 \mathcal{W}_7+\mathcal{W}_5\right) \left(3 a^4 \mathcal{W}_4^2+2 a^2 \mathcal{W}_1 \mathcal{W}_5\right)}{2 a^2 \left(\mathcal{W}_4^2-4 \mathcal{W}_1 \mathcal{W}_7\right)}\,,
\ee
which corresponds to the case $\mathcal{A}_2=\mathcal{A}_4=0$. The above expression must be positive in order to guarantee that the theory does not exhibit ghost instabilities. Then, the  speed of propagation can be easily obtained from action~(\ref{bGLPVonefield}) once the terms proportional to $\dot{\zeta}\zeta$ have been integrated by parts and it reads:
\ba
c_s^2(t,k)&=&\frac{1}{ \left(\mathcal{W}_4^2-4 \mathcal{W}_1 \mathcal{W}_7\right) \left(3 a^2 \mathcal{W}_4{}^2+2 \mathcal{W}_1 \mathcal{W}_5\right) \left(6 a^2 \mathcal{W}_7+\mathcal{W}_5\right)}\l\{30 a^2 \mathcal{W}_4 \mathcal{W}_6 \mathcal{W}_7 \left(\mathcal{W}_4{}^2-4 \mathcal{W}_1 \mathcal{W}_7\right) H \r.\nn\\
&&\l.+3 \mathcal{W}_4 \mathcal{W}_5 \mathcal{W}_6 \left(\mathcal{W}_4^2-4 \mathcal{W}_1 \mathcal{W}_7\right) H-\mathcal{W}_6 \mathcal{W}_4^2 \mathcal{W}_5 \dot{\mathcal{W}}_4-4 \mathcal{W}_1 \mathcal{W}_6 \mathcal{W}_7 \mathcal{W}_5 \dot{\mathcal{W}}_4\r.\nn\\
&&\l.+\mathcal{W}_4^3 \left(\mathcal{W}_6 \dot{\mathcal{W}}_5+\mathcal{W}_5 \dot{\mathcal{W}}_6\right)+4 \mathcal{W}_4 \left[\mathcal{W}_5 \left(\mathcal{W}_6 \left(\mathcal{W}_7 \dot{\mathcal{W}}_1+\mathcal{W}_1 \dot{\mathcal{W}}_7\right)-\mathcal{W}_1 \mathcal{W}_7 \dot{\mathcal{W}}_6\right)-\mathcal{W}_1 \mathcal{W}_6 \mathcal{W}_7 \dot{\mathcal{W}}_5\right]\r.\nn\\
&&\l.+2 \mathcal{W}_0 \left(\mathcal{W}_4^2-4 \mathcal{W}_1 \mathcal{W}_7\right)^2+6 a^2 \left[\mathcal{W}_4^3 \left(\mathcal{W}_7 \dot{\mathcal{W}}_6+\mathcal{W}_6 \dot{\mathcal{W}}_7\right)+4 \mathcal{W}_7^2 \mathcal{W}_4 \left(\mathcal{W}_6 \dot{\mathcal{W}}_1-\mathcal{W}_1 \dot{\mathcal{W}}_6\right)\r.\r.\nn\\
&&\l.\l.-4 \mathcal{W}_1 \mathcal{W}_6 \mathcal{W}_7^2 \dot{\mathcal{W}}_4-\mathcal{W}_4^2 \mathcal{W}_6 \mathcal{W}_7 \dot{\mathcal{W}}_4\right]-2k^2a\mathcal{W}_6^2\mathcal{W}_7(\mathcal{W}_4^2-4\mathcal{W}_1\mathcal{W}_7)\r\},
\ea
where the k-dependence of the speed is due to $W_7\neq0$.  Moreover, in this case, the final action is of the form~(\ref{action1}) with $\bar{M}=0$.  Since the kinetic terms is free from any k-dependence there is no ambiguity in defining the mass term which, after the normalization~(\ref{redefinition}), ends up being of the same form as in Eq.~(\ref{effective}) where, in this case, $\mathcal{L}_{\dot{\zeta}\dot{\zeta}}$ is given by Eq.~(\ref{ghost1case}). Finally, the effective speed of propagation remains invariant under the field redefinition.

\item Second case: $3 a^2 \mathcal{W}_4^2+2 \mathcal{W}_1 \mathcal{W}_5 = 0$ and $m_2^2 \neq 0$. In this case the kinetic term reduces to: 
\be\label{ghost2case}
\mathcal{L}_{\dot{\zeta}\dot{\zeta}}(t,k)= \frac{4 m_{2}^2 \mathcal{W}_5^2 \left(6 a^2 \mathcal{W}_7+\mathcal{W}_5\right)\f{k^2}{a^2}}{ \mathcal{W}_4^2\l(6 a^2 \mathcal{W}_7+ \mathcal{W}_5\r)-16 \f{k^2}{a^2} m_{2}^2 \mathcal{W}_5 \mathcal{W}_7}\,,
\ee
which corresponds to $\mathcal{A}_1=0$ and $\mathcal{A}_2(t)$,  $\mathcal{A}_4(t)$ both being functions of time. From the action~(\ref{bGLPVonefield}) it follows that there is an overall factor $k^2$ in front of the Lagrangian which can be reabsorbed by redefining the field as $\tilde{\zeta}= k\zeta$. As a result we obtain an action of the form~(\ref{actiondef2}). Let us notice that, in this case, $\mathcal{V}_2=0$. After integrating by parts the term $\sim \dot{\zeta}\zeta $, we end up with an action as in~(\ref{action1}) where $\bar{M}\neq0$, and both the friction and  dispersive coefficients in the field equation are functions of time and k.  Now we can compute the  speed of propagation which is:
\be
c_s^2(t,k)=\frac{\mathcal{V}_1\dot{\mathcal{A}}_2+\mathcal{A}_2 (2 k^2 \mathcal{B}_2-\dot{\mathcal{V}}_1+2 \mathcal{B}_1)+2\mathcal{A}_3 \mathcal{B}_2-3  H \mathcal{A}_2\mathcal{V}_1}{2 \mathcal{A}_4 \left(k^2 \mathcal{A}_2+\mathcal{A}_3\right)}\,.
\ee
In conclusion, we give the expressions for the effective mass and speed of propagation: 
\ba
&&m_{\rm{eff}}^2(t,k)= \frac{6 \mathcal{A}_2 \mathcal{A}_3 H (\mathcal{A}_2 \dot{\mathcal{A}}_3-\mathcal{A}_3 \dot{\mathcal{A}}_2)+ \mathcal{A}_3 \mathcal{A}_2 (2\dot{\mathcal{A}}_2 \dot{\mathcal{A}}_3-2\mathcal{A}_3 \ddot{\mathcal{A}}_2+\mathcal{G}_1)+\mathcal{A}_2{}^2 (2 \mathcal{A}_3 \ddot{\mathcal{A}}_3-3 \dot{\mathcal{A}}_3{}^2)+\mathcal{A}_3{}^2 \dot{\mathcal{A}}_2{}^2}{8  \mathcal{A}_4{}^2 \left(k^2 \mathcal{A}_2+\mathcal{A}_3\right){}^2}\nn\\
&&c^2_{s,\rm{eff}}(t,k)=\l\{6 \mathcal{A}_4 H \left[\mathcal{A}_2 \left(\mathcal{A}_4 \left(k^2 \dot{\mathcal{A}}_2+\dot{\mathcal{A}}_3\right)-2 \mathcal{A}_3 \dot{\mathcal{A}}_4\right)+\mathcal{A}_3 \mathcal{A}_4 \dot{\mathcal{A}}_2-k^2 \mathcal{A}_2^2 \dot{\mathcal{A}}_4\right]
+2 \mathcal{A}_2 \left[\mathcal{A}_4 \left(k^2 \dot{\mathcal{A}}_2 \dot{\mathcal{A}}_4\right.\right.\right.\nonumber\\
&&\hspace{0.9cm}\left.\left.\left.+2 k^2 \mathcal{G}_2+\dot{\mathcal{A}}_3 \dot{\mathcal{A}}_4-2 \mathcal{A}_3 \ddot{\mathcal{A}}_4+2 \mathcal{G}_1\right)+\mathcal{A}_4{}^2 \left(k^2 \ddot{\mathcal{A}}_2+\ddot{\mathcal{A}}_3\right)+\mathcal{A}_3 \dot{\mathcal{A}}_4^2\right]+\mathcal{A}_4 \left[2 \mathcal{A}_3 \left(\dot{\mathcal{A}}_2 \dot{\mathcal{A}}_4+\mathcal{A}_4 \ddot{\mathcal{A}}_2+2 \mathcal{G}_2\right)\right.\right.\nonumber\\
&&\hspace{0.9cm}\left.\left.-3 \mathcal{A}_4 \dot{\mathcal{A}}_2 \left(k^2 \dot{\mathcal{A}}_2+2 \dot{\mathcal{A}}_3\right)\right]+k^2 \mathcal{A}_2^2 \left(\dot{\mathcal{A}}_4^2-2 \mathcal{A}_4 \ddot{\mathcal{A}}_4\right)\r\}/\l[8 \mathcal{A}_4^2 \left(k^2 \mathcal{A}_2+\mathcal{A}_3\right)^2\r]\,,
\ea
where the function $\mathcal{G}_i(i=1,2)$ can be read from:
\be
G(t,k)=\frac{\mathcal{V}_1\dot{\mathcal{A}}_2+\mathcal{A}_2 (-\dot{\mathcal{V}}_1+2 \mathcal{B}_1)+2\mathcal{A}_3 \mathcal{B}_2-3  H \mathcal{A}_2\mathcal{V}_1+2k^2\mathcal{A}_2 \mathcal{B}_2}{2 \left(k^2 \mathcal{A}_2+\mathcal{A}_3\right)^2}.
\ee 
\end{itemize}

Finally, let us notice that in the case $\bar{M} \neq 0$, one may wonder if the conservation of the  curvature perturbation is preserved on super-horizon scales. It is not so trivial to draw a general conclusion about the behaviour of $\zeta$ in such limit, because the EFT functions involved in the $\bar{M}$ term are all unknown functions of time.  Therefore, we can conclude that in the general field equation for $\zeta$ on super-horizon scales such term might be non zero, possibly leading to a non conserved curvature perturbation. However,  we expect that well behaved DE/MG models will have either $\bar{M}=0$ or that such term will contribute a  decaying mode, thus leaving the conservation of $\zeta$ unaffected. In this regard, we will argument our last statement  by using an explicit example,  which is not conclusive but can give an insight on how  $\bar{M}$  can behave in the low k regime when theoretical models are considered.  Considering the mapping (\ref{horavamapping}), it is easy to verify that  the low energy Ho\v rava gravity falls in the special case under analysis and that the corresponding $\bar{M}\neq 0$. However, when considering the super-horizon limit  the $\bar{M}$ term goes to zero and the equation for $\zeta$ reduces to
\be
\ddot{\zeta}+H\dot{\zeta}=0,
\ee
which solution is $\zeta\to \zeta_c -\frac{c_1 e^{-\sqrt{2} t \sqrt{\frac{\xi \Lambda }{9 \lambda -3}}}}{\sqrt{2} \sqrt{\frac{\xi \Lambda }{9 \lambda -3}}}$. $\zeta_c, c_1$ are constant and the second term is a decaying mode. Hence, the conservation of $\zeta$ is preserved.

Let us conclude by saying that  the cases treated in this Section are only few examples of ``special'' cancellations that might happen.

\section{An extended basis for theories with higher spatial derivatives} \label{Sec:basis}

In Ref.~\cite{Bellini:2014fua}, the authors proposed a new  basis to describe  Horndeski theories, in terms of four free functions of time which parametrize the departure from GR. Specifically, these functions are: $\{\alpha_B,\alpha_M, \alpha_K, \alpha_T\}$, hereafter referred to as ReParametrized Horndeski (RPH). They are equivalent and an alternative to the EFT functions needed to describe the dynamics of perturbations in the Horndeski class, i.e. $\{\Omega, M^4_2, \bar{M}^2_2, \bar{M}^3_1\}$. In both cases one needs to supply also the Hubble parameter, $H(a)$. The latest publicly released version of EFTCAMB contains also the RPH basis as a built-in alternative~\cite{Hu:2014oga}. RPH is also the building block at the basis of HiCLASS~\cite{Bellini:2015xja}.

The RPH basis was constructed in order to encode departures from GR in terms of some key properties of the (effective) DE component. As discussed in details in Ref.~\cite{Bellini:2014fua}, the braiding function $\alpha_B$ is connected to the clustering of DE,  $\alpha_M$ parametrizes the time-dependence of the Planck mass and, along with $\alpha_T$, is related to the anisotropic stress while large values of the kinetic function, $\alpha_K$ correspond to suppressed values of the speed of propagation of the scalar mode. In Ref.~\cite{Gleyzes:2014qga}, the RPH basis has been extended to include the GLPV class of theories by adding the function $\alpha_H$, which parametrizes the deviation from the Horndeski class. 

In this Section we introduce an extended version of the RPH basis which generalizes the original one~\cite{Bellini:2014fua}, as well as its extension to GLPV~\cite{Gleyzes:2014qga}, by encompassing the higher order spatial derivatives terms appearing in action~(\ref{EFTgeneral}). We also present the explicit mapping  between this new basis and the  EFT functions in the extended action~(\ref{EFTgeneral}),  in order to facilitate the link between phenomenological properties and the theory which is responsible for them.

Let us start  with tensor perturbations of the EFT action~(\ref{EFTADM}) analysed in Section~\ref{Sec:stability}. Here, for completeness we rewrite its compact form: 
\be
\mathcal{S}_{EFT}^{T\,(2)}=\f{1}{(2\pi)^3}\int{}d^3kdt\,a^3\f{A_T(t)}{8}\l[(\dot{h}_{ij}^T)^2-\f{c^2_T(t,k)}{a^2}k^2(h_{ij}^T)^2\r]\,.
\ee
Now, following Ref.~\cite{Bellini:2014fua},  we  define the deviation from GR of the tensor speed of propagation as:
\be
c_T^2(t,k)=1+\tilde{\alpha}_T(t,k),
\ee
where:
\be
\tilde{\alpha}_T(t,k)=\alpha_T(t)+\alpha_{T_2}(t)\f{k^2}{a^2}+\alpha_{T_6}(t)\f{k^4}{a^4}\,,
\ee
with:
\be
\alpha_T(t)=\frac{\bar{M}^2_3}{m_0^2(1+\Omega)-\bar{M}_3^2}\equiv \bar{c}_T^2-1\,,\qquad\alpha_{T_2}(t)= -8\f{\lambda_2}{m_0^2(1+\Omega)-\bar{M}^2_3}\,, \qquad\alpha_{T_6}(t)=-8\f{\lambda_6}{m_0^2(1+\Omega)-\bar{M}^2_3}.
\ee
As expected, the additional higher order operators will contribute by adding a k-dependence in the original definition of the $\alpha_T$ function introduced in Ref.~\cite{Bellini:2014fua}.  Moreover,  we can define the rate of evolution of the mass function $M^2(t)\equiv A_T(t)$ (defined in Eq.~(\ref{tensordefinition})) as:
\be
\alpha_M(t)=\f{1}{H(t)}\f{d}{dt}\l(\ln{M^2(t)}\r).
\ee
It is clear that $\alpha_T$ and $\alpha_M$  differ from the ones in Ref.~\cite{Bellini:2014fua} since, in general, $\bar{M}^2_3(t)\neq -\bar{M}^2_2(t)$  for theories with higher spatial derivatives. It is important to notice that the EFT functions which are involved in the definition of $\alpha_M$ and $\alpha_T$ are $\{\Omega, \bar{M}^2_3\}$. Therefore, the class of theories which can contribute to a time dependent Planck mass and modify the tensor speed of propagation, are the ones which are non-minimally coupled with  gravity and/or contain the $\mathcal{S}$-term in the action; specifically, Horndeski models with non zero $L_4^{GG}, L_5^{GG}$, GLPV models with non zero $L_4^{GLPV}, L_5^{GLPV}$ and Ho\v rava gravity. Moreover, the k-dependence in the speed of propagation is related to the $\alpha_{T2}, \alpha_{T6}$ functions which are present in Ho\v rava gravity. Finally, let us notice that, since $M^2$ appears in the denominator of $c^2_T$,  high values of $M^2$ will generally suppress the speed of propagation and  in case only background EFT functions are at play or theories for which $\{\bar{M}^2_3(t),\lambda_{2,6}\}=0$ are considered, $c_T^2$ is identically one. Therefore, it would be not possible to discriminate between minimally and non-minimally coupled models.

Let us now focus on the scalar perturbations.  Collecting terms with the same perturbations, the second order action~(\ref{EFTADM}) can be written as follows: 
\ba\label{alphageneralized}
\mathcal{S}_{EFT}^{(2)}&=&\f{1}{(2\pi)^3}\int{}d^3kdt\,a^3\,\f{M^2}{2}\l\{\l(1+\tilde{\alpha}_H\r)\delta N\delta_1 \tilde{\mathcal{R}}-4H\alpha_B\delta N\delta \tilde{K}+\delta \tilde{K}^\mu_\nu\delta \tilde{K}^\nu_\mu- (\alpha^{GLPV}_B+1)(\delta \tilde{K})^2+\tilde{\alpha}_KH^2(\delta N)^2\r.\nn\\
&-&\l.\f{1}{4}\l(\alpha_{T_2}+\alpha_{T_6}\f{k^2}{a^2}\r)\delta \tilde{\mathcal{R}}_{ij}\delta \tilde{\mathcal{R}}^{ij}+(1+\alpha_T)\delta_2 \tilde{\mathcal{R}} +(1+\alpha_T)\delta_1\tilde{\mathcal{R}}\delta\tilde{(\sqrt{h})}+\l(\alpha_1+\alpha_5\f{k^2}{a^2}\r)(\delta \tilde{\mathcal{R}})^2 +\bar{\alpha}_5 \delta_1 \tilde{\mathcal{R}}\delta \tilde{K}\r\}\,, \nn\\
&&
\ea
where the geometrical quantities with tildes are the Fourier transform of the corresponding quantities in Eq.~(\ref{scalarperturbations}), moreover we have identified the following functions:
\ba\label{alphageneralizeddef}
&&\alpha_B(t)=\frac{m_0^2\dot{\Omega}+\bar{M}^3_1}{2H M^2}\,, \qquad \alpha^{GLPV}_B(t)=\f{\bar{M}^2_3+\bar{M}^2_2}{M^2}\,,\nn \\
&&\tilde{\alpha}_K(t,k)=\alpha_K (t)+\alpha_{K_2}(t)\f{k^2}{a^2}+\alpha_{K_4}(t)\f{k^4}{a^4}+\alpha_{K_7}(t)\f{k^6}{a^6}\,,\nn\\
&&\hspace{0.5cm}\textrm{where}\,\, \qquad\alpha_K(t)=\frac{2c+4M_2^4}{H^2 M^2}\,,\qquad \alpha_{K_2}(t)=\frac{8m_2^2}{M^2H^2}\,,\qquad\alpha_{K_4}(t)=-\frac{8\lambda_4}{M^2H^2}\,, \qquad \alpha_{K_7}(t)=\frac{8\lambda_7}{H^2 M^2}\,,\nn\\&&\tilde{\alpha}_H(t,K)=\alpha_H(t)+\alpha_{H_3}(t)\f{k^2}{a^2}+\alpha_{H_8}(t)\f{k^4}{a^4}\,,\nn\\
&&\hspace{0.5cm}\textrm{where}\,\, \qquad\alpha_H(t)=\f{2\hat{M}^2+\bar{M}^2_3}{M^2}\,,\qquad \alpha_{H_3}(t)=-\f{4\lambda_3}{M^2}\,, \qquad \alpha_{H_8}(t)=\f{4\lambda_8}{M^2}\,,\nn\\
&&\alpha_1(t)=\f{2\lambda_1}{M^2}\,,\qquad \alpha_5(t)=\f{2\lambda_5}{M^2}\,, \qquad \bar{\alpha}_5(t)=\f{\bar{m_5}}{M^2}.
\ea
The relations between the $\mathcal{W}$-functions introduced in Section~\ref{Sec:stability} and the above $\alpha$-functions are the following:
\ba\label{Walpha}
&&\mathcal{W}_0\equiv -\f{M^2}{a^2}\l(\alpha_T+1+3H\bar{\alpha}_5+3\dot{\bar{\alpha}}_5+3\bar{\alpha}_5H\alpha_M \r)\,,\qquad \mathcal{W}_1\equiv \f{M^2H^2}{2}\alpha_K+\frac{3}{2}a^2H\mathcal{W}_4-3H^2 M^2 \alpha_B\,, \nn \\
&&\mathcal{W}_{2} \equiv \f{M^2}{a^6}\l(-8\alpha_5+\f{3}{4}\alpha_{T_6}\r) \,,\qquad \mathcal{W}_{3} \equiv \f{M^2}{a^4}\l(-8\alpha_1+\f{3}{4}\alpha_{T_2}\r)\,,
\qquad \mathcal{W}_4\equiv -\f{HM^2}{a^2}\l(2+2\alpha_B+3\alpha_B^{GLPV}\r)\,,\nn\\
&& \mathcal{W}_5\equiv \f{M^2}{a^2}\l(2+3\alpha_B^{GLPV}\r)\,, \qquad \mathcal{W}_6\equiv -\f{2M^2}{a^2}\l(1+\alpha_H+3H\bar{\alpha}_5\r)\,,\qquad \mathcal{W}_7\equiv -\f{M^2}{2a^4}\alpha_B^{GLPV}\,.
\ea
Before discussing in details the  meaning of the $\alpha$-functions and how they contribute to the evolution of the propagating DoF, we  introduce the  perturbed  linear equations which will help us in the discussion. The variation of the action~(\ref{alphageneralized}) w.r.t to $\psi$  and $\delta N$  gives: 
\ba\label{auxiliaryequations}
&&H\l[2(1+\alpha_B)+3\alpha^{GLPV}_B\r]\delta N-(2+3\alpha_B^{GLPV})\dot{\zeta}-\alpha^{GLPV}_B\f{k^2\psi}{a^2}-2\bar{\alpha}_5\f{k^2\zeta}{a^2}=0\,, \nn\\
%%%%%%%%%%%%%%%
&&\Big[ 3H^2 \left(2-4\alpha_B-3 \alpha_B^{GLPV}\r)+ H^2 \tilde{\alpha}_K\Big]\delta N+2 H\Big[ 2\alpha _B+3 \alpha _B^{GLPV}+2\Big]\f{k^2}{a^2}\psi +\Big[3  H \left(2+2\alpha _B+3 \alpha _B^{GLPV}\right)\Big]\dot{\zeta}\nn \\
&&+ 2\Big[ 1+H \bar{\alpha}_5 +\tilde{\alpha}_H\Big]\f{k^2}{a^2}\zeta=0.
\ea
These equations allow us to eliminate the auxiliary fields $\delta N$ and $\psi$ from the action, yielding an action solely in terms of the dynamical field $\zeta$. A detailed description of how to eliminate the auxiliary fields  was the subject of the previous Section~\ref{Sec:stability}, indeed the above equations are equivalent to Eqs.~(\ref{fieldeqsauxiliaryfields}), once the relations~(\ref{Walpha}) have been considered. At this point, we can describe the meaning of the different $\alpha$-functions in terms of the phenomenology of $\zeta$. 

Let us now focus on the definition of the $\alpha$-functions which characterize the new basis, $\{\alpha_M, \tilde{\alpha_T},\alpha_B, \alpha_B^{GLPV}, \tilde{\alpha_H}, \tilde{\alpha}_K, \bar{\alpha}_5, \alpha_1,\alpha_5\}$, extending and generalizing the RPH one. A first difference that can be noticed w.r.t. the RPH parametrization, is the presence of $\{\tilde{\alpha}_H,\tilde{\alpha}_K\}$  which are now functions of k, since  they contain the contributions from operators with higher spatial derivatives.  Let us now describe the new basis in details with the help of the definitions~(\ref{alphageneralizeddef}) and Eqs.~(\ref{auxiliaryequations}):
 \begin{itemize}
\item $\{\alpha_B,\alpha_B^{GLPV} \}$: $\alpha_B$ is the braiding function as defined in Ref.~\cite{Bellini:2014fua}. \footnote{The definition of $\alpha_B$ presented here differs from the one in Ref.~\cite{Bellini:2014fua} by a minus sign and a factor 2.} Its role is clear by looking at Eqs.~(\ref{auxiliaryequations}), indeed $\alpha_B$ regulates the relation between the auxiliary field $\delta N$ and the dynamical DoF $\zeta$. Analogously,  we define $\alpha^{GLPV}_B$, which   contributes to the braiding since it mediates the relationship of  $\psi$ and $\delta N$ with $\zeta$. The effects of  these braiding coefficients on the kinetic term and the speed of propagation is more involved.  Indeed, by looking at the action~(\ref{alphageneralized}) we can notice that $\alpha^{GLPV}_B$ has a direct contribution to the kinetic term since it is the pre-factor of $(\delta K)^2$, which contains $\dot{\zeta}^2$. 
Moreover, both $\alpha_B$ and $\alpha^{GLPV}_B$  affect indirectly the kinetic term: the $\delta N$ term in Eq.~(\ref{auxiliaryequations}), whose pre-factor contains the braiding functions, turns out to be proportional to $\dot{\zeta}$, then substituting it back to action~(\ref{alphageneralized}), the term in $(\delta N)^2$ will generate a contribution to the kinetic term. Furthermore, their involvement in the speed of propagation of the scalar DoF comes in two ways: 1) from the kinetic term as previously mentioned. Indeed through Eq.~(\ref{speeddef}) they enter in the denominator of the definition of the propagating speed; 2) because they multiply both the $\delta N$ and $\psi$ terms in Eq.~(\ref{auxiliaryequations}) which result to be proportional to $k^2\zeta$ which contributes to $G$ in Eq.~(\ref{speeddef}). Moreover, analogously to the definition of $\alpha_H$, which parametrizes the deviation w.r.t. Horndeski/GG theories,  $\alpha^{GLPV}_B$ is defined such as to parametrize the deviation from GLPV theories; indeed the latter are characterized by the condition $\alpha^{GLPV}_B=0$, hence the name. If $\alpha^{GLPV}_B\neq 0$, higher spatial derivatives appear in the $\zeta$ equation. Finally, $\alpha_B$ is different from zero for all the theories showing non-minimal coupling to gravity and/or possessing the $\delta N \delta K$ operator in the action, i.e. $f(R), L_3^{GG}, L_4^{GG}, L_5^{GG}, L_4^{GLPV}, L_5^{GLPV}$. This operator does not appear when one considers  quintessence and k-essence models ($L_2^{GG}$) and  Ho\v rava gravity.  $\alpha_B^{GLPV}$ is non zero for the low-energy Ho\v rava gravity action.
\item $\tilde{\alpha}_K(t,k)$: it is the generalization of the purely kinetic function $\alpha_K(t)$ and it describes the extension of the kinetic term to higher order spatial derivatives in the case of non zero $\{\alpha_{K2}, \alpha_{K4}, \alpha_{K7}\}$. 
It is easy to see that $\tilde{\alpha}_K(t,k)$ is related to the kinetic term of the scalar DoF since it appears in action~(\ref{alphageneralized}) as a coefficient of  the operator $(\delta N)^2$ and,  through the linear perturbed equations~(\ref{auxiliaryequations}),  $\delta N\sim\dot{\zeta}$. Since it describes the kinetic term, it will affect the speed of propagation of $\zeta$ as well as the condition for the absence of a scalar ghost. The last point is easy to understand because as we extensively discussed in Section~\ref{Sec:stability} the kinetic terms goes in the denominator of the speed of propagation of scalar perturbation (see Eq.~(\ref{speeddef})). 
The $\alpha_K$ function is characteristic of theories belonging to GLPV, while for Ho\v rava gravity it is identically zero. On the other hand, Ho\v rava gravity  contributes non zero $\{\alpha_{K2}, \alpha_{K4}, \alpha_{K7}\}$.  Finally, let us note that when considering theories beyond GLPV the braiding coefficient discussed in the previous point, $\alpha^{GLPV}_B$, gives a direct contribution to the kinetic term through the operator $(\delta K)^2$.  
\item $\{\alpha_1,\alpha_5, \bar{\alpha}_5,\tilde{\alpha}_H\}$: from the constraint equations~(\ref{auxiliaryequations}), it can be noticed that $\tilde{\alpha}_H$ and $\bar{\alpha}_5$ contribute to the speed of propagation of the scalar DoF since they multiply the term $k^2\zeta$. In particular, if $\bar{\alpha}_5$ and the k-dependent parts of  $\tilde{\alpha}_H$  are different from zero, the dispersion relation of $\zeta$ will be modified and the speed of propagation will depend on k.  The functions $\{\alpha_1, \alpha_5\}$ have a similar impact since they are the pre-factors of $\delta_1\mathcal{R}$ in the action which, once expressed in terms of the perturbations of the metric, gives a term  proportional to $k^2 \zeta $. In this case by looking at Eq.~(\ref{speeddef}) these functions will enter in the definition of $G$. The theories where these functions are present are  GLPV and Ho\v rava gravity models. In particular, in the case of Ho\v rava gravity the functions associated with  higher order spatial derivatives terms are present.
\end{itemize} 

The above represents an interesting extension and generalization of the original RPH parametrization~\cite{Bellini:2014fua}, carefully built while considering the different phenomenological aspects of the dark energy fluid. However, let us notice that the desired correspondence between the $\alpha$-functions and actual observables becomes weaker as we go beyond the Horndeski class. Indeed, due to the high number of $\alpha$-functions involved, their dependence on many EFT functions and the way they enter in the actual physical quantities, such as the speed of sound and the kinetic term,  identifying exactly the underlying theory of gravity responsible for a specific effect is a hard task.

\section{Conclusions}\label{Sec:conclusion}
 
Cosmic acceleration still represents an open problem for modern cosmology and a plethora of theories of gravity have been proposed to account for it.  In the light of current and upcoming data it has become imperative to identify efficient ways of testing these models. This led to the investigation of unifying frameworks, of which a  recent and very promising proposal is the EFT for DE/MG models introduced in Refs.~\cite{Gubitosi:2012hu,Bloomfield:2013efa}. This formalism offers a unified and 
model independent way to study the dynamics of linear perturbations in a wide range of theories  which display an additional scalar DoF, besides the usual tensorial one, and have  a well defined Jordan frame. Interestingly, the implementation of this framework in the Einstein-Boltzmann solver CAMB, offers an universal tool to solve accurately the dynamics of linear perturbations. This has been done in what is known as EFTCAMB~\cite{Hu:2013twa,Hu:2014oga}~(\url{http://wwwhome.lorentz.leidenuniv.nl/~hu/codes/}), and its applications have been demonstrated in Refs.~\cite{Raveri:2014cka,Hu:2014sea,Frusciante:2015maa,Ade:2015rim}.

In this paper we have  generalized the original EFT action for DE/MG, including operators up to sixth order in spatial derivatives. This was motivated by the recent rise of theories containing a (sub)set of these operators with higher-order spatial derivatives, like Ho\v rava gravity.  Indeed, such theories were not covered by the operators included in the first proposal of the EFT action as presented in Refs.~\cite{Kase:2014cwa,Frusciante:2015maa}. From there on the extended Lagrangian~(\ref{EFTgeneral}) became the basis of the rest of the paper as the new operators play a central role.

Starting from the extended Lagrangian~(\ref{EFTgeneral}) we first proceeded to obtain an efficient recipe which allows one  to efficiently map theories of gravity, expressed in terms of geometrical quantities, into the EFT language. By considering an equivalent action in ADM formalism,  we have derived a general  mapping between the ADM and the EFT formalism for such an extended Lagrangian.  Additionally,  we illustrated this systematic procedure of mapping models of DE/MG, with an additional scalar DoF, into the EFT formalism, by providing a vast set of worked out examples. These include  minimally coupled quintessence, f(R), Hornedski/GG, GLPV and  Ho\v rava gravity. The preliminary step of writing the theories in the ADM formalism has also been presented as it is an integral part of the procedure. Therefore we created a very useful guide for the theoretical steps necessary in order to implement a given model of DE/MG into EFTCAMB and a "dictionary" for many of the existing DE/MG models. To this extent, we have been very careful and explicit about the conventions which lie at the basis of the EFT formalism, specific to EFTCAMB. These becomes obvious when comparing with the equivalent approaches in the literature as there are some clear differences. Thus the take-home message is that the user should be careful with the conventions when implementing a given model into EFTCAMB.

An ongoing field of research regarding the EFT of DE/MG is the determination of the parameter space corresponding to physically healthy theories. This is vital  from a theoretical as well as from a numerical point of view. As such it was natural to subject  our extended Lagrangian to a thorough stability analysis while considering only the gravity sector. In fact, since the EFT formalism is based on an action, we were able to determine general conditions of theoretical viability which are model independent and can, a priori, greatly reduce the parameter space. The most common criteria would be the absence of ghosts and gradient instabilities in the scalar and tensor sector, the exclusion of  tachyonic instabilities and  
positive (squared) speeds  of propagation.
Regarding the first two criteria,  one can find results in the literature either with or without the inclusion of a matter sector~\cite{DeFelice:2011bh,Gubitosi:2012hu,Bloomfield:2012ff,Gleyzes:2013ooa,Piazza:2013pua,Kase:2014cwa,Gao:2014soa,Gleyzes:2014qga}. In this work the study of the physical stability is particularly interesting due to the appearance of operators with higher order spatial derivatives. We proceeded, without including a matter sector, to study the stability of different sets of theories, leaving the analysis of the matter backreactions to future investigation~\cite{generalstability}. After integrating out the auxiliary fields we obtained an EFT action describing only the dynamics of the  propagating DoF. From this action, we identified  the kinetic term and  the speed of propagation  which have now become functions of scale, besides the usual dependence on time, due to the presence of higher derivative operators. We required both to be positive in order to guarantee a viable theory free from ghost and gradient instabilities. Subsequently we identified, at the level of the equations of motion, the friction and dispersive coefficients. We did this both for the scalar and tensor DoF. Finally, we normalized the scalar DoF in order to obtain an action in the canonical form. This form allowed us to identify the effective mass term on which we imposed conditions in order to avoid the appearance of tachyonic instabilities in the scalar sector. As a result, we obtained a set of very general stability conditions which must be imposed in order to ensure theoretical viability of models with operators containing up to sixth order in spatial derivatives, in absence of matter. It is worth noting that due to the complicated nature of some classes of theories, when written in the EFT formalism, we had to divide the treatment and the resulting conditions in different subsets.

Finally, we have  built an extended and generalized version of the phenomenological parametrization  in terms of $\alpha$ functions introduced in Ref.~\cite{Bellini:2014fua}, to which we refer as ReParametrized Horndeski (RPH). This parametrization was originally built to include all models in the Horndeski class, and was afterwards extended to encompass
beyond Horndeski models known as GLPV, in Ref.~\cite{Gleyzes:2014qga}. This was achieved by introducing an additional function which parametrizes the deviation from Horndeski theories. From this point we proceeded  to introduce  new functions and generalize the definition of the original ones, in order to account for all the beyond GLPV models described by the higher order operators that we have included in our extended EFT action~(\ref{EFTgeneral}).  In particular, we have found a new function parametrizing the braiding, which also contributes to the kinetic term; we have generalized the definitions of the kinetic and tensor speed excess functions, the latter one now being both time and scale dependent; finally, we have identified four extra functions entering in the definition of the speed of propagation of the scalar DoF. It is important to notice that the structure of this extended phenomenological basis in terms of $\alpha$ functions  becomes quite cumbersome when higher order operators are considered and the correspondence between the different functions and cosmological 
observables becomes weaker.

\begin{acknowledgments}
We are grateful to Antonio De Felice and Daniele Vernieri for useful discussions and comments on the manuscript.
The research of NF has received funding from the European Research Council under the European Community's Seventh Framework Programme (FP7/2007-2013, Grant Agreement No.~307934). GP acknowledges support from the D-ITP consortium, a program of the Netherlands Organisation for Scientific Research (NWO) that is funded by the Dutch Ministry of Education, Culture and Science (OCW).
AS acknowledges support from The Netherlands Organization for Scientific Research (NWO/OCW), and also from the D-ITP consortium, a program of the Netherlands Organisation for Scientific Research (NWO) that is funded by the Dutch Ministry of Education, Culture and Science (OCW).
\end{acknowledgments}

\appendix

\section{On $\delta K$ and $\delta S$ perturbations}\label{KSperturbations}

In this Section we explicitly work out the perturbations associated to $\delta K$ and $\delta S$ used in Section~\ref{generallagrangian} and show the difference with previous approaches~\cite{Gleyzes:2013ooa,Kase:2014cwa}. For this purpose, we consider the following terms of the Lagrangian~(\ref{firstADMlag}):
\be
\delta L \supset L_K\delta K+L_S\delta S= \mathcal{F}\delta K+ L_S \delta K^\mu_\nu\delta K_\mu^\nu\equiv \mathcal{F}(K+3H)+ L_S \delta K^\mu_\nu\delta K_\mu^\nu\,,
\ee
where  we have defined:
\be\label{Fdef}
\mathcal{F}\equiv L_K-2HL_S.
\ee 
Now, let us prove a relation which is useful in order to obtain action~(\ref{actionexpanded}):
\begin{align}\label{DerTrick}
\int d^4x\sqrt{-g} \mathcal{F} K=\int d^4x\sqrt{-g} \mathcal{F} \nabla_{\mu}n^{\mu}=-\int d^4x\sqrt{-g} \nabla _{\mu}\mathcal{F}n^{\mu}=\int d^4x\sqrt{-g}\frac{\dot{\mathcal{F}}}{N}\,. 
\end{align}
Using the above relation and the expansion of the lapse function:
\be
N=1+\delta N+\delta N^2+\mathcal{O}(3),
\ee
finally, we obtain:
\be\label{finalperturbation}
L_K\delta K+L_S\delta S= 3H\mathcal{F}+\dot{\mathcal{F}}\l(1-\delta N +(\delta N)^2\r)+ L_S \delta K^\mu_\nu\delta K_\mu^\nu.
\ee
The differences with previous works are due to the different convention on the normal vector, $n^\mu$ (see Eq.~(\ref{convention})), which is responsible of the different sign in Eq.~(\ref{DerTrick}) w.r.t. the definition used in Refs.~\cite{Gleyzes:2013ooa,Kase:2014cwa} and then in the final results~(\ref{finalperturbation}). Moreover, the difference in the definition of the extrinsic curvature, see Eq.~(\ref{convention}), which is a consequence of the convention adopted for the normal vector, leads to the minus sign in Eq.~(\ref{Fdef}) because its background value is $K^{i(0)}_j=-H\delta_j^i$.

\section{On $\delta \mathcal{U}$ perturbation}\label{Uperturbations}

Due to the different convention for $n^\mu$ we adopted here (see Eq.~(\ref{convention})),  the result obtained in Refs.~\cite{Gleyzes:2013ooa,Kase:2014cwa} concerning the perturbation associated to $\mathcal{U}=\mathcal{R}_{\mu\nu}K^{\mu\nu}$,  can not be directly applied to our Lagrangian~(\ref{firstADMlag}).  Therefore, we need to derive again such result, which is crucial in order to obtain the coefficients of the action~(\ref{actionexpanded}). Then, let us prove the following relation: 
\be\label{UR}
\int d^4x\sqrt{g}\lambda(t) \mathcal{R}_{\mu\nu}K^{\mu\nu}=\int d^4 x\sqrt{g}\l(\frac{\lambda(t)}{2}\mathcal{R}K-\frac{\dot{\lambda}(t)}{2N}\mathcal{R}\r),
\ee
where $\lambda(t)$ is a generic function of time. We notice that in Ref.~\cite{Gleyzes:2013ooa} the above relation is defined with a plus in front of the second term in the last expression.  Using the relation $K=\nabla^{\mu}n_{\mu}$ we obtain:
\begin{align}
\int d^4 x\sqrt{-g}\l(\lambda(t)\mathcal{R}_{\mu\nu}K^{\mu\nu}-\frac{\lambda(t)}{2}\mathcal{R}\nabla_{\mu}n^{\mu}+\frac{\dot{\lambda}(t)}{2N}\mathcal{R}\r)=0\,.
\end{align}
Now, after integration by parts of the second term and using $n^{\mu}=\l(-1/N,N^i/N\r)$, the last term cancels and we are left with:
\be\label{U1}
\int d^4 x\sqrt{-g}\l(\lambda(t)\mathcal{R}_{\mu\nu}K^{\mu\nu}+\frac{\lambda(t)}{2}n^{\mu}\nabla_{\mu}\mathcal{R}\r)=0\,.
\ee
The first term can be rewritten using the expression for the extrinsic curvature in the ADM formalism:
\be
K_{ij}=-\frac{1}{2N}\big[\partial_t h_{ij}-\nabla_i N_j-\nabla_j N_i\big],
\ee
where covariant derivative is w.r.t. the spatial metric $h_{ij}$. The overall minus sign which appears in the above definition  makes the expression to differ from the one usually encountered that follows from the definition of $n^\mu$ we employed.  After substituting this expression into Eq.~(\ref{U1}) we get:
\begin{align}
\int d^4 x\sqrt{h}\lambda(t)\l[-\f{1}{2}\l(\mathcal{R}_{ij}h^{il}h^{jk}\dot{h}_{lk}+\dot{\mathcal{R}}\r)+\nabla^iN^j \mathcal{R}_{ij}+\f{1}{2}N^i\nabla_i \mathcal{R}\r]=0\,.
\end{align}
From here on the subsequent steps follows Ref.~\cite{Gleyzes:2013ooa}, indeed the last two terms vanish due to the Bianchi identity and the first two can be combined as a total divergence. Hence, the relation~(\ref{UR}) holds.

Finally, using the above relation we can now  compute the perturbations coming from $\mathcal{U}=\mathcal{R}_{\mu\nu}K^{\mu\nu}$. Indeed we have:
\ba
\int{}d^4x\sqrt{-g}L_\mathcal{U}\mathcal{R}_{\mu\nu}K^{\mu\nu}&=&\int{}d^4x\sqrt{-g}\l[\f{1}{2}L_\mathcal{U}\mathcal{R}K-\f{1}{2N}\dot{L}_{\mathcal{U}}\mathcal{R}\r]\nn \\
&=&\int{}d^4x\sqrt{-g}\l[\f{1}{2}L_\mathcal{U}\l(K^{(0)}\delta\mathcal{R}+\delta K\delta \mathcal{R}\r)-\f{1}{2}\dot{L}_{\mathcal{U}}\mathcal{R}\l(1-\delta N\r)\r]\,,
\ea
then we get:
\be
L_\mathcal{U}\delta \mathcal{U}=-\f{1}{2}\l(3L_\mathcal{U}+\f{1}{2}\dot{L}_{\mathcal{U}}\r)\delta \mathcal{R}+\l(\f{1}{2}L_\mathcal{U}\delta K+\f{1}{2}\dot{L}_{\mathcal{U}}\delta N\r)\delta \mathcal{R}\,.
\ee

\section{Conformal EFT functions for Generalized Galileon and GLPV}\label{Conformal}

In this Appendix we collect the results of Sections~\ref{galileons} and~\ref{Sec:GLPVmapping}, and convert them to functions of the scale factor; the Hubble parameter and its time derivative are defined  in terms of the conformal time, still they need to be considered functions of the scale factor. This further step is important for a direct implementation in EFTCAMB of Horndeski/GG and GLPV theories. \emph{In this Section only}, primes indicate derivatives w.r.t. the scale factor. Furthermore, $\hub \equiv d\ln{a}/d\tau$ and $\dot{\hub}\equiv d\hub/d\tau$, where $\tau$ is the conformal time.  In order to get the correct results  $\{\mathcal{K}, G_i, \tilde{F}_i\}$ have to be considered functions of the scale factor.

First, we consider the EFT functions derived in Section~\ref{galileons} for Horndeski/GG theories:
\begin{itemize}
\item $L_2$-Lagrangian
\end{itemize}
\begin{align}
\Lambda(a)&=\mathcal{K},\nn \\
 c(a)&=\mathcal{K}_X X_0\,,\nn\\
M^4_2(a)&=\mathcal{K}_{XX}X_0^2\,,
\end{align}
where $X_0$ is: 
\be
X_0=-\hub^2\phi_0^{\prime 2}\,.
\ee
\begin{itemize}
\item $L_3$-Lagrangian
\end{itemize}
\begin{align}
\Lambda(a)&=\mathcal{H}^2\phi_0'^2\l[ G_{3\phi}-2G_{3X}\l(\frac{\dot{\mathcal{H}}}{a}\phi_0^\prime+\mathcal{H}^2\phi_0''\r)\r]\,,\nn\\
c(a)&=\mathcal{H}^2\phi_0'^2\l[G_{3X}\l(\l(3\mathcal{H}^2-\dot{\mathcal{H}}\r)\frac{\phi_0'}{a}-\mathcal{H}^2\phi_0''\r)+G_{3\phi}\r]\,,\nn\\
M^4_2(a)&=\frac{G_{3X}}{2}\mathcal{H}^2\phi_0'^2\l(\l(3\mathcal{H}^2+\dot{\mathcal{H}}\r)\frac{\phi_0'}{a}+\mathcal{H}^2\phi_0''\r)-3\frac{\mathcal{H}^6}{a}G_{3XX}\phi_0'^5-\frac{G_{3\phi X}}{2}\mathcal{H}^4\phi_0'^4\,,\nn\\
\bar{M}_1^3(a)&=-2\mathcal{H}^3G_{3X}\phi_0'^3\,.
\end{align}
\begin{itemize}
\item $L_4$-Lagrangian
\end{itemize}
\begin{align}
\Omega(a)&=-1+\f{2}{m_0^2}G_4\,,\nn\\
%%%%%%%%%%%%%%%%%%%%%%%%%%
c(a)&=G_{4X}\left[2\l(\dot{\hub}^2+\hub\ddot{\hub}+2\hub^2\dot{\hub}-5\hub^4\r)\f{\phi^{\prime\,2}_0}{a^2}+2\l(5\hub^2\dot{\hub}+\hub^4\r)\f{\phi^\prime_0}{a}\phi^{\prime\prime}_0+2\hub^4\phi^{\prime\prime\,2}_0+2\hub^4\phi^\prime_0\phi'''_0\right]\nn\\
&+G_{4X\phi}\l[2\mathcal{H}^2\phi_0'^2\l(\frac{\dot{\mathcal{H}}}{a}\phi_0'+\mathcal{H}^2\phi_0''\r)+10\frac{\mathcal{H}^4}{a}\phi_0'^3\r]+G_{4XX}\left[ 12\frac{\mathcal{H}^6}{a^2}\phi_0'^4-8\frac{\mathcal{H}^4}{a}\phi_0'^3\l(\frac{\dot{\mathcal{H}}}{a}\phi_0'+\mathcal{H}^2\phi_0''\r)\nn\r.\\&\l.-4\mathcal{H}^2\phi_0'^2\left(\frac{\dot{\mathcal{H}}^2}{a^2}\phi_0'^2+2\frac{\dot{\mathcal{H}}\mathcal{H}^2}{a}\phi_0'\phi_0''+\mathcal{H}^4\phi_0''^2\right) \right]\,,\nn\\
%%%%%%%%%%%%%%%%%%%%%%%%%%
\Lambda(a)&=G_{4X}\l[4\l(\hub^4+5\hub^2\dot{\hub}+\dot{\hub}^2+\hub\ddot{\hub}\r)\f{\phi^{\prime\,2}_0}{a^2}+4\l(4\hub^4+5\hub^2\dot{\hub}\r)\f{\phi^{\prime}_0}{a}\phi^{\prime\prime}_0+4\hub^4\phi^{\prime\prime\,2}_0+4\hub^4\phi^\prime_0\phi'''_0 \r]\nn\\
&+8\frac{\mathcal{H}^4}{a}G_{4X\phi}\phi_0'^3-8G_{4XX}\hub^2\phi^{\prime\,2}_0\l(\dot{\hub}\f{\phi^\prime_0}{a}+\hub^2\phi^{\prime\prime}_0\r)\l(2\hub^2\f{\phi^\prime_0}{a}+\dot{\hub}\f{\phi^\prime_0}{a}+\hub^2\phi_0^{\prime\prime}\r) \,,\nn\\
%%%%%%%%%%%%%
M^4_2(a)&=G_{4X\phi}\l[4\frac{\mathcal{H}^4}{a}\phi_0'^3-\mathcal{H}^2\phi_0^{\prime\,2}\l(\frac{\dot{\mathcal{H}}}{a}\phi_0'+\mathcal{H}^2\phi_0''\r)\r]-6\frac{\mathcal{H}^6}{a}\phi_0'^5G_{4\phi XX}-12\frac{\mathcal{H}^8}{a^2}G_{4XXX}\phi_0'^6\nn\\
&+G_{4XX}\hub^2\phi^{\prime\,2}_0\l[2\l(9\hub^4+\dot{\hub}^2+2\hub^2\dot{\hub}\r)\f{\phi^{\prime\,2}_0}{a^2}+2\l(2\hub^2\dot{\hub}+2\hub^4\r)\f{\phi^{\prime}_0}{a}\phi''_0+2\hub^4\phi^{\prime\prime\,2}\r]\nn\\
&+G_{4X}\l[\l(-2\dot{\hub}\hub^2+2\hub^4-\dot{\hub}^2-\hub\ddot{\hub}\r)\f{\phi^{\prime\,2}_0}{a^2}-\l(\hub^4+5\hub^2\dot{\hub}\r)\f{\phi^{\prime}_0}{a}\phi^{\prime\prime}_0-\hub^4\phi^{\prime\prime\,2}-\hub^4\phi^\prime_0\phi'''_0\r]\,,\nn \\
%%%%%%%%%%%%%%%%%%%%%%%%%%%%%%%%%
\bar{M}^3_1(a)&=4G_{4X}\hub\phi_0^\prime\l[\l(\dot{\hub}+2\hub^2\r)\f{\phi_0^\prime}{a}+\mathcal{H}^2\phi_0''\r]-16G_{4XX}\frac{\mathcal{H}^5}{a}\phi_0'^5-4G_{4X\phi}\mathcal{H}^3\phi_0'^3\,,\nn\\
%%%%%%%%%%%%%%%
\bar{M}^2_2(a)&=4\mathcal{H}^2 G_{4X}\phi_0'^2=-\bar{M}^2_3(a)=2\hat{M}^2(a)\,.
\end{align}
%%%%%%%%%%%%%%%%%%%%%%%%%%%%%%%%%%
\begin{itemize}
\item $L_5$-Lagrangian
\end{itemize}
\begin{align}
\Omega(a)&=\frac{2\mathcal{H}^2}{m_0^2}\phi_0'^2\l[G_{5X}\l(\frac{\dot{\mathcal{H}}}{a}\phi_0'+\mathcal{H}^2\phi_0''\r)-\f{G_{5\phi}}{2}\r]-1\,,\nn\\
%%%%%%%%%%%%%%%%%%%%%
c(a)&=\frac{\mathcal{H}}{2}\tilde{\mathcal{F}}'+\frac{3}{2}\frac{\mathcal{H}^2}{a}m_0^2\Omega'-3\frac{\hub^4}{a^2}\phi_0'^2G_{5\phi}+\frac{3\mathcal{H}^6}{a^2}\phi_0'^4G_{5\phi X}-3\frac{\mathcal{H}^6}{a^3}\phi_0'^3 G_{5X}+2\frac{\mathcal{H}^8}{a^3}\phi_0'^5G_{5XX}\,,\nn\\
%%%%%%%%%%%%%%%%%
\Lambda(a)&=\tilde{\mathcal{F}}-3m_0^2\frac{\mathcal{H}^2}{a^2}(1+\Omega)+4G_{5X}\frac{\mathcal{H}^6}{a^3}\phi_0'^3+3\frac{\mathcal{H}^3}{a}G_{5\phi}\phi_0'^2\,,\nn\\
%%%%%%%%%%%%%%%%%%
M^2_4(a)&=-\frac{\tilde{\mathcal{F}}}{4}-\frac{3}{4}\frac{\mathcal{H}^2}{a}m_0^2\Omega'-2\frac{\mathcal{H}^{10}}{a^3}\phi_0'^7G_{5XXX}-3\frac{\mathcal{H}^8}{a^2}\phi_0'^6G_{5\phi XX}+6G_{5XX}\frac{\mathcal{H}^8}{a^3}\phi_0'^5+6\frac{\mathcal{H}^6}{a^2}\phi_0'^4 G_{5\phi X}-\frac{3}{2}\frac{\mathcal{H}^6}{a^3}\phi_0'^3 G_{5X}\,,\nn\\
%%%%%%%%%%%%%%%%%%%%%%%%%%%%%%
\bar{M}^2_2(a)&=2\l[ \mathcal{H}^2\phi_0'^2G_{5\phi}-G_{5X} \l[ -\frac{\mathcal{H}^4}{a}\phi_0'^3+\mathcal{H}^2\phi_0'^2\l(\frac{\dot{\mathcal{H}}}{a}\phi_0'+\mathcal{H}^2\phi_0''\r)   \r] \r]=-\bar{M}^2_3(a)=2\hat{M}^2(a)\,,\nn\\
%%%%%%%%%%%%%%%%%%%%%%%%%%
\bar{M}^3_1(a)&=-\mathcal{H}m_0^2\Omega'+4\frac{\mathcal{H}^3}{a}\phi_0'^2G_{5\phi}-4\frac{\mathcal{H}^5}{a}\phi_0'^4 G_{5\phi X}-4\frac{\mathcal{H}^7}{a^2}\phi_0'^5G_{5XX}+6\frac{\mathcal{H}^5}{a^2}\phi_0'^3 G_{5X}\,,
\end{align}
where $\tilde{\mathcal{F}}(a)=\mathcal{F}-m_0^2\hub\Omega^\prime-2\f{\hub}{a}m_0^2(1+\Omega)$ and $\mathcal{F}(\tau)=2\frac{\mathcal{H}^4}{a}G_{5X}\phi_0^3+2\frac{\mathcal{H}^3}{a}G_{5\phi}\phi_0'^2$.
\vspace{1cm}

Let us now consider the two Lagrangians which extend the Horndeski/GG theories to the GLPV ones introduced in Section~\ref{Sec:GLPVmapping}:
\begin{itemize}
\item $L_4^{GLPV}$-Lagrangian
\end{itemize}
\begin{align}
c(a)&=2\frac{\mathcal{H}^4}{a^2}\phi_0'^4(\dot{\mathcal{H}}-\mathcal{H}^2)\tilde{F}_4+8\frac{\mathcal{H}^4}{a}\phi_0'^3\tilde{F}_4\l(\frac{\dot{\mathcal{H}}}{a}\phi_0'+\mathcal{H}^2\phi_0''\r)-4\frac{\mathcal{H}^6}{a}\tilde{F}_{4X}\l( \frac{\dot{\mathcal{H}}}{a}\phi_0'+\mathcal{H}^2\phi_0''  \r)\phi_0'^5+2\mathcal{H}^6\frac{\tilde{F}_{4\phi}}{a}\phi_0'^5\nn\\
&-12\frac{\mathcal{H}^6}{a^2}\phi_0'^4\tilde{F}_4\,,\nn\\
%%%%%%
\Lambda(a)&=6\frac{\mathcal{H}^6}{a^2}\tilde{F}_4\phi_0'^4+4\frac{\mathcal{H}^4}{a^2}(\dot{\mathcal{H}}-\mathcal{H}^2)\phi_0'^4\tilde{F}_4+16\frac{\mathcal{H}^4}{a}\phi_0'^3\tilde{F}_4\l(\frac{\dot{\mathcal{H}}}{a}\phi_0'+\mathcal{H}^2\phi_0''\r)-8\frac{\mathcal{H}^6}{a}\tilde{F}_{4X}\l(\frac{\dot{\mathcal{H}}}{a}\phi_0'+\mathcal{H}^2\phi_0''\r)\phi_0'^5\nn\\
&+4\frac{\mathcal{H}^6}{a}\tilde{F}_{4\phi}\phi_0'^5\,,\nn\\
%%%%%%%%%%%%%%%%%
M^4_2(a)&=-18\frac{\mathcal{H}^6}{a^2}\phi_0'^4\tilde{F}_4-\frac{\mathcal{H}^4}{a^2}\phi_0'^4(\dot{\mathcal{H}}-\mathcal{H}^2)\tilde{F}_4-4\frac{\mathcal{H}^4}{a}\phi_0'^3\tilde{F}_4\l(\frac{\dot{\mathcal{H}}}{a}\phi_0'+\mathcal{H}^2\phi_0''\r)+2\frac{\mathcal{H}^6}{a}\phi_0'^5\tilde{F}_{4X}\l(\frac{\dot{\mathcal{H}}}{a}\phi_0'+\mathcal{H}\phi_0''\r)\nn\\
&-\frac{\mathcal{H}^6}{a}\phi_0^{\prime\,5} \tilde{F}_{4\phi}+6\frac{\mathcal{H}^6}{a^2}\phi_0'^4\tilde{F}_4\,,\nn\\
%%%%%%%%%%%%%%
\bar{M}^2_2(a)&=2\mathcal{H}^4\phi_0'^4\tilde{F}_4=-\bar{M}^2_3(a)\,,\nn\\
%%%%%%%%%%%%
\bar{M}^3_1(a)&=16\frac{\mathcal{H}^5}{a}\phi_0'^4\tilde{F}_4\,.
\end{align}
\begin{itemize}
\item $L_5^{GLPV}$-Lagrangian
\end{itemize}
\begin{align}
\Lambda(a)&=-3\frac{\mathcal{H}^8}{a^3}\phi_0'^5\tilde{F}_5-12\frac{\mathcal{H}^6}{a^3}\phi_0'^5(\dot{\mathcal{H}}-\mathcal{H}^2)\tilde{F}_5-30\frac{\mathcal{H}^6}{a^2}\tilde{F}_5\l(\frac{\dot{\mathcal{H}}}{a}\phi_0'+\mathcal{H}^2\phi_0''\r)\phi_0'^4+12\frac{\mathcal{H}^8}{a^2}\tilde{F}_{5X}\l(\frac{\dot{\mathcal{H}}}{a}\phi_0'+\mathcal{H}^2\phi_0''\r)\phi_0'^6\nn\\&-6\frac{\mathcal{H}^8}{a^2}\tilde{F}_{5\phi}\phi_0'^6\,,\nn\\
c(a)&=6\frac{\mathcal{H}^8}{a^2}\phi_0'^6\tilde{F}_{5X}\l(\frac{\dot{\mathcal{H}}}{a}\phi_0'+\mathcal{H}^2\phi_0''\r)-6\frac{\mathcal{H}^6}{a^3}(\dot{\mathcal{H}}-\mathcal{H}^2)\phi_0'^5\tilde{F}_5-15\frac{\mathcal{H}^6}{a^2}\phi_0'^4\l(\frac{\dot{\mathcal{H}}}{a}\phi_0'+\mathcal{H}^2\phi_0''\r)-3\frac{\mathcal{H}^8}{a^2}\phi_0'^6 \tilde{F}_{5\phi}+15\frac{\mathcal{H}^8}{a^3}\tilde{F}_5\phi_0'^5\,,\nn\\
M^4_2(a)&=\frac{45}{2}\frac{\mathcal{H}^8}{a^3}\phi_0'^5\tilde{F}_5+3\frac{\mathcal{H}^6}{a^3}(\dot{\mathcal{H}}-\mathcal{H}^2)\phi_0'^5\tilde{F}_5+\frac{15}{2}\frac{\mathcal{H}^6}{a^2}\phi_0'^4\tilde{F}_5\l(\frac{\dot{\mathcal{H}}}{a}\phi_0'+\mathcal{H}^2\phi_0''\r)-3\frac{\mathcal{H}^8}{a^2}\phi_0'^6\l(\frac{\dot{\mathcal{H}}}{a}\phi_0'+\mathcal{H}^2\phi_0''\r)\tilde{F}_{5X}\nn\\&+\frac{3}{2}\frac{\mathcal{H}^8}{a^2}\phi_0'^6\tilde{F}_{5\phi}\,,\nn\\
\bar{M}^2_2(a)&=-6\frac{\mathcal{H}^6}{a}\phi_0'^5\tilde{F}_5=-\bar{M}^2_3(a)\,,\nn\\
\bar{M}^3_1(a)&=-30\frac{\mathcal{H}^7}{a^2}\tilde{F}_5\phi_0'^5\,.
\end{align}
\vspace{0.6cm}

Finally, we write the EFT functions obtained from the GLPV action~(\ref{GLPVADM}) in Section~\ref{Sec:GLPVmapping} in the appropriate form adopted in EFTCAMB : 
\begin{align}
\Omega(a)&=\frac{2}{m_0^2}\Big( \bar{B}_4-\frac{\mathcal{H}}{2}\bar{B}_5' \Big)-1\,, \nn\\
%%%%%%%%%%
\Lambda(a)&=\bar{A}_2-6\frac{\mathcal{H}^2}{a}\bar{A}_4+12\frac{\mathcal{H}^3}{a^3}\bar{A}_5+\mathcal{H}\bar{A}_3'-\frac{4}{a^2}(\dot{\mathcal{H}}-\mathcal{H}^2)\bar{A}_4-4\frac{\mathcal{H}^2}{a}\bar{A}_4'+6\frac{\mathcal{H}^3}{a^2}\bar{A}_5'+12\frac{\mathcal{H}}{a^3}(\dot{\mathcal{H}}-\mathcal{H}^2)\bar{A}_5\nonumber\\
&-\l[ \f{2}{a^2}\l(\mathcal{H}^2+2\dot{\mathcal{H}}\r)\bar{B}_4 +\f{2}{a}\l(\dot{\hub}+2\hub^2\r)\bar{B}_4'+2\hub^2\bar{B}_4''-\f{\hub}{a^2}\l(\hub^2+3\dot{\hub}+\f{\ddot{\mathcal{H}}}{\mathcal{H}}\r)\bar{B}_5'-\f{\hub}{a}\l(3\dot{\hub}+2\hub^2\r)\bar{B}_5''-\hub^3\bar{B}_5'''\r]\,, \nn\\
 %%%%%%%%%
c(a)&=\frac{1}{2}\Big( \mathcal{H}\bar{A}_3'-\frac{4}{a^2}(\dot{\mathcal{H}}-\mathcal{H}^2)\bar{A}_4-4\frac{\mathcal{H}^2}{a}\bar{A}_4'
+6\frac{\mathcal{H}^3}{a^2}\bar{A}_5'+12\frac{\mathcal{H}}{a^3}(\dot{\mathcal{H}}-\mathcal{H}^2)\bar{A}_5-\bar{A}_{2N}  +3\mathcal{H}\bar{A}_{3N}
-6\frac{\mathcal{H}^2}{a^2}\bar{A}_{4N}+6\frac{\mathcal{H}^3}{a^3}\bar{A}_{5N}\Big)\nonumber\\
&+\f{1}{a}\l(\hub^2-\dot{\hub}\r)\bar{B}_4'+\f{\hub}{2a}\l(3\dot{\hub}-\hub^2\r)\bar{B}_5'' -\mathcal{H}^2\bar{B}_4''+\frac{\mathcal{H}^3}{2}\bar{B}_5'''+\frac{1}{2a^2}\l( \ddot{\mathcal{H}}-2\mathcal{H}^3 \r)\bar{B}_5'-\frac{2}{a^2}(\dot{\mathcal{H}}-\mathcal{H}^2)\bar{B}_4\,,\nn\\
%%%%%%%%%%%%%
M^4_2(a)&=\frac{1}{4}\Big(  \bar{A}_{2NN}    -3\frac{\mathcal{H}}{a}\bar{A}_{3NN}+6\frac{\mathcal{H}^2}{a^2}\bar{A}_{4NN}-6\frac{\mathcal{H}^3}{a^3}\bar{A}_{5NN} \Big)-\frac{1}{4} \l[  \mathcal{H}\bar{A}_3'-4\frac{\bar{A}_4}{a^2}(\dot{\mathcal{H}}-\mathcal{H}^2) -4\frac{\mathcal{H}^2}{a}\bar{A}_4'
+6\frac{\mathcal{H}^3}{a^2}\bar{A}_5'\r.\nn\\
&\l.+12\bar{A}_5 \frac{\mathcal{H}}{a^3}(\dot{\mathcal{H}}-\mathcal{H}^2)  \r]  +\frac{3}{4}\Big(  \bar{A}_{2N}-3\frac{\mathcal{H}}{a}\bar{A}_{3N}+6\frac{\mathcal{H}^2}{a^2}\bar{A}_{4N}-6\frac{\mathcal{H}^3}{a^3}\bar{A}_{5N} \Big)-\frac{1}{2}\l[-\frac{2}{a^2}(\dot{\mathcal{H}}-\mathcal{H}^2)\bar{B}_4 \r.\nn\\
&\l. +\f{1}{a}\l(\hub^2-\dot{\hub}\r)\bar{B}_4'-\mathcal{H}^2 \bar{B}_4''+\f{1}{a^2}\l(\ddot{\hub}-\hub^3\r)\bar{B}_5'+\f{\hub}{2a}\l(3\dot{\hub}-\hub^2\r)\bar{B}_5''+\f{\mathcal{H}^3}{2} \bar{B}_5'''\r]\,,\nn\\
%%%%%%%%%%%%%%%
\bar{M}^2_2(a)&=-2\bar{A}_4+6\frac{\mathcal{H}}{a}\bar{A}_5-2\bar{B}_4+\mathcal{H}\bar{B}_5'=-\bar{M}^2_3(a)\,,\nn\\
%%%%%%%%%%%%%%%%
\bar{M}^3_1(a)&=-\bar{A}_{3N}+4\frac{\mathcal{H}}{a}\bar{A}_{4N}-6\frac{\mathcal{H}^2}{a^2}\bar{A}_{5N}-2\bar{B}_4'\mathcal{H}+\frac{\dot{\mathcal{H}}}{a}\bar{B}_5'+\mathcal{H}^2 \bar{B}_5''\,,\nn\\
%%%%%%%%%%%%%%%%%%%
\hat{M}^2(a)&=\bar{B}_{4N}+\frac{\mathcal{H}}{2a}\bar{B}_{5N}+\frac{\mathcal{H}}{2}\bar{B}_5'\,.
\end{align}

\section{On the $\mathcal{J}$ coefficient in the $L_5$ Lagrangian}\label{Jcoefficient}

In this Appendix we will show the details of the calculation regarding the $J$ coefficient in the $L_5$ Lagrangian~(\ref{L5lagrangian}). 
Let us  consider the following term:
\begin{align}
G_{5X}\mathcal{J}&=G_{5X}\Big(-\frac{1}{2}\phi^{;\rho}X_{;\rho}(K^2-S)+2\gamma^{-3}(\gamma^2\frac{h^{\rho}_{\mu}}{2}X_{;\rho})(K\dot{n}_{\mu}-K^{\mu\nu}\dot{n}_{\nu})\Big)  \nonumber\\
&=-\frac{1}{2}\Big(\gamma\nabla_{\rho}(\gamma^{-1}F_5)-F_{5\phi}\gamma^{-1}n_{\rho}\Big)(K^2-S)\phi^{;\rho}+\gamma^{-1}(K\dot{n}^{\mu}-K^{\mu\nu}\dot{n}_{\nu})h^{\mu}_{\rho}\Big( \gamma\nabla_{\rho}(\gamma^{-1}F_5)+F_{5\phi}\gamma^{-1}n^{\rho}\Big)\,.
\end{align}
The last parenthesis contains a quantity which  is orthogonal to the quantities that multiply it, hence it vanishes. Therefore, we have: 
\begin{align}
G_{5X}\mathcal{J}&=\frac{F_{5\phi}}{2}n_{\rho}n^{\rho}(K^2-S)-\frac{1}{2}n^{\rho}\nabla_{\rho}(\gamma^{-1}F_5)(K^2-S)+h^{\rho}_{\mu}\nabla_{\rho}(\gamma^{-1}F_5)(K\dot{n}^{\mu}-K^{\mu\nu}\dot{n}_{\nu})\nonumber\\
&=-\frac{F_{5\phi}}{2}(K^2-S)+\frac{F_5}{\gamma}\l[\frac{1}{2}\nabla_{\rho}(n^{\rho}K^2-n^{\rho}K_{\mu\nu}K^{\mu\nu})-(K\dot{n}^{\mu}-K^{\mu\nu}\dot{n}_{\nu})_{;\mu}\r]\nonumber\\
&=\frac{F_5}{\gamma}\l(  \frac{K^3}{2}+n^{\rho}K\nabla_{\rho} K -\frac{K}{2}K_{\mu\nu}K^{\mu\nu}-n^{\rho}K^{\mu\nu}\nabla_{\rho}K_{\mu\nu}-\dot{n}^{\rho}\nabla_{\rho}K-K\nabla_{\rho}\dot{n}^{\rho}+\dot{n}_{\nu}\nabla_{\rho}K^{\rho\nu}+K^{\rho\nu}\nabla_{\rho}\dot{n}_{\nu}       \r)  \nonumber\\
 & -\frac{F_{5\phi}}{2}(K^2-S),
\end{align}
where in the second line we have used the fact that $n_{\mu}$ is orthogonal to $\dot{n}_{\mu}$  and $K^{\mu\nu}$.  Now, employing the following geometrical quantities: 
\begin{align}%\label{FinRnn}
R_{\mu\nu}n^{\mu}n^{\nu}&=-n^{\mu}\nabla_{\mu}K+\nabla_{\mu}\dot{n}^{\mu}+n^{\mu}\nabla^{\nu}K_{\mu\nu}\,,\nn\\
%\label{Rnnd}
R_{\mu\nu}n^{\nu}\dot{n}^{\mu}&=\dot{n}^{\mu}\nabla^{\nu}K_{\mu\nu}-\dot{n}^{\mu}\dot{n}_{\nu}\nabla^{\nu}n_{\mu}-\dot{n}^{\mu}\nabla_{\mu}K\,,\nn\\
%\label{KReiman}
K^{\mu\nu}n^{\rho}n^{\sigma}R_{\mu\sigma\nu\rho}&=K^{\gamma\alpha}n^{\beta}(\nabla_{\alpha}K_{\beta\gamma})-K^{\gamma\alpha}n^{\beta}(\nabla_{\beta}K_{\alpha\gamma})+K^{\gamma\alpha}(\nabla_{\alpha}\dot{n}_{\gamma})+K^{\gamma\alpha}\dot{n}_{\gamma}\dot{n}_{\alpha}\,,
\end{align}
we obtain:
\begin{align}
G_{5X}\mathcal{J}&=\frac{F_5}{\gamma}\Big(  \frac{K^3}{2}+n^{\rho}K\nabla_{\rho} K -\frac{K}{2}K_{\mu\nu}K^{\mu\nu}-n^{\rho}K^{\mu\nu}\nabla_{\rho}K_{\mu\nu}-\dot{n}^{\rho}\nabla_{\rho}K-K\nabla_{\rho}\dot{n}^{\rho}+\dot{n}_{\nu}\nabla_{\rho}K^{\rho\nu}+K^{\rho\nu}\nabla_{\rho}\dot{n}_{\nu}       \Big)  \nonumber\\
 & -\frac{F_{5\phi}}{2}(K^2-S)\nn\\
&=\frac{F_5}{\gamma}\Big(  \frac{K^3}{2}-\frac{K}{2}K_{\mu\nu}K^{\mu\nu}-K R_{\mu\nu}n^{\mu}n^{\nu}+n^{\mu}K(\nabla^{\nu}K_{\mu\nu})+K^{\mu\nu}n^{\rho}n^{\sigma}+K^{\mu\nu}n^{\sigma}n^{\rho}R_{\mu\sigma\nu\rho}-K^{\gamma\alpha}n^{\beta}(\nabla_{\alpha}K_{\beta\gamma})\nonumber\\
&-K^{\gamma\alpha}\dot{n}_{\gamma}\dot{n}_{\alpha}+R_{\mu\nu}n^{\mu}\dot{n}^{\nu}+\dot{n}^{\mu}\dot{n}^{\nu}\nabla_{\nu}n_{\mu} \Big)  -\frac{F_{5\phi}}{2}(K^2-S)\nn\\
&=\frac{F_5}{\gamma}\Big(  \frac{K^3}{2}-\frac{K}{2}K_{\mu\nu}K^{\mu\nu}-K R_{\mu\nu}n^{\mu}n^{\nu}-K K^{\mu\nu}K_{\mu\nu}+K^{\mu\nu}n^{\rho}n^{\sigma}+K^{\mu\nu}n^{\sigma}n^{\rho}R_{\mu\sigma\nu\rho}+K^{\gamma\alpha}K^{\beta}_{\alpha}K_{\beta\gamma}\nonumber\\
&-K^{\gamma\alpha}\dot{n}_{\gamma}\dot{n}_{\alpha}+R_{\mu\nu}n^{\mu}\dot{n}^{\nu}+\dot{n}^{\mu}\dot{n}^{\nu}\nabla_{\nu}n_{\mu} \Big)  -\frac{F_{5\phi}}{2}(K^2-S)\,,
\end{align}
where we have dropped a total derivative term. Finally, we use the definition $\tilde{\mathcal{K}}$ in Eq.~(\ref{defKJ}) and we obtain the final result used in Section~\ref{galileons}:
\begin{align}
\label{Jterm}
G_{5X}\mathcal{J}&=F_5\gamma^{-1}\Big[ \frac{\tilde{\mathcal{K}}}{2}+K^{\mu\nu}n^{\sigma}n^{\rho}R_{\mu\sigma\nu\rho}+\dot{n}^{\sigma}n^{\rho}R_{\sigma\rho}-Kn^{\sigma}n^{\rho}R_{\sigma\rho}  \Big]-\frac{F_{5\phi}}{2}(K^2-S)\,.
\end{align}


\begin{thebibliography}{99}


%%%DE/MG REVIEWS %%%%%%%

%\cite{Sotiriou:2008rp}
\bibitem{Sotiriou:2008rp} 
  T.~P.~Sotiriou and V.~Faraoni,
  %``f(R) Theories Of Gravity,''
  Rev.\ Mod.\ Phys.\  {\bf 82}, 451 (2010)
  %doi:10.1103/RevModPhys.82.451
  [arXiv:0805.1726 [gr-qc]].
  %%CITATION = doi:10.1103/RevModPhys.82.451;%%
  %1424 citations counted in INSPIRE as of 25 Dec 2015

%\cite{Silvestri:2009hh}
\bibitem{Silvestri:2009hh} 
  A.~Silvestri and M.~Trodden,
 %``Approaches to Understanding Cosmic Acceleration,''
  Rept.\ Prog.\ Phys.\  {\bf 72}, 096901 (2009)
  [arXiv:0904.0024 [astro-ph.CO]].
  %%CITATION = ARXIV:0904.0024;%%
  %145 citations counted in INSPIRE as of 19 Oct 2014

%\cite{DeFelice:2010aj}
\bibitem{DeFelice:2010aj} 
  A.~De Felice and S.~Tsujikawa,
  %``f(R) theories,''
  Living Rev.\ Rel.\  {\bf 13}, 3 (2010)
  %doi:10.12942/lrr-2010-3
  [arXiv:1002.4928 [gr-qc]].
  %%CITATION = doi:10.12942/lrr-2010-3;%%
  %951 citations counted in INSPIRE as of 25 Dec 2015

%\cite{Clifton:2011jh}
\bibitem{Clifton:2011jh} 
  T.~Clifton, P.~G.~Ferreira, A.~Padilla and C.~Skordis,
  %``Modified Gravity and Cosmology,''
  Phys.\ Rept.\  {\bf 513}, 1 (2012)
  %doi:10.1016/j.physrep.2012.01.001
  [arXiv:1106.2476 [astro-ph.CO]].
  %%CITATION = doi:10.1016/j.physrep.2012.01.001;%%
  %834 citations counted in INSPIRE as of 25 Dec 2015
	
	%\cite{Tsujikawa:2013fta}
\bibitem{Tsujikawa:2013fta} 
  S.~Tsujikawa,
  %``Quintessence: A Review,''
  Class.\ Quant.\ Grav.\  {\bf 30}, 214003 (2013)
  %doi:10.1088/0264-9381/30/21/214003
  [arXiv:1304.1961 [gr-qc]].
  %%CITATION = doi:10.1088/0264-9381/30/21/214003;%%
  %64 citations counted in INSPIRE as of 20 Nov 2015
	
	%\cite{Deffayet:2013lga}
\bibitem{Deffayet:2013lga} 
  C.~Deffayet and D.~A.~Steer,
  %``A formal introduction to Horndeski and Galileon theories and their generalizations,''
  Class.\ Quant.\ Grav.\  {\bf 30}, 214006 (2013)
  %doi:10.1088/0264-9381/30/21/214006
  [arXiv:1307.2450 [hep-th]].
  %%CITATION = doi:10.1088/0264-9381/30/21/214006;%%
  %36 citations counted in INSPIRE as of 07 Jan 2016
	
	%\cite{Joyce:2014kja}
\bibitem{Joyce:2014kja} 
  A.~Joyce, B.~Jain, J.~Khoury and M.~Trodden,
  %``Beyond the Cosmological Standard Model,''
  Phys.\ Rept.\  {\bf 568}, 1 (2015)
  %doi:10.1016/j.physrep.2014.12.002
  [arXiv:1407.0059 [astro-ph.CO]].
  %%CITATION = doi:10.1016/j.physrep.2014.12.002;%%
  %132 citations counted in INSPIRE as of 25 Dec 2015

	%\cite{Koyama:2015vza}
\bibitem{Koyama:2015vza} 
  K.~Koyama,
  %``Cosmological Tests of Gravity,''
  arXiv:1504.04623 [astro-ph.CO].
  %%CITATION = ARXIV:1504.04623;%%
  %28 citations counted in INSPIRE as of 07 Jan 2016
	
	
	%\cite{Bull:2015stt}
\bibitem{Bull:2015stt} 
  P.~Bull {\it et al.},
  %``Beyond $\Lambda$CDM: Problems, solutions, and the road ahead,''
  arXiv:1512.05356 [astro-ph.CO].
  %%CITATION = ARXIV:1512.05356;%%
  %1 citations counted in INSPIRE as of 12 Jan 2016

%%%%%%%%%%%%% EFT DE/MG  %%%%%%


	%\cite{Gubitosi:2012hu}
\bibitem{Gubitosi:2012hu} 
  G.~Gubitosi, F.~Piazza and F.~Vernizzi,
  %``The Effective Field Theory of Dark Energy,''
  JCAP {\bf 1302}, 032 (2013)
  [JCAP {\bf 1302}, 032 (2013)]
  %doi:10.1088/1475-7516/2013/02/032
  [arXiv:1210.0201 [hep-th]].
  %%CITATION = doi:10.1088/1475-7516/2013/02/032;%%
  %81 citations counted in INSPIRE as of 19 Nov 2015

%\cite{Bloomfield:2012ff}
\bibitem{Bloomfield:2012ff} 
  J.~K.~Bloomfield, E.~E.~Flanagan, M.~Park and S.~Watson,
  %``Dark energy or modified gravity?  An effective field theory approach,''
  JCAP {\bf 1308}, 010 (2013)
  %doi:10.1088/1475-7516/2013/08/010
  [arXiv:1211.7054 [astro-ph.CO]].
  %%CITATION = doi:10.1088/1475-7516/2013/08/010;%%
  %70 citations counted in INSPIRE as of 19 Nov 2015
	
%\cite{Gleyzes:2013ooa}
\bibitem{Gleyzes:2013ooa} 
  J.~Gleyzes, D.~Langlois, F.~Piazza and F.~Vernizzi,
  %``Essential Building Blocks of Dark Energy,''
  JCAP {\bf 1308}, 025 (2013)
  %doi:10.1088/1475-7516/2013/08/025
  [arXiv:1304.4840 [hep-th]].
  %%CITATION = doi:10.1088/1475-7516/2013/08/025;%%
  %75 citations counted in INSPIRE as of 19 Nov 2015
	
		%\cite{Bloomfield:2013efa}
\bibitem{Bloomfield:2013efa} 
  J.~Bloomfield,
  %``A Simplified Approach to General Scalar-Tensor Theories,''
  JCAP {\bf 1312}, 044 (2013)
  %doi:10.1088/1475-7516/2013/12/044
  [arXiv:1304.6712 [astro-ph.CO]].
  %%CITATION = doi:10.1088/1475-7516/2013/12/044;%%
  %39 citations counted in INSPIRE as of 20 Nov 2015
	
		%\cite{Piazza:2013coa}
\bibitem{Piazza:2013coa} 
  F.~Piazza and F.~Vernizzi,
  %``Effective Field Theory of Cosmological Perturbations,''
  Class.\ Quant.\ Grav.\  {\bf 30}, 214007 (2013)
  %doi:10.1088/0264-9381/30/21/214007
  [arXiv:1307.4350 [hep-th]].
  %%CITATION = doi:10.1088/0264-9381/30/21/214007;%%
  %41 citations counted in INSPIRE as of 25 Dec 2015
	
	%\cite{Frusciante:2013zop}
\bibitem{Frusciante:2013zop} 
  N.~Frusciante, M.~Raveri and A.~Silvestri,
  %``Effective Field Theory of Dark Energy: a Dynamical Analysis,''
  JCAP {\bf 1402}, 026 (2014)
  %doi:10.1088/1475-7516/2014/02/026
  [arXiv:1310.6026 [astro-ph.CO]].
  %%CITATION = doi:10.1088/1475-7516/2014/02/026;%%
  %22 citations counted in INSPIRE as of 12 Jan 2016
	
		%\cite{Gleyzes:2014rba}
\bibitem{Gleyzes:2014rba} 
  J.~Gleyzes, D.~Langlois and F.~Vernizzi,
 %``A unifying description of dark energy,''
  Int.\ J.\ Mod.\ Phys.\ D {\bf 23}, no. 13, 1443010 (2015)
  [arXiv:1411.3712 [hep-th]].
  %%CITATION = ARXIV:1411.3712;%%
  %13 citations counted in INSPIRE as of 10 juil. 2015
	
	%\cite{Perenon:2015sla}
\bibitem{Perenon:2015sla} 
  L.~Perenon, F.~Piazza, C.~Marinoni and L.~Hui,
  %``Phenomenology of dark energy: general features of large-scale perturbations,''
  JCAP {\bf 1511}, no. 11, 029 (2015)
  %doi:10.1088/1475-7516/2015/11/029
  [arXiv:1506.03047 [astro-ph.CO]].
  %%CITATION = doi:10.1088/1475-7516/2015/11/029;%%
  %9 citations counted in INSPIRE as of 25 Dec 2015

	
	%%%%%EFT INFLATION & 5e %%%%%%

%\cite{Creminelli:2006xe}
\bibitem{Creminelli:2006xe} 
  P.~Creminelli, M.~A.~Luty, A.~Nicolis and L.~Senatore,
  %``Starting the Universe: Stable Violation of the Null Energy Condition and Non-standard Cosmologies,''
  JHEP {\bf 0612}, 080 (2006)
  %doi:10.1088/1126-6708/2006/12/080
  [hep-th/0606090].
  %%CITATION = doi:10.1088/1126-6708/2006/12/080;%%
  %219 citations counted in INSPIRE as of 12 Jan 2016

%\cite{Cheung:2007st}
\bibitem{Cheung:2007st} 
  C.~Cheung, P.~Creminelli, A.~L.~Fitzpatrick, J.~Kaplan and L.~Senatore,
  %``The Effective Field Theory of Inflation,''
  JHEP {\bf 0803}, 014 (2008)
  [arXiv:0709.0293 [hep-th]].
  %%CITATION = ARXIV:0709.0293;%%
  %376 citations counted in INSPIRE as of 07 Jul 2015

%\cite{Weinberg:2008hq}
\bibitem{Weinberg:2008hq} 
  S.~Weinberg,
  %``Effective Field Theory for Inflation,''
  Phys.\ Rev.\ D {\bf 77}, 123541 (2008)
  [arXiv:0804.4291 [hep-th]].
  %%CITATION = ARXIV:0804.4291;%%
  %180 citations counted in INSPIRE as of 07 juil. 2015
	
	%\cite{Creminelli:2008wc}
\bibitem{Creminelli:2008wc} 
  P.~Creminelli, G.~D'Amico, J.~Norena and F.~Vernizzi,
  %``The Effective Theory of Quintessence: the w<-1 Side Unveiled,''
  JCAP {\bf 0902}, 018 (2009)
  [arXiv:0811.0827 [astro-ph]].
  %%CITATION = ARXIV:0811.0827;%%
  %106 citations counted in INSPIRE as of 07 juil. 2015
	
	
	
	%%%%%%%%%%%%%%%%%% EFT LSS %%%%%%%%%%%%%%%%%%%%%%%%%%%%%%
	%\cite{Park:2010cw}
\bibitem{Park:2010cw} 
  M.~Park, K.~M.~Zurek and S.~Watson,
  %``A Unified Approach to Cosmic Acceleration,''
  Phys.\ Rev.\ D {\bf 81}, 124008 (2010),
   [arXiv:1003.1722 [hep-th]].
  %%CITATION = ARXIV:1003.1722;%%
  %16 citations counted in INSPIRE as of 29 Oct 2013 
	
	  %\cite{Jimenez:2011nn}
\bibitem{Jimenez:2011nn} 
  R.~Jimenez, P.~Talavera and L.~Verde,
  %``An effective theory of accelerated expansion,''
  Int.\ J.\ Mod.\ Phys.\ A {\bf 27}, 1250174 (2012),
  [arXiv:1107.2542 [astro-ph.CO]].
  %%CITATION = ARXIV:1107.2542;%%
  %9 citations counted in INSPIRE as of 29 Oct 2013
  
	  %\cite{Carrasco:2012cv}
\bibitem{Carrasco:2012cv} 
  J.~J.~M.~Carrasco, M.~P.~Hertzberg and L.~Senatore,
 % \emph{The Effective Field Theory of Cosmological Large Scale Structures},
JHEP {\bf 1209}, 082  (2012), 
 [arXiv:1206.2926 [astro-ph.CO]].  
 
  %\cite{Hertzberg:2012qn}
\bibitem{Hertzberg:2012qn}
  M.~P.~Hertzberg,
  %``The Effective Field Theory of Dark Matter and Structure Formation: Semi-Analytical Results,''
  Phys.\ Rev.\ D {\bf 89}, 043521 (2014),
  [arXiv:1208.0839 [astro-ph.CO]].
  %%CITATION = ARXIV:1208.0839;%%
  %11 citations counted in INSPIRE as of 01 May 2014
	
	%\cite{Carrasco:2013mua}
\bibitem{Carrasco:2013mua} 
  J.~J.~M.~Carrasco, S.~Foreman, D.~Green and L.~Senatore,
  %``The Effective Field Theory of Large Scale Structures at Two Loops,''
  JCAP {\bf 1407}, 057 (2014)
  %doi:10.1088/1475-7516/2014/07/057
  [arXiv:1310.0464 [astro-ph.CO]].
  %%CITATION = doi:10.1088/1475-7516/2014/07/057;%%
  %49 citations counted in INSPIRE as of 12 Jan 2016
	
	%\cite{Porto:2013qua}
\bibitem{Porto:2013qua} 
  R.~A.~Porto, L.~Senatore and M.~Zaldarriaga,
  %``The Lagrangian-space Effective Field Theory of Large Scale Structures,''
  JCAP {\bf 1405}, 022 (2014)
  %doi:10.1088/1475-7516/2014/05/022
  [arXiv:1311.2168 [astro-ph.CO]].
  %%CITATION = doi:10.1088/1475-7516/2014/05/022;%%
  %53 citations counted in INSPIRE as of 12 Jan 2016
	
	%\cite{Senatore:2014vja}
\bibitem{Senatore:2014vja} 
  L.~Senatore and M.~Zaldarriaga,
  %``Redshift Space Distortions in the Effective Field Theory of Large Scale Structures,''
  arXiv:1409.1225 [astro-ph.CO].
  %%CITATION = ARXIV:1409.1225;%%
  %11 citations counted in INSPIRE as of 12 Jan 2016
	
		
	%%%%%%%%%%%%%%%%%%%%%%%%%%%
	

		%\cite{Horndeski:1974wa}
\bibitem{Horndeski:1974wa} 
  G.~W.~Horndeski,
  %``Second-order scalar-tensor field equations in a four-dimensional space,''
  Int.\ J.\ Theor.\ Phys.\  {\bf 10}, 363 (1974).
  %%CITATION = IJTPB,10,363;%%
  %253 citations counted in INSPIRE as of 27 Nov 2014
	
	
	%\cite{Deffayet:2009mn}
\bibitem{Deffayet:2009mn} 
  C.~Deffayet, S.~Deser and G.~Esposito-Farese,
  %``Generalized Galileons: All scalar models whose curved background extensions maintain second-order field equations and stress-tensors,''
  Phys.\ Rev.\ D {\bf 80}, 064015 (2009)
  [arXiv:0906.1967 [gr-qc]].
  %%CITATION = ARXIV:0906.1967;%%
  %247 citations counted in INSPIRE as of 27 Nov 2014
	
	%\cite{Gleyzes:2014dya}
\bibitem{Gleyzes:2014dya} 
  J.~Gleyzes, D.~Langlois, F.~Piazza and F.~Vernizzi,
  %``Healthy theories beyond Horndeski,''
  Phys.\ Rev.\ Lett.\  {\bf 114}, no. 21, 211101 (2015)
  [arXiv:1404.6495 [hep-th]].
  %%CITATION = ARXIV:1404.6495;%%
  %68 citations counted in INSPIRE as of 27 Oct 2015
	
	%\cite{Horava:2008ih}
\bibitem{Horava:2008ih}
  P.~Ho\v rava,
  %``Membranes at Quantum Criticality,''
  JHEP {\bf 0903} (2009) 020
  [arXiv:0812.4287 [hep-th]].
  %%CITATION = ARXIV:0812.4287;%%
  %439 citations counted in INSPIRE as of 06 Mar 2015
	
	%\cite{Horava:2009uw}
\bibitem{Horava:2009uw}
  P.~Ho\v rava,
  %``Quantum Gravity at a Lifshitz Point,''
  Phys.\ Rev.\ D {\bf 79} (2009) 084008
  [arXiv:0901.3775 [hep-th]].
  %%CITATION = ARXIV:0901.3775;%%
  %1077 citations counted in INSPIRE as of 06 mar 2015	
	
	
	%%%%%%%%%%%%CAMB CosmoMC %%%%%
	
	\bibitem{CAMB}
\url{http://camb.info} \,.

  %\cite{Lewis:1999bs}
\bibitem{Lewis:1999bs} 
  A.~Lewis, A.~Challinor and A.~Lasenby,
  %``Efficient computation of CMB anisotropies in closed FRW models,''
  Astrophys.\ J.\  {\bf 538}, 473 (2000),
  [astro-ph/9911177].
  %%CITATION = ASTRO-PH/9911177;%%
  %1277 citations counted in INSPIRE as of 06 Dec 2013
  
%\cite{Lewis:2002ah}
\bibitem{Lewis:2002ah}
  A.~Lewis and S.~Bridle,
  %``Cosmological parameters from CMB and other data: A Monte Carlo approach,''
  Phys.\ Rev.\ D {\bf 66}, 103511 (2002)
  [astro-ph/0205436].
  %%CITATION = ASTRO-PH/0205436;%%
  %1312 citations counted in INSPIRE as of 25 Mar 2014 
	

%%%%% EFTCAMB %%%%%%%%%%%%%%

%\cite{Hu:2013twa}
\bibitem{Hu:2013twa} 
  B.~Hu, M.~Raveri, N.~Frusciante and A.~Silvestri,
  %``Effective Field Theory of Cosmic Acceleration: an implementation in CAMB,''
  Phys.\ Rev.\ D {\bf 89}, 103530 (2014)
  [arXiv:1312.5742 [astro-ph.CO]].
  %%CITATION = ARXIV:1312.5742;%%
  %15 citations counted in INSPIRE as of 21 Oct 2014
	
%\cite{Raveri:2014cka}
\bibitem{Raveri:2014cka} 
  M.~Raveri, B.~Hu, N.~Frusciante and A.~Silvestri,
  %``Effective Field Theory of Cosmic Acceleration: constraining dark energy with CMB data,''
  Phys.\ Rev.\ D {\bf 90}, 043513 (2014)
  [arXiv:1405.1022 [astro-ph.CO]].
  %%CITATION = ARXIV:1405.1022;%%
  %5 citations counted in INSPIRE as of 21 Oct 2014

%\cite{Hu:2014sea}
\bibitem{Hu:2014sea} 
  B.~Hu, M.~Raveri, A.~Silvestri and N.~Frusciante,
  %``Exploring massive neutrinos in dark cosmologies with $\scriptsize{EFTCAMB}$/  EFTCosmoMC,''
  Phys.\ Rev.\ D {\bf 91}, no. 6, 063524 (2015)
  %doi:10.1103/PhysRevD.91.063524
  [arXiv:1410.5807 [astro-ph.CO]].
  %%CITATION = doi:10.1103/PhysRevD.91.063524;%%
  %10 citations counted in INSPIRE as of 12 Jan 2016
	
%\cite{Hu:2014oga}
\bibitem{Hu:2014oga} 
  B.~Hu, M.~Raveri, N.~Frusciante and A.~Silvestri,
  %``EFTCAMB/EFTCosmoMC: Numerical Notes v2.0,''
  arXiv:1405.3590 [astro-ph.IM].
  %%CITATION = ARXIV:1405.3590;%%

%\cite{Frusciante:2015maa}
\bibitem{Frusciante:2015maa} 
  N.~Frusciante, M.~Raveri, D.~Vernieri, B.~Hu and A.~Silvestri,
  %``Hořava Gravity in the Effective Field Theory formalism: From cosmology to observational constraints,''
  Phys.\ Dark Univ.\  {\bf 13}, 7 (2016)
  %doi:10.1016/j.dark.2016.03.002
  [arXiv:1508.01787 [astro-ph.CO]].
  %%CITATION = doi:10.1016/j.dark.2016.03.002;%%
  %6 citations counted in INSPIRE as of 16 Jun 2016

%%%%%%%%%%%%%%%%%%%%%%%%%%%%%%%%%%%%%
	
	%\cite{Bellini:2014fua}
\bibitem{Bellini:2014fua} 
  E.~Bellini and I.~Sawicki,
  %``Maximal freedom at minimum cost: linear large-scale structure in general modifications of gravity,''
  JCAP {\bf 1407}, 050 (2014)
  %doi:10.1088/1475-7516/2014/07/050
  [arXiv:1404.3713 [astro-ph.CO]].
  %%CITATION = doi:10.1088/1475-7516/2014/07/050;%%
  %38 citations counted in INSPIRE as of 25 Nov 2015
	
	%\cite{Lesgourgues:2011re}
\bibitem{Lesgourgues:2011re} 
  J.~Lesgourgues,
  %``The Cosmic Linear Anisotropy Solving System (CLASS) I: Overview,''
  arXiv:1104.2932 [astro-ph.IM].
  %%CITATION = ARXIV:1104.2932;%%
  %104 citations counted in INSPIRE as of 12 Jan 2016
	
	%\cite{Bellini:2015xja}
\bibitem{Bellini:2015xja} 
  E.~Bellini, A.~J.~Cuesta, R.~Jimenez and L.~Verde,
  %``Constraints on deviations from {\Lambda}CDM within Horndeski gravity,''
  arXiv:1509.07816 [astro-ph.CO].
  %%CITATION = ARXIV:1509.07816;%%
  %1 citations counted in INSPIRE as of 12 Jan 2016
	
	
	%\cite{Kase:2014cwa}
\bibitem{Kase:2014cwa} 
  R.~Kase and S.~Tsujikawa,
  %``Effective field theory approach to modified gravity including Horndeski theory and Hořava–Lifshitz gravity,''
  Int.\ J.\ Mod.\ Phys.\ D {\bf 23}, no. 13, 1443008 (2015)
  %doi:10.1142/S0218271814430081
  [arXiv:1409.1984 [hep-th]].
  %%CITATION = doi:10.1142/S0218271814430081;%%
  %15 citations counted in INSPIRE as of 19 Nov 2015

	
	%\cite{Gao:2014soa}
\bibitem{Gao:2014soa} 
  X.~Gao,
  %``Unifying framework for scalar-tensor theories of gravity,''
  Phys.\ Rev.\ D {\bf 90}, 081501 (2014)
  %doi:10.1103/PhysRevD.90.081501
  [arXiv:1406.0822 [gr-qc]].
  %%CITATION = doi:10.1103/PhysRevD.90.081501;%%
  %50 citations counted in INSPIRE as of 18 Dec 2015
	



%\cite{Calcagni:2009ar}
\bibitem{Calcagni:2009ar} 
 G.~Calcagni,
  %``Cosmology of the Lifshitz universe,''
JHEP {\bf 0909}, 112 (2009)
[arXiv:0904.0829 [hep-th]].
  %%CITATION = ARXIV:0904.0829;%%
  %293 citations counted in INSPIRE as of 09 juil. 2015

%\cite{Kiritsis:2009sh}
\bibitem{Kiritsis:2009sh} 
 E.~Kiritsis and G.~Kofinas,
  %``Horava-Lifshitz Cosmology,''
 Nucl.\ Phys.\ B {\bf 821}, 467 (2009)
[arXiv:0904.1334 [hep-th]].
  %%CITATION = ARXIV:0904.1334;%%
  %320 citations counted in INSPIRE as of 09 juil. 2015

%\cite{Brandenberger:2009yt}
\bibitem{Brandenberger:2009yt} 
 R.~Brandenberger,
  %``Matter Bounce in Horava-Lifshitz Cosmology,''
Phys.\ Rev.\ D {\bf 80}, 043516 (2009)
[arXiv:0904.2835 [hep-th]].
  %%CITATION = ARXIV:0904.2835;%%
  %248 citations counted in INSPIRE as of 09 juil. 2015



%\cite{Mukohyama:2009gg}
\bibitem{Mukohyama:2009gg} 
 S.~Mukohyama,
  %``Scale-invariant cosmological perturbations from Horava-Lifshitz gravity without inflation,''
JCAP {\bf 0906}, 001 (2009)
[arXiv:0904.2190 [hep-th]].
  %%CITATION = ARXIV:0904.2190;%%
  %198 citations counted in INSPIRE as of 09 juil. 2015
  
%\cite{Cai:2009dx}
\bibitem{Cai:2009dx} 
  R.~G.~Cai, B.~Hu and H.~B.~Zhang,
  %``Dynamical Scalar Degree of Freedom in Horava-Lifshitz Gravity,''
  Phys.\ Rev.\ D {\bf 80}, 041501 (2009)
  [arXiv:0905.0255 [hep-th]].
  %%CITATION = ARXIV:0905.0255;%%
  %133 citations counted in INSPIRE as of 03 août 2015


	
	%\cite{Chen:2009jr}
\bibitem{Chen:2009jr} 
  B.~Chen, S.~Pi and J.~Z.~Tang,
  %``Scale Invariant Power Spectrum in Horava-Lifshitz Cosmology without Matter,''
  JCAP {\bf 0908}, 007 (2009)
  [arXiv:0905.2300 [hep-th]].
  %%CITATION = ARXIV:0905.2300;%%
  %102 citations counted in INSPIRE as of 19 août 2015

	
  
 %\cite{Cai:2010hi}
\bibitem{Cai:2010hi} 
  R.~G.~Cai, B.~Hu and H.~B.~Zhang,
  %``Scalar graviton in the healthy extension of Ho\v{r}ava-Lifshitz theory,''
  Phys.\ Rev.\ D {\bf 83}, 084009 (2011)
  [arXiv:1008.5048 [hep-th]].
  %%CITATION = ARXIV:1008.5048;%%
  %10 citations counted in INSPIRE as of 03 Aug 2015   




%\cite{Carroll:2004ai}
\bibitem{Carroll:2004ai} 
  S.~M.~Carroll and E.~A.~Lim,
  %``Lorentz-violating vector fields slow the universe down,''
  Phys.\ Rev.\ D {\bf 70}, 123525 (2004)
  %doi:10.1103/PhysRevD.70.123525
  [hep-th/0407149].
  %%CITATION = doi:10.1103/PhysRevD.70.123525;%%
  %190 citations counted in INSPIRE as of 07 Jan 2016

%\cite{Zuntz:2008zz}
\bibitem{Zuntz:2008zz} 
  J.~A.~Zuntz, P.~G.~Ferreira and T.~G.~Zlosnik,
  %``Constraining Lorentz violation with cosmology,''
  Phys.\ Rev.\ Lett.\  {\bf 101}, 261102 (2008)
  %doi:10.1103/PhysRevLett.101.261102
  [arXiv:0808.1824 [gr-qc]].
  %%CITATION = doi:10.1103/PhysRevLett.101.261102;%%
  %34 citations counted in INSPIRE as of 07 Jan 2016

	%\cite{Gao:2009ht}
\bibitem{Gao:2009ht} 
 X.~Gao, Y.~Wang, R.~Brandenberger and A.~Riotto,
  %``Cosmological Perturbations in Horava-Lifshitz Gravity,''
Phys.\ Rev.\ D {\bf 81}, 083508 (2010)
[arXiv:0905.3821 [hep-th]].
  %%CITATION = ARXIV:0905.3821;%%
  %120 citations counted in INSPIRE as of 09 Jul 2015


%\cite{Wang:2009yz}
\bibitem{Wang:2009yz} 
 A.~Wang and R.~Maartens,
  %``Linear perturbations of cosmological models in the Horava-Lifshitz theory of gravity without detailed balance,''
Phys.\ Rev.\ D {\bf 81}, 024009 (2010)
[arXiv:0907.1748 [hep-th]].
  %%CITATION = ARXIV:0907.1748;%%
  %146 citations counted in INSPIRE as of 09 juil. 2015

%\cite{Kobayashi:2009hh}
\bibitem{Kobayashi:2009hh} 
 T.~Kobayashi, Y.~Urakawa and M.~Yamaguchi,
  %``Large scale evolution of the curvature perturbation in Horava-Lifshitz cosmology,''
JCAP {\bf 0911}, 015 (2009)
[arXiv:0908.1005 [astro-ph.CO]].
  %%CITATION = ARXIV:0908.1005;%%
  %58 citations counted in INSPIRE as of 09 juil. 2015

	%\cite{Dutta:2009jn}
\bibitem{Dutta:2009jn} 
  S.~Dutta and E.~N.~Saridakis,
  %``Observational constraints on Horava-Lifshitz cosmology,''
  JCAP {\bf 1001}, 013 (2010)
  [arXiv:0911.1435 [hep-th]].
  %%CITATION = ARXIV:0911.1435;%%
  %72 citations counted in INSPIRE as of 17 Aug 2015
	
	


	%\cite{Kobayashi:2010eh}
\bibitem{Kobayashi:2010eh} 
 T.~Kobayashi, Y.~Urakawa and M.~Yamaguchi,
  %``Cosmological perturbations in a healthy extension of Horava gravity,''
JCAP {\bf 1004}, 025 (2010)
[arXiv:1002.3101 [hep-th]].
  %%CITATION = ARXIV:1002.3101;%%
  %39 citations counted in INSPIRE as of 09 Jul 2015
	
	
	%\cite{Dutta:2010jh}
\bibitem{Dutta:2010jh} 
  S.~Dutta and E.~N.~Saridakis,
  %``Overall observational constraints on the running parameter $\lambda$ of Horava-Lifshitz gravity,''
  JCAP {\bf 1005}, 013 (2010)
  [arXiv:1002.3373 [hep-th]].
  %%CITATION = ARXIV:1002.3373;%%
  %48 citations counted in INSPIRE as of 17 Aug 2015
	
			%\cite{Mukohyama:2010xz}
\bibitem{Mukohyama:2010xz}
  S.~Mukohyama,
  %``Horava-Lifshitz Cosmology: A Review,''
  Class.\ Quant.\ Grav.\  {\bf 27} (2010) 223101
  [arXiv:1007.5199 [hep-th]].
  %%CITATION = ARXIV:1007.5199;%%
  %123 citations counted in INSPIRE as of 15 Nov 2014
	
		%\cite{Blas:2012vn}
\bibitem{Blas:2012vn} 
  D.~Blas, M.~M.~Ivanov and S.~Sibiryakov,
  %``Testing Lorentz invariance of dark matter,''
  JCAP {\bf 1210}, 057 (2012)
 % doi:10.1088/1475-7516/2012/10/057
  [arXiv:1209.0464 [astro-ph.CO]].
  %%CITATION = doi:10.1088/1475-7516/2012/10/057;%%
  %22 citations counted in INSPIRE as of 07 Jan 2016
	
		%\cite{Audren:2013dwa}
\bibitem{Audren:2013dwa} 
  B.~Audren, D.~Blas, J.~Lesgourgues and S.~Sibiryakov,
  %``Cosmological constraints on Lorentz violating dark energy,''
  JCAP {\bf 1308}, 039 (2013)
  %doi:10.1088/1475-7516/2013/08/039
  [arXiv:1305.0009 [astro-ph.CO]].
  %%CITATION = doi:10.1088/1475-7516/2013/08/039;%%
  %21 citations counted in INSPIRE as of 07 Jan 2016
	

	
	%\cite{Audren:2014hza}
\bibitem{Audren:2014hza} 
  B.~Audren, D.~Blas, M.~M.~Ivanov, J.~Lesgourgues and S.~Sibiryakov,
  %``Cosmological constraints on deviations from Lorentz invariance in gravity and dark matter,''
  JCAP {\bf 1503}, no. 03, 016 (2015)
  %doi:10.1088/1475-7516/2015/03/016
  [arXiv:1410.6514 [astro-ph.CO]].
  %%CITATION = doi:10.1088/1475-7516/2015/03/016;%%
  %8 citations counted in INSPIRE as of 07 Jan 2016
	


%\cite{Sotiriou:2010wn}
\bibitem{Sotiriou:2010wn} 
  T.~P.~Sotiriou,
  %``Horava-Lifshitz gravity: a status report,''
  J.\ Phys.\ Conf.\ Ser.\  {\bf 283}, 012034 (2011)
  %doi:10.1088/1742-6596/283/1/012034
  [arXiv:1010.3218 [hep-th]].
  %%CITATION = doi:10.1088/1742-6596/283/1/012034;%%
  %119 citations counted in INSPIRE as of 07 Jan 2016
	
		%\cite{Visser:2011mf}
\bibitem{Visser:2011mf} 
  M.~Visser,
  %``Status of Horava gravity: A personal perspective,''
  J.\ Phys.\ Conf.\ Ser.\  {\bf 314}, 012002 (2011)
  %doi:10.1088/1742-6596/314/1/012002
  [arXiv:1103.5587 [hep-th]].
  %%CITATION = doi:10.1088/1742-6596/314/1/012002;%%
  %35 citations counted in INSPIRE as of 07 Jan 2016
	
		%\cite{Barvinsky:2015kil}
\bibitem{Barvinsky:2015kil} 
  A.~O.~Barvinsky, D.~Blas, M.~Herrero-Valea, S.~M.~Sibiryakov and C.~F.~Steinwachs,
  %``Renormalization of Horava Gravity,''
  arXiv:1512.02250 [hep-th].
  %%CITATION = ARXIV:1512.02250;%%
  %1 citations counted in INSPIRE as of 22 Dec 2015
	
	%\cite{Visser:2009fg}
\bibitem{Visser:2009fg} 
  M.~Visser,
  %``Lorentz symmetry breaking as a quantum field theory regulator,''
  Phys.\ Rev.\ D {\bf 80}, 025011 (2009)
  %doi:10.1103/PhysRevD.80.025011
  [arXiv:0902.0590 [hep-th]].
  %%CITATION = doi:10.1103/PhysRevD.80.025011;%%
  %210 citations counted in INSPIRE as of 07 Jan 2016
	
		
		%\cite{Visser:2009ys}
\bibitem{Visser:2009ys} 
  M.~Visser,
  %``Power-counting renormalizability of generalized Horava gravity,''
  arXiv:0912.4757 [hep-th].
  %%CITATION = ARXIV:0912.4757;%%
  %42 citations counted in INSPIRE as of 21 Nov 2014
	
			
	
%\cite{Blas:2009qj}
\bibitem{Blas:2009qj} 
  D.~Blas, O.~Pujolas and S.~Sibiryakov,
  %``Consistent Extension of Horava Gravity,''
  Phys.\ Rev.\ Lett.\  {\bf 104}, 181302 (2010)
  [arXiv:0909.3525 [hep-th]].
  %%CITATION = ARXIV:0909.3525;%%
  %288 citations counted in INSPIRE as of 27 Oct 2015	
	
	
	%\cite{Gleyzes:2014qga}
\bibitem{Gleyzes:2014qga} 
  J.~Gleyzes, D.~Langlois, F.~Piazza and F.~Vernizzi,
  %``Exploring gravitational theories beyond Horndeski,''
  JCAP {\bf 1502}, 018 (2015)
  %doi:10.1088/1475-7516/2015/02/018
  [arXiv:1408.1952 [astro-ph.CO]].
  %%CITATION = doi:10.1088/1475-7516/2015/02/018;%%
  %56 citations counted in INSPIRE as of 23 Nov 2015


%\cite{Piazza:2013pua}
\bibitem{Piazza:2013pua} 
  F.~Piazza, H.~Steigerwald and C.~Marinoni,
  %``Phenomenology of dark energy: exploring the space of theories with future redshift surveys,''
  JCAP {\bf 1405}, 043 (2014)
  %doi:10.1088/1475-7516/2014/05/043
  [arXiv:1312.6111 [astro-ph.CO]].
  %%CITATION = doi:10.1088/1475-7516/2014/05/043;%%
  %27 citations counted in INSPIRE as of 08 Jan 2016

	  
%\cite{Gourgoulhon:2007ue}
\bibitem{Gourgoulhon:2007ue} 
  E.~Gourgoulhon,
  %``3+1 formalism and bases of numerical relativity,''
  gr-qc/0703035 [GR-QC].
  %%CITATION = GR-QC/0703035;%%
  %136 citations counted in INSPIRE as of 15 Nov 2014
	
	
		

	 %\cite{Song:2006ej}
\bibitem{Song:2006ej} 
  Y.~-S.~Song, W.~Hu and I.~Sawicki,
  %``The Large Scale Structure of f(R) Gravity,''
  Phys.\ Rev.\ D {\bf 75}, 044004 (2007),
  [astro-ph/0610532].
  %%CITATION = ASTRO-PH/0610532;%%
  %251 citations counted in INSPIRE as of 04 Dec 2013
  
  %\cite{Pogosian:2007sw}
\bibitem{Pogosian:2007sw} 
  L.~Pogosian and A.~Silvestri,
  %``The pattern of growth in viable f(R) cosmologies,''
  Phys.\ Rev.\ D {\bf 77}, 023503 (2008),
  [Erratum-ibid.\ D {\bf 81}, 049901 (2010)],\,\,\,\,\,\,\,\,\,\,\,\,\,
  [arXiv:0709.0296[astro-ph]].
  %%CITATION = ARXIV:0709.0296;%%
  %163 citations counted in INSPIRE as of 04 Dec 2013  	
  
  
	%\cite{Hu:2016zrh}
\bibitem{Hu:2016zrh} 
  B.~Hu, M.~Raveri, M.~Rizzato and A.~Silvestri,
  %`Testing Hu-Sawicki f(R) gravity with the Effective Field Theory approach,''
  %doi:10.1093/mnras/stw775
  arXiv:1601.07536 [astro-ph.CO].
  %%CITATION = doi:10.1093/mnras/stw775;%%
  %1 citations counted in INSPIRE as of 08 Jul 2016

	
	%\cite{Nicolis:2008in}
\bibitem{Nicolis:2008in} 
  A.~Nicolis, R.~Rattazzi and E.~Trincherini,
  %``The Galileon as a local modification of gravity,''
  Phys.\ Rev.\ D {\bf 79}, 064036 (2009)
  %doi:10.1103/PhysRevD.79.064036
  [arXiv:0811.2197 [hep-th]].
  %%CITATION = doi:10.1103/PhysRevD.79.064036;%%
  %768 citations counted in INSPIRE as of 25 Dec 2015
	
	%\cite{Dvali:2000hr}
\bibitem{Dvali:2000hr} 
  G.~R.~Dvali, G.~Gabadadze and M.~Porrati,
  %``4-D gravity on a brane in 5-D Minkowski space,''
  Phys.\ Lett.\ B {\bf 485}, 208 (2000)
  %doi:10.1016/S0370-2693(00)00669-9
  [hep-th/0005016].
  %%CITATION = doi:10.1016/S0370-2693(00)00669-9;%%
	
	%\cite{Deffayet:2009wt}
\bibitem{Deffayet:2009wt} 
  C.~Deffayet, G.~Esposito-Farese and A.~Vikman,
  %``Covariant Galileon,''
  Phys.\ Rev.\ D {\bf 79}, 084003 (2009)
  %doi:10.1103/PhysRevD.79.084003
  [arXiv:0901.1314 [hep-th]].
  %%CITATION = doi:10.1103/PhysRevD.79.084003;%%
  %414 citations counted in INSPIRE as of 25 Dec 2015
	
	
		%\cite{Creminelli:2010ba}
\bibitem{Creminelli:2010ba} 
  P.~Creminelli, A.~Nicolis and E.~Trincherini,
  %``Galilean Genesis: An Alternative to inflation,''
  JCAP {\bf 1011}, 021 (2010)
  %doi:10.1088/1475-7516/2010/11/021
  [arXiv:1007.0027 [hep-th]].
  %%CITATION = doi:10.1088/1475-7516/2010/11/021;%%
  %167 citations counted in INSPIRE as of 25 Dec 2015
	
	%\cite{Kobayashi:2010cm}
\bibitem{Kobayashi:2010cm} 
  T.~Kobayashi, M.~Yamaguchi and J.~Yokoyama,
  %``G-inflation: Inflation driven by the Galileon field,''
  Phys.\ Rev.\ Lett.\  {\bf 105}, 231302 (2010)
  %doi:10.1103/PhysRevLett.105.231302
  [arXiv:1008.0603 [hep-th]].
  %%CITATION = doi:10.1103/PhysRevLett.105.231302;%%
  %183 citations counted in INSPIRE as of 25 Dec 2015
	
	
%\cite{Burrage:2010cu}
\bibitem{Burrage:2010cu} 
  C.~Burrage, C.~de Rham, D.~Seery and A.~J.~Tolley,
  %``Galileon inflation,''
  JCAP {\bf 1101}, 014 (2011)
  %doi:10.1088/1475-7516/2011/01/014
  [arXiv:1009.2497 [hep-th]].
  %%CITATION = doi:10.1088/1475-7516/2011/01/014;%%
  %139 citations counted in INSPIRE as of 25 Dec 2015
	
	%\cite{Kamada:2010qe}
\bibitem{Kamada:2010qe} 
  K.~Kamada, T.~Kobayashi, M.~Yamaguchi and J.~Yokoyama,
  %``Higgs G-inflation,''
  Phys.\ Rev.\ D {\bf 83}, 083515 (2011)
  %doi:10.1103/PhysRevD.83.083515
  [arXiv:1012.4238 [astro-ph.CO]].
  %%CITATION = doi:10.1103/PhysRevD.83.083515;%%
  %91 citations counted in INSPIRE as of 25 Dec 2015
	
	%\cite{Creminelli:2010qf}
\bibitem{Creminelli:2010qf} 
  P.~Creminelli, G.~D'Amico, M.~Musso, J.~Norena and E.~Trincherini,
  %``Galilean symmetry in the effective theory of inflation: new shapes of non-Gaussianity,''
  JCAP {\bf 1102}, 006 (2011)
  %doi:10.1088/1475-7516/2011/02/006
  [arXiv:1011.3004 [hep-th]].
  %%CITATION = doi:10.1088/1475-7516/2011/02/006;%%
  %107 citations counted in INSPIRE as of 25 Dec 2015
	
		
	
%\cite{Kobayashi:2011nu}
\bibitem{Kobayashi:2011nu} 
  T.~Kobayashi, M.~Yamaguchi and J.~Yokoyama,
  %``Generalized G-inflation: Inflation with the most general second-order field equations,''
  Prog.\ Theor.\ Phys.\  {\bf 126}, 511 (2011)
  %doi:10.1143/PTP.126.511
  [arXiv:1105.5723 [hep-th]].
  %%CITATION = doi:10.1143/PTP.126.511;%%
  %236 citations counted in INSPIRE as of 10 Dec 2015
	
	%\cite{Gao:2011qe}
\bibitem{Gao:2011qe} 
  X.~Gao and D.~A.~Steer,
  %``Inflation and primordial non-Gaussianities of 'generalized Galileons',''
  JCAP {\bf 1112}, 019 (2011)
  %doi:10.1088/1475-7516/2011/12/019
  [arXiv:1107.2642 [astro-ph.CO]].
  %%CITATION = doi:10.1088/1475-7516/2011/12/019;%%
  %79 citations counted in INSPIRE as of 25 Dec 2015
	
	%\cite{DeFelice:2013ar}
\bibitem{DeFelice:2013ar} 
  A.~De Felice and S.~Tsujikawa,
  %``Shapes of primordial non-Gaussianities in the Horndeski's most general scalar-tensor theories,''
  JCAP {\bf 1303}, 030 (2013)
  %doi:10.1088/1475-7516/2013/03/030
  [arXiv:1301.5721 [hep-th]].
  %%CITATION = doi:10.1088/1475-7516/2013/03/030;%%
  %24 citations counted in INSPIRE as of 25 Dec 2015
	
		%\cite{Takamizu:2013gy}
\bibitem{Takamizu:2013gy} 
  Y.~i.~Takamizu and T.~Kobayashi,
  %``Nonlinear superhorizon curvature perturbation in generic single-field inflation,''
  PTEP {\bf 2013}, no. 6, 063E03 (2013)
  %doi:10.1093/ptep/ptt033
  [arXiv:1301.2370 [gr-qc]].
  %%CITATION = doi:10.1093/ptep/ptt033;%%
  %4 citations counted in INSPIRE as of 07 Jan 2016
	
	%\cite{Frusciante:2013haa}
\bibitem{Frusciante:2013haa} 
  N.~Frusciante, S.~Y.~Zhou and T.~P.~Sotiriou,
  %``Gradient expansion of superhorizon perturbations in G-inflation,''
  JCAP {\bf 1307}, 020 (2013)
  %doi:10.1088/1475-7516/2013/07/020
  [arXiv:1303.6628 [astro-ph.CO]].
  %%CITATION = doi:10.1088/1475-7516/2013/07/020;%%
  %3 citations counted in INSPIRE as of 07 Jan 2016
	
		
	%\cite{Chow:2009fm}
\bibitem{Chow:2009fm} 
  N.~Chow and J.~Khoury,
  %``Galileon Cosmology,''
  Phys.\ Rev.\ D {\bf 80}, 024037 (2009)
  %doi:10.1103/PhysRevD.80.024037
  [arXiv:0905.1325 [hep-th]].
  %%CITATION = doi:10.1103/PhysRevD.80.024037;%%
  %161 citations counted in INSPIRE as of 25 Dec 2015
	
	%\cite{Silva:2009km}
\bibitem{Silva:2009km} 
  F.~P.~Silva and K.~Koyama,
  %``Self-Accelerating Universe in Galileon Cosmology,''
  Phys.\ Rev.\ D {\bf 80}, 121301 (2009)
  %doi:10.1103/PhysRevD.80.121301
  [arXiv:0909.4538 [astro-ph.CO]].
  %%CITATION = doi:10.1103/PhysRevD.80.121301;%%
  %125 citations counted in INSPIRE as of 25 Dec 2015
	

	
	%\cite{DeFelice:2010pv}
\bibitem{DeFelice:2010pv} 
  A.~De Felice and S.~Tsujikawa,
  %``Cosmology of a covariant Galileon field,''
  Phys.\ Rev.\ Lett.\  {\bf 105}, 111301 (2010)
  %doi:10.1103/PhysRevLett.105.111301
  [arXiv:1007.2700 [astro-ph.CO]].
  %%CITATION = doi:10.1103/PhysRevLett.105.111301;%%
  %128 citations counted in INSPIRE as of 10 Dec 2015
	
	
	%\cite{Deffayet:2010qz}
\bibitem{Deffayet:2010qz} 
  C.~Deffayet, O.~Pujolas, I.~Sawicki and A.~Vikman,
  %``Imperfect Dark Energy from Kinetic Gravity Braiding,''
  JCAP {\bf 1010}, 026 (2010)
  %doi:10.1088/1475-7516/2010/10/026
  [arXiv:1008.0048 [hep-th]].
  %%CITATION = doi:10.1088/1475-7516/2010/10/026;%%
  %234 citations counted in INSPIRE as of 22 Jan 2016
	
	%\cite{Pujolas:2011he}
\bibitem{Pujolas:2011he} 
  O.~Pujolas, I.~Sawicki and A.~Vikman,
  %``The Imperfect Fluid behind Kinetic Gravity Braiding,''
  JHEP {\bf 1111}, 156 (2011)
  %doi:10.1007/JHEP11(2011)156
  [arXiv:1103.5360 [hep-th]].
  %%CITATION = doi:10.1007/JHEP11(2011)156;%%
  %83 citations counted in INSPIRE as of 22 janv. 2016
	
	
	
	
	

%\cite{Vainshtein:1972sx}
\bibitem{Vainshtein:1972sx} 
  A.~I.~Vainshtein,
  %``To the problem of nonvanishing gravitation mass,''
  Phys.\ Lett.\ B {\bf 39}, 393 (1972).
  %doi:10.1016/0370-2693(72)90147-5
  %%CITATION = doi:10.1016/0370-2693(72)90147-5;%%
  %728 citations counted in INSPIRE as of 10 Dec 2015
	
		
	%\cite{Babichev:2013usa}
\bibitem{Babichev:2013usa} 
  E.~Babichev and C.~Deffayet,
  %``An introduction to the Vainshtein mechanism,''
  Class.\ Quant.\ Grav.\  {\bf 30}, 184001 (2013)
  %doi:10.1088/0264-9381/30/18/184001
  [arXiv:1304.7240 [gr-qc]].
  %%CITATION = doi:10.1088/0264-9381/30/18/184001;%%
  %96 citations counted in INSPIRE as of 07 Jan 2016
	
	
		
	%\cite{Brax:2011sv}
\bibitem{Brax:2011sv} 
  P.~Brax, C.~Burrage and A.~C.~Davis,
  %``Laboratory Tests of the Galileon,''
  JCAP {\bf 1109}, 020 (2011)
  %doi:10.1088/1475-7516/2011/09/020
  [arXiv:1106.1573 [hep-ph]].
  %%CITATION = doi:10.1088/1475-7516/2011/09/020;%%
  %34 citations counted in INSPIRE as of 07 Jan 2016
	
	%\cite{Burrage:2010rs}
\bibitem{Burrage:2010rs} 
  C.~Burrage and D.~Seery,
  %``Revisiting fifth forces in the Galileon model,''
  JCAP {\bf 1008}, 011 (2010)
  %doi:10.1088/1475-7516/2010/08/011
  [arXiv:1005.1927 [astro-ph.CO]].
  %%CITATION = doi:10.1088/1475-7516/2010/08/011;%%
  %44 citations counted in INSPIRE as of 07 Jan 2016
	
	%\cite{Bloomfield:2014zfa}
\bibitem{Bloomfield:2014zfa} 
  J.~K.~Bloomfield, C.~Burrage and A.~C.~Davis,
  %``Shape dependence of Vainshtein screening,''
  Phys.\ Rev.\ D {\bf 91}, no. 8, 083510 (2015)
  %doi:10.1103/PhysRevD.91.083510
  [arXiv:1408.4759 [gr-qc]].
  %%CITATION = doi:10.1103/PhysRevD.91.083510;%%
  %5 citations counted in INSPIRE as of 07 Jan 2016
	
	%\cite{Kase:2013uja}
\bibitem{Kase:2013uja} 
  R.~Kase and S.~Tsujikawa,
  %``Screening the fifth force in the Horndeski's most general scalar-tensor theories,''
  JCAP {\bf 1308}, 054 (2013)
  %doi:10.1088/1475-7516/2013/08/054
  [arXiv:1306.6401 [gr-qc]].
  %%CITATION = doi:10.1088/1475-7516/2013/08/054;%%
  %23 citations counted in INSPIRE as of 07 Jan 2016
	

	%\cite{DeFelice:2011th}
\bibitem{DeFelice:2011th} 
  A.~De Felice, R.~Kase and S.~Tsujikawa,
  %``Vainshtein mechanism in second-order scalar-tensor theories,''
  Phys.\ Rev.\ D {\bf 85}, 044059 (2012)
  %doi:10.1103/PhysRevD.85.044059
  [arXiv:1111.5090 [gr-qc]].
  %%CITATION = doi:10.1103/PhysRevD.85.044059;%%
  %43 citations counted in INSPIRE as of 10 Dec 2015

	
	
	%\cite{Ostrogradski}
\bibitem{Ostrogradski}
M.~Ostrogradski, Mem.~Ac.~St.~Petersburg VI 4, 385 (1850).


%\cite{Sbisa:2014pzo}
\bibitem{Sbisa:2014pzo} 
  F.~Sbisa,
  %``Classical and quantum ghosts,''
  Eur.\ J.\ Phys.\  {\bf 36}, 015009 (2015)
  %doi:10.1088/0143-0807/36/1/015009
  [arXiv:1406.4550 [hep-th]].
  %%CITATION = doi:10.1088/0143-0807/36/1/015009;%%
  %9 citations counted in INSPIRE as of 06 Jan 2016
	
	  
  \bibitem{generalstability}
 A.~De~Felice, N.~Frusciante, G.~Papadomanolakis,
 work in progress.


%\cite{Ade:2015rim}
\bibitem{Ade:2015rim} 
  P.~A.~R.~Ade {\it et al.} [Planck Collaboration],
  %``Planck 2015 results. XIV. Dark energy and modified gravity,''
  arXiv:1502.01590 [astro-ph.CO].
  %%CITATION = ARXIV:1502.01590;%%
  %97 citations counted in INSPIRE as of 06 Jan 2016
  
	

  
%\cite{DeFelice:2011bh}
\bibitem{DeFelice:2011bh} 
  A.~De Felice and S.~Tsujikawa,
  %``Conditions for the cosmological viability of the most general scalar-tensor theories and their applications to extended Galileon dark energy models,''
  JCAP {\bf 1202}, 007 (2012)
  %doi:10.1088/1475-7516/2012/02/007
  [arXiv:1110.3878 [gr-qc]].
  %%CITATION = doi:10.1088/1475-7516/2012/02/007;%%
  %62 citations counted in INSPIRE as of 08 Jan 2016
	
	
	


\end{thebibliography}
\end{document}